\documentclass[preprint,12pt]{elsarticle}

\usepackage[numbers]{natbib}
\usepackage{dsfont}
\usepackage{amsmath,amssymb,amsfonts,dsfont}
\usepackage{algorithmic}
\usepackage{graphicx}
\usepackage{textcomp}
\usepackage{caption}
\usepackage{color,soul}
\usepackage{url}
\usepackage{subfigure}
\usepackage{epstopdf}
\usepackage{multirow}
\usepackage{bm}
\usepackage{ulem}
\usepackage{cancel}
\usepackage[ruled,linesnumbered]{algorithm2e}
\usepackage{makecell}

\begin{document}

\begin{frontmatter}



\title{Network structural perturbation against interlayer link prediction}


\author{Rui Tang$^{1}$, Shuyu Jiang$^{1}$, Xingshu Chen$^{1,2*}$, Wenxian Wang$^{2}$, and Wei Wang$^{2*}$}
\cortext[cor1]{Corresponding Author: chenxsh@scu.edu.cn, wwzqbx@hotmail.com}

\address{1. School of Cyber Science and Engineering, Sichuan University,
Chengdu 610065, China}
\address{2. Cyber Science Research Institute, Sichuan University,
Chengdu 610065, China}

\begin{abstract}
Interlayer link prediction aims at matching the same entities across different layers of the multiplex network. Existing studies attempt to predict more accurately, efficiently, or generically from the aspects of network structure, attribute characteristics, and their combination. Few of them analyze the effects of intralayer links. Namely, few works study the backbone structures which can effectively preserve the predictive accuracy while dealing with a smaller number of intralayer links. It can be used to investigate what types of intralayer links are most important for correct prediction. Are there any intralayer links whose presence leads to worse predictive performance than their absence, and how to attack the prediction algorithms at the minimum cost? To this end, two kinds of network structural perturbation methods are proposed. For the scenario where the structural information of the whole network is completely known, we offer a global perturbation strategy that gives different perturbation weights to different types of intralayer links and then selects a predetermined proportion of intralayer links to remove according to the weights. In contrast, if these information cannot be obtained at one time, we design a biased random walk procedure, local perturbation strategy, to execute perturbation. Four kinds of interlayer link prediction algorithms are carried out on different real-world and artificial perturbed multiplex networks. We find out that the intralayer links connected with small degree nodes have the most significant impact on the prediction accuracy. The intralayer links connected with large degree nodes may have side effects on the interlayer link prediction. To download the data and code, see https://github.com/MollyShuu/NSP-KBS2022

\end{abstract}


\end{frontmatter}



\section{Introduction}
Multiplex networks can be used to describe complex systems with multiple kinds of interactions~\cite{watts1998collective,Barabasi1999-BA,li2022competing,zhang2018scalable}. For instance, people are often involved in numerous social network applications~\cite{tang2021learning}. The friendships in each application can create a layer of the multiplex social network. The accounts are nodes, and the friendships in an application are intralayer links. The relationships that account across different applications belonging to the same user are represented as interlayer links. A language expression can be represented by a network where words are represented as nodes and co-occurrence of words in the same sentences are represented as links~\cite{cancho2001small,xuan2009node}. Multiple language networks can constitute a multiplex language network where interlayer links exist between words describing the same meaning in different languages. In addition, multiple transportation networks, communication networks, and information spreading networks~\cite{pan2021optimal,wang2021generalized,tang2021susceptible} can also be represented as a multiplex network.

Interlayer link prediction in the multiplex network, which is also called network alignment~\cite{kelley2003conserved,singh2008global}, graph matching~\cite{xu2019gromov,fey2020deep}, anchor link prediction~\cite{ManTong2016-IJCAI}, etc., aims at matching the same entities across different layers based on the structural or feature information. It has been widely applied in many fields, including bioinformatics, online social network analysis, cross-lingual knowledge graph construction, and so forth. For example, in bioinformatics, describing and analyzing protein-protein interactions of different species is the foundation to understand biological processes. At the same time, interlayer link prediction can be used to guide the transfer of biological knowledge from well-known to poorly-studied species across aligned protein-protein interaction networks~\cite{vijayan2017alignment}. In online social network analysis, we can leverage interlayer link prediction to identify whether the accounts in different social network applications belong to the same person~\cite{tang2020interlayer,tangrui2021interlayer}. This study has been favored by researchers, platform holders, e-commerce people, and others since it could help them to improve the understanding of information diffusion across distinct social network platforms, to construct more complete online personality profiles, and to recommend products or friends that can better meet customers' needs and wants.

Existing studies focus on predicting interlayer links more accurately, efficiently, or generically with or without prior interlayer links from the aspects of network structure, attribute characteristics, and the combination of them~\cite{tang2020interlayer}. These methods usually focus on improving the capability of capturing predicting clues~\cite{ZhouFan2018,fu2020deep}, adapting to the case with more than two layers~\cite{zhang2019reconciling,ChuXiaokai2019-www}, or optimizing the efficiency~\cite{WangYongqing2019-www,chen2020multi}, etc. Few of them have analyzed the effects of intralayer links on interlayer link prediction. Namely, few works study the backbone structures~\cite{serrano2009extracting,zhang2013extracting} which can effectively preserve the predictive accuracy while dealing with a smaller number of intralayer links. A common intuition is that more structural information results in more successful interlayer link prediction. Is this intuition correct?
Meanwhile, intralayer links have different characteristics. Some of them are connected to the nodes with high degree, while some others are connected to the nodes with low degree. What types of intralayer links are most important for correct prediction? Are there any intralayer links whose presence leads to worse predictive performance than their absence? On the one hand, the research on these problems will help to find the most effective backbone information in the network and effectively improve the accuracy of the interlayer link prediction algorithms. Furthermore, it is helpful for us to remove the redundant and noise information in the network and get a pure network for interlayer link prediction. In addition, from the perspective of attacking prediction algorithms to protect privacy, the most effective intralayer links can be removed or hidden in some ways to interfere with the algorithm's prediction at the minimum cost.

In this paper, two kinds of network structural perturbation methods are proposed to study the influence of structural information on interlayer link prediction and find ways for disturbing interlayer link prediction algorithms hence protecting user privacy. The contributions are as follows:

\begin{itemize}
\item A global perturbation strategy is proposed when the global information of the network structure can be obtained. By controlling the parameters, the strategy gives different perturbation weights to different types of intralayer links. It then selects a predetermined proportion of intralayer links to remove according to the probability distribution of the weights.

\item For the scenario where the structural information of the whole network cannot be obtained at one time, we design a biased random walk procedure called local perturbation strategy. This strategy randomly selects a node as the starting node and then calculates the perturbation weights of all intralayer links linked to it based on the degree property of the possible next-hop node. According to the weights, the local perturbation strategy will walk to the next node and remove the intralayer link connected between these two nodes. This strategy will perform the same steps until a predetermined proportion of intralayer links are removed.

\item Experiments are carried out on real-world network datasets and artificial network datasets. We find that the intralayer links connected with low degree nodes have the most significant impact on the prediction accuracy of interlayer links. Moreover, the intralayer links connected with large degree nodes may have side effects on the interlayer link prediction. Therefore, it is not that the more intralayer links are, the better the accuracy of interlayer link prediction is.

\end{itemize}

\section{Related Works}
This section introduces the related works from the two aspects of interlayer link prediction and network structure perturbation.

\subsection{Interlayer link prediction}
Interlayer link prediction in multiplex networks aims at matching the same entities across different layers, which is a common task in many areas ranging from bioinformatics, online social network analysis to cross-lingual knowledge graph construction. It is typically addressed by using feature information of the nodes or structure information of the whole network. According to the information used, it is mainly divided into feature-based methods, structure-based methods, and a combination of them.

The feature-based methods predict interlayer links by calculating the similarity of nodes' features in different layers or using the machine learning models trained by the features of the observed interlayer node pairs. Such methods are mainly applied to identifying whether the accounts in different social network applications belong to the same person. The features are extracted from the user profile, user-generated contents, and check-in data, etc. The username is the most basic information in a user profile. Refs.~\cite{Zafarani2009,Perito2011,Zafarani2013,li2019matching} explored how to use it to matching accounts of the same user across multiple social network applications. Another feature, the publicly available photo, matches the same individuals in Ref.~\cite{Acquisti2014}. Since only a few people tend to put personal photos on multiple social network applications, using images is not widespread. The authors of Refs.~\cite{Iofciu2011,Carmagnola2009,MuXin-KDD2016} tried to leveraging other types of profile information such as gender, address, and work experience to improve the performance of identifying. Instead of extracting features from the user profile, Goga et al.~\cite{Goga2013a} focus on capture characteristics from user-generated contents. They examined the geographic locations, time patterns, and the writing style of user-generated content on three famous social network applications. Liu et al.~\cite{LiuSiyuan2014} extracted the users' topical interest and language style from the user-generated contents to measure the similarity of accounts across social network applications. Mobile social network application users can share their locations by the service of "check-in". Refs.~\cite{Riederer2016,chen2018effective,feng2019dplink} explored ways of obtaining trajectory information from the "check-in" data to linking the interlayer nodes.

The structure-based methods can be categorized into local-structure-based and global-structure-based methods, according to the completeness and connectivity of network structures~\cite{ShuKai2017}. Local-structure-based methods study the contribution of neighbors within k-hop. Narayanan et al.~\cite{narayanan2009anonymizing} proposed a one-hop graph-theoretic model to re-identify the same entities across two social networks. Xuan et al.~\cite{xuan2009node} described an algorithm to calculate the similarities between nodes through their links to several pairs of preliminarily revealed matched nodes. They found that the structure of the layers may significantly influence the final results. Furthermore, in Ref~\cite{xuan2010iterative}, the authors proposed an iterative version to improve the performance and two matched nodes' selection strategies to reduce time complexity. Zhou et al.~\cite{ZhouXiaoping2016} proved a large number of users shared the same friends across social network applications and proposed a semisupervised scheme that counts the number of shared friends iteratively as the matching degree of the unmatched pairs. The pair possessing the maximum matching degree would be chosen as matched pair. All of the above studies calculated the matching degree from the 1-hop neighbors. To investigate the contribution of $k$-hop ($k>1$) neighbors, Li et al.~\cite{li2020exploiting} crawled ground-truth relationship networks across three social network applications and analyzed the similarities of $k$-hop neighbors. Similar studies could be found in Refs~\cite{tang2020interlayer,korula2014efficient}.

Unlike local-structure-based methods, global-structure-based methods determine whether a node pair has an interlayer link based on the information of the entire network. Singh et al.~\cite{singh2007pairwise} assumed that node $v_i^\alpha$ in layer $G^\alpha$ should be matched with node $v_j^\beta$ in layer $G^\beta$ if and only if the neighbors of nodes $v_i^\alpha$ and $v_j^\beta$ are good match. Based on this assumption, they designed an algorithm analogous to PageRank, named as IsoRank, to align two networks. Later, they expanded IsoRank to IsoRankN in Ref.~\cite{singh2008global} to deal with the situation that the number of layers is more than two. Bayati et al.~\cite{bayati2009algorithms} modified IsoRank to a distributed version for sparse networks. In Ref.~\cite{ZhuYuanyuan2012}, the interlayer link prediction problem was transformed into a maximum common subgraph problem. Zhang et al.~\cite{ZhangSi2016-KDD} developed a family of unsupervised algorithms to align the network with node and edge attribute information. A node pair can be aligned if the topology consistency, node attribute consistency, and edge attribute consistency are held simultaneously. Many researchers used network embedding techniques~\cite{TangJian2015,mo2021effective} to predict the interlayer links. The nodes in each layer are embedded and unified into a density, latent representation space as a low-dimensional vector. The interlayer links are predicted based on the vector distance of unmatched nodes. PALE~\cite{ManTong2016-IJCAI} and IONE~\cite{LiuLi2016} were the typical two works of this type. Then, further studies such as Refs.~\cite{ZhouFan2018,WangYongqing2019-www,ChuXiaokai2019-www,li2019adversarial} were carried out from the aspects of improving accuracy, saving time, and applying on more layer scenarios, etc.

Since structure and feature describe different aspects of nodes, it is critical to integrate them. Kong et al.~\cite{KongXiangnan2013} predicted interlayer links by extracting heterogeneous features from structural, spatial, temporal, and text information. Zhou et al.~\cite{zhou2019translink} proposed an approach called TransLink to capture information of network structure, behavior, and contents and embedded nodes into a unified representation space. Ren et al.~\cite{ren2020banana} designed an end-to-end framework to leverage location and structure data for the network alignment.
\subsection{Network Structural Perturbation}
The perturbation problem of network structure is proposed in the single complex network. Two reasons motivate researchers to pay attention to it. On the one hand, different link prediction methods would achieve different precisions in the same network. Thus, the link predictability associated with the intrinsic regularity of the network can not be reflected by the precision~\cite{lu2015toward}. The network structural perturbation approaches were developed to measure the link predictability, which characterizes the inherent difficulty of link prediction. L\"{u} et al.~\cite{lu2015toward} proposed the structural consistency index to reveal the intrinsic link predictability and designed an eigenvalue perturbation method to quantify the consistency level.
Similarly, Xian et al.~\cite{xian2020netsre} proposed a structural perturbation algorithm to measure and regulate link predictability. On the other hand, sophisticated and widespread online network analysis tools raise security and privacy-related concerns of the general public and some organizations~\cite{waniek2019hide}. For example, cybercriminals can leverage link prediction algorithms to predict possible financial connections, while competitors can analyze commercial ties. To mitigate such threats and protect link privacy, it is urgent to develop network structure perturbation methods systematically. Yu et al.~\cite{yu2019target} proposed two perturbation methods from the perspective of heuristics and evolution respectively to hide the sensitive links from being predicted. Lin et al.~\cite{lin2020adversarial} presented a graph neural network-based framework to attack against link prediction algorithms. It makes the predictive algorithms worse than random guesses. Xian et al.~\cite{xian2021deepec} proposed a deep architecture-based framework to analyze the vulnerability of link prediction. The framework perturbed the network structure to adversarial attack the link prediction methods. The adversarial attack models were not only used to attack the link prediction algorithms, Zhang et al.~\cite{zhang2020adversarial} used it to perturb the network structure to attack network matching.

\section{Preliminaries and problem}
In general, we can use a graph $G(V,E)$ to represent a network, where $V$ is a set of nodes and $E\subseteq V \times V$ is a set of edges. Networks can be partitioned into directed and undirected by whether or not the edges have direction and weighted and unweighted by whether or not the edges have weights. We only consider undirected unweighted networks in this paper. We can use $\bm {\mathrm{A}}\in\mathbb{R}^{n\times n}$ to represent the adjacency matrix of $G$. If there exists an edge between nodes $v_i$ and $v_j$, $\bm {\mathrm{A}}(i,j)=1$. Other wise, $\bm {\mathrm{A}}(i,j)=0$. Obviously, $\bm {\mathrm{A}}(i,j)=\bm {\mathrm{A}}(j,i)$ in undirected networks.

\textbf{Multiplex network: }
A multiplex network is a pair $\mathcal{M}=(g,c)$, where $g={\{ G^\alpha | \alpha \in \{ 1,\dots,n_m \} \}} $ is the set of different networks $G^\alpha(V^\alpha,E^\alpha)$ referring to as layers of $\mathcal{M}$ and
\begin{equation}
 c={\{E^{\alpha \beta}\subseteq V^\alpha \times V^\beta | \alpha,\beta\in\{1,\dots,n_m\},\alpha\neq\beta\}}
\end{equation}
is the set of interrelationships between nodes across different layers. The edges in a layer $G^\alpha(V^\alpha,E^\alpha)$, i.e. the elements of $E^\alpha$ are called intralayer links while the edges across different layers, i.e. the elements of $E^{\alpha \beta}$ are referred as interlayer links or interlayer node pairs.

\begin{figure} [!t]
    \centering
    \includegraphics[width=0.5\textwidth]{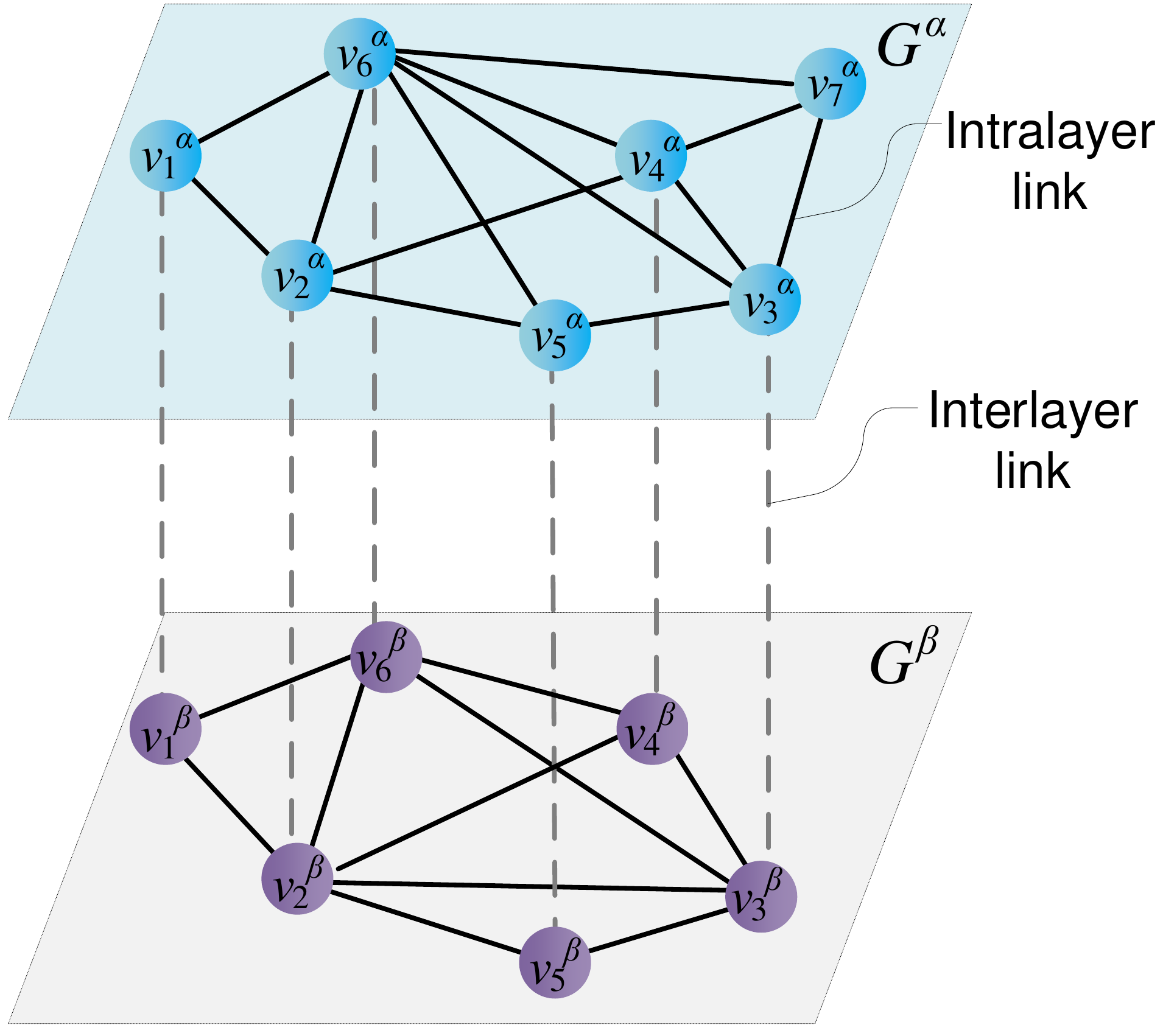}
    \caption{Example of the multiplex network. There is a two-layer multiplex network. Layer $G^\alpha$ has seven nodes while layer $G^\beta$ has six nodes. The solid black lines between nodes in the same layer are the intralayer links, which are used to represent the interactions between nodes. The dotted gray lines between nodes across different layers are the interlayer links, which are used to represent that each pair of nodes correspond to the same underlying entity.}
    \label{pic:multiplexnetwork}
\end{figure}

\textbf{Interlayer link prediction: }
Supposing that we have an observed multiplex network $\mathcal{M^T}=(g,c^T)$ with fixed nodes and intralayer links in each layer $G^\alpha \in g$, interlayer link prediction aims at finding the unobserved interlayer links based on $\mathcal{M^T}=(g,c^T)$ to generate a predicted multiplex network for approximating the true underlying multiplex network $\mathcal{M}$. Namely, the interlayer link prediction is to determine whether any two unmatched nodes $v^\alpha_i$ and $v^\beta_j$ across different layers have an interlayer link, which can be defined as
\begin{equation}
f(v^\alpha_i,v^\beta_j) = \left\{
\begin{array}{ll}
1, & \mbox{if $e^{\alpha\beta}_{ij}$ exists} \\
0, & \mbox{otherwise}
\end{array},
\right.
\label{eq:objectivefunction}
\end{equation}
where $f(v^\alpha_i,v^\beta_j) = 1$ represents that an iterlayer link exists between unmatched nodes $v^\alpha_i$ and $v^\beta_j$.

\textbf{Network structural perturbation: }
Network structural perturbation is inspired by the perturbation theory in quantum mechanics, where an additional perturbing Hamiltonian representing a weak disturbance is added to a known simple system~\cite{cao2020link}. Recently, this theory is leveraged to characterize complex networks. Given a network $G$ with its adjacency matrix $\bm {\mathrm{A}}$, network structural perturbation aims to obtain a perturbed network $G^p$ with its adjacency matrix $\bm {\mathrm{A}}^p=\bm {\mathrm{A}}\pm \Delta \bm {\mathrm{A}}$. The perturbation methods contain removing existing links, adding non-existing links, or switching links~\cite{xian2021towards}.

\textbf{Interlayer link prediction perturbation: }
For a given multiplex network $\mathcal{M}=(g,c)$, interlayer link prediction perturbation aims to generate a perturbed multiplex network $\mathcal{M}^p$ to replace $\mathcal{M}$ to make interlayer link prediction algorithm $f$ fail to predict unobserved interlayer links in $c^p$ accurately.

This paper focuses on perturbing any one layer of the multiplex network by removing the existing intralayer links with different structural properties since intralayer links only exist between nodes belonging to the same layer.

\section{Methodology}
In the process of perturbation, two situations may be faced as follows: (i) Global information about the network to be perturbed is available, and perturbation strategy can be formulated based on the global information available. For example, for social network platform holders they can analyze the whole network on the platform and perturb the specified intralayer links based on the complete network information. (ii) The executor of perturbation can only obtain local information about the network to be perturbed and formulate perturbation strategies based on the local information. For example, third-party organizations that do not belong to the platform holder cannot obtain the intralayer links of the entire network at one time. They can only start from a node in the network and gradually get the intralayer links of surrounding nodes.

We design a global perturbation strategy and a local perturbation strategy to implement perturbation of the network in both cases, respectively.

\subsection{Global Perturbation Strategy}
Suppose the executor of perturbation has global information about the network to be perturbed. In that case, it can analyze the whole network using certain strategies and perturb the specified intralayer links. Many real-world networks exhibit a scale-free property~\cite{Newman2003}. The degree distribution of the nodes in these networks follows a power law. This means that many nodes with a low degree and a small number of nodes with a high degree in these real-world networks. Zhou et al.~\cite{ZhouTao2011} investigate Flickr, Delicious, Twitter, and YouTube, four famous social network applications, and revealed that all of these applications exhibit the scale-free property. Based on this, to distinguish different types of intralayer links based on the network structure, we simply classify the nodes as (i) \textbf{n}ode \textbf{t}ype \textbf{(NT) 1}: node with low degree; (ii) \textbf{NT2}: node with medium degree; (iii) \textbf{NT3}: node with high degree. Besides, the intralayer links are simply classified as (i) intralayer \textbf{l}ink \textbf{t}ype in \textbf{g}lobal strategy \textbf{(GLT) 1}: intralayer link between a node with high degree and a node with low degree (NT3-NT1); (ii) \textbf{GLT2}: intralayer link between a node with high degree and a node with medium degree (NT3-NT2); (iii) \textbf{GLT3}: intralayer link between a node with high degree and a node with high degree (NT3-NT3); (iv) \textbf{GLT4}: intralayer link between a node with medium degree and a node with low degree (NT2-NT1); (v) \textbf{GLT5}: intralayer link between a node with medium degree and a node with medium degree (NT2-NT2); (vi) \textbf{GLT6}: intralayer link between a node with low degree and a node with low degree (NT1-NT1). The example of different types of nodes and intralayer links are shown in Fig.~\ref{pic:intralayer_link_types}. Two questions need to be answered once the categories have been identified:
\begin{figure} [!t]
    \centering
    \includegraphics[width=0.8\textwidth]{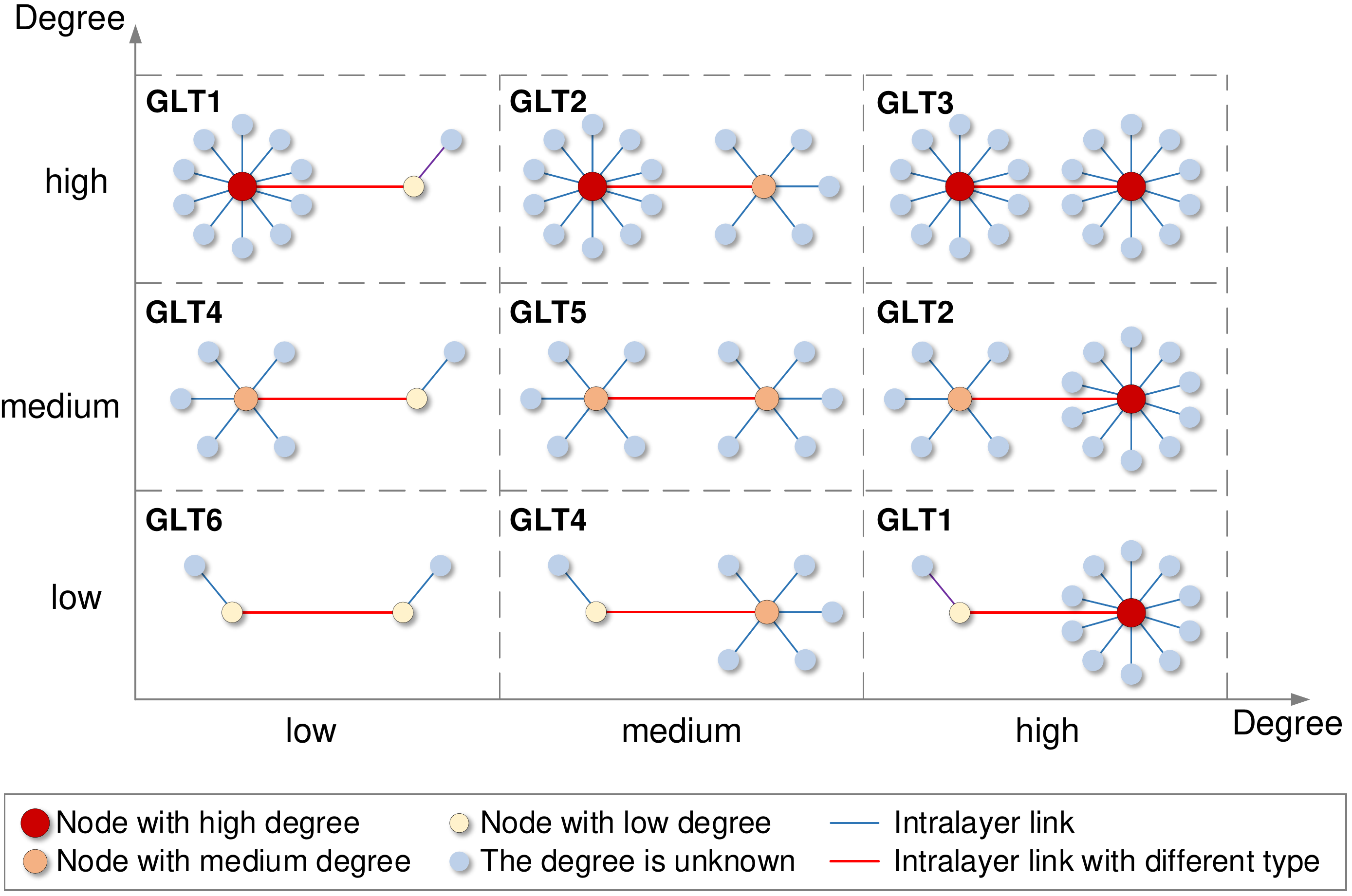}
    \caption{Different types of nodes and intralayer links in a network}
    \label{pic:intralayer_link_types}
\end{figure}

What types of intralayer links contribute more to the prediction of interlayer links? Do all intralayer links play a positive role in correct prediction? We propose a global perturbation strategy to investigate these problems.

\begin{algorithm}
\caption{Global perturbation strategy}
\KwIn{adjacent matrix $\bm {\mathrm{A}}$, parameters $\eta$ and $\xi$, number of nodes $n$}
\KwOut{perturbed adjacent matrix $\bm {\mathrm{A}}$}
{
    $\bm {\mathrm{D}}[n][n] \gets \{0\}$;\\
    \For{$i = 1 \to n$}{
        $\bm {\mathrm{D}}[i][i] \gets \sum\limits_{a=1}^{n}{\bm {\mathrm{A}}[i][a]}$;\\
    }
    $\bm {\mathrm{I}}[n][n]\gets \{1\}$;\\
    $\bm{\mathrm{B}}=\bm{\mathrm{A}}\odot((\bm{\mathrm{D}}\bm{\mathrm{I}})^\eta+((\bm{\mathrm{D}}\bm{\mathrm{I}})^\eta)^T)$;\\
    $\bm{\mathrm{W}}\gets\frac{\bm{\mathrm{B}}}{\sum\limits_{a=1}^{n}\sum\limits_{b=1}^{n} {\bm{\mathrm{B}}[a][b]}}$;\\
    \For{$i = 1 \to n\xi$}{
        Choosing a position $(i,j)$ based on the perturbation wight matrix $\bm{\mathrm{W}}$;\\
        $\bm {\mathrm{A}}[i][j] \gets 0$;\\
        $\bm {\mathrm{A}}[j][i] \gets 0$;\\
    }
    return $\bm {\mathrm{A}}$\\
}
\end{algorithm}
Intralayer link is used to describe the relationship or interaction between two nodes within a same layer. It only exists between nodes belonging to the same layer. To clearly investigate the effects of intralayer links to interlayer link prediction, we randomly choose a layer to be perturbed, leaving the other layer unchanged. For ease of description, later on, we denote the layer to be perturbed directly by the symbol $G(V,E)$. Denoting $n=|V|$ denote the total number of nodes in the network $G$ , for any intralayer link $e_{ij}$ in $G$, we compute its perturbation weight $w_{ij}$ by the following formula:
\begin{equation}
w_{ij}=\frac{ \bm {\mathrm{A}}(i,j) \cdot ({d_i}^\eta + {d_j}^\eta)}{\sum\limits_{a=1}^{n}\sum\limits_{b=1}^{n} {\bm {\mathrm{A}}(a,b) \cdot ({d_a}^\eta + {d_b}^\eta)}}.
\label{eq_wight_global}
\end{equation}

In Eq.~(\ref{eq_wight_global}), ${d_i}^\eta$ and ${d_j}^\eta$ are the degrees of node $v_i$ and $v_j$ to the power $\eta$, respectively. $({d_i}^\eta + {d_j}^\eta)$ is the sum of these two values, which is called degree index for the convenience of description. The purpose of $\bm {\mathrm{A}}(i,j)$ multiplying $({d_i}^\eta + {d_j}^\eta)$ is to make the weights between nodes without intralayer links equaling to 0. The denominator is the sum of the degree indices for all intralayer links, ensuring the sum of weights for all intralayer links equaling to 1. After obtaining the perturbation weights for all intralayer links, we can perturb the network by selecting and removing a certain proportion of $\xi$ intralayer links based on the weights. The backbone structure for interlayer link prediction can be detected by changing $\eta$ since different intralayer links will achieve different weights under different $\eta$. For example, when $\eta$ is a small negative number, the intralayer link connected to nodes with a low degree would obtain a larger weight while obtaining a smaller weight when $\eta$ is a large positive number.

Eq.~(\ref{eq_wight_global}) introduces the calculation of the weight of a single intralayer link. Is it possible to calculate the weights of all the intralayer links at once? How to implement? We can solve this problem by the following method. Denoting the diagonal matrix $\bm {\mathrm{D}}\in\mathbb{R}^{n\times n}$ be the degree matrix of network $G$, its diagonal element $\bm {\mathrm{D}}(i,i)$ represents the degree of node $v_i$ and can be computed by
\begin{equation}
\bm {\mathrm{D}}(i,i)=\sum\limits_{a=1}^{n}{\bm {\mathrm{A}}(i,a)}.
\label{eq_degree_node}
\end{equation}
In matrix $\bm {\mathrm{D}}$, the values of other elements except for the diagonal elements are 0. Denoting $\bm {\mathrm{I}}\in\mathbb{R}^{n\times n}$ be the matrix with all elements having value 1, the matrix of perturbation weights for the entire network can be expressed as
\begin{equation}
\bm{\mathrm{W}}=\frac{\bm{\mathrm{A}}\odot((\bm{\mathrm{D}}\bm{\mathrm{I}})^\eta+((\bm{\mathrm{D}}\bm{\mathrm{I}})^\eta)^T)}{\sum\limits_{a=1}^{n}\sum\limits_{b=1}^{n}{\bm{\mathrm{A}}(a,b) \cdot (\bm{\mathrm{D}}(a,a)^\eta + \bm{\mathrm{D}}(b,b)^\eta)}},
\label{eq_wight_mat}
\end{equation}
where $\odot$ is the Hadamard product. Denoting $\bm{\mathrm{B}}=\bm{\mathrm{A}}\odot((\bm{\mathrm{D}}\bm{\mathrm{I}})^\eta+((\bm{\mathrm{D}}\bm{\mathrm{I}})^\eta)^T)$, Eq.~\ref{eq_wight_mat} can be abbreviated by
\begin{equation}
\bm{\mathrm{W}}=\frac{\bm{\mathrm{B}}}{\sum\limits_{a=1}^{n}\sum\limits_{b=1}^{n} {\bm{\mathrm{B}}(a,b)}}.
\label{eq_wight_mat_simple}
\end{equation}
By Eq.~(\ref{eq_wight_mat_simple}), the perturbation weights for all intralayer links in the network $G$ can be directly obtained. The pseudo-code for the global perturbation strategy is shown in Algorithm 1.

\subsection{Local perturbation strategy}
Suppose the executor of perturbation could only obtain the local information of the network to be perturbed starting from a node. In that case, the global perturbation strategy will not be able to complete the perturbation. We design a biased random walk procedure called the local perturbation strategy to deal with this situation. It randomly selects a node as the starting node. It then calculates the perturbation weights, which can be seen as the transition probabilities of all intralayer links linked to the current node based on the degree property of the possible next-hop nodes. The walker will walk to the next-hop node according to the perturbation weights. After arriving in the next-hop node, the local perturbation strategy will remove the intralayer link between the current and the last hop node. This strategy will perform the same steps until a predetermined proportion of intralayer links are removed.

\begin{figure} [!t]
    \centering
    \includegraphics[width=0.8\textwidth]{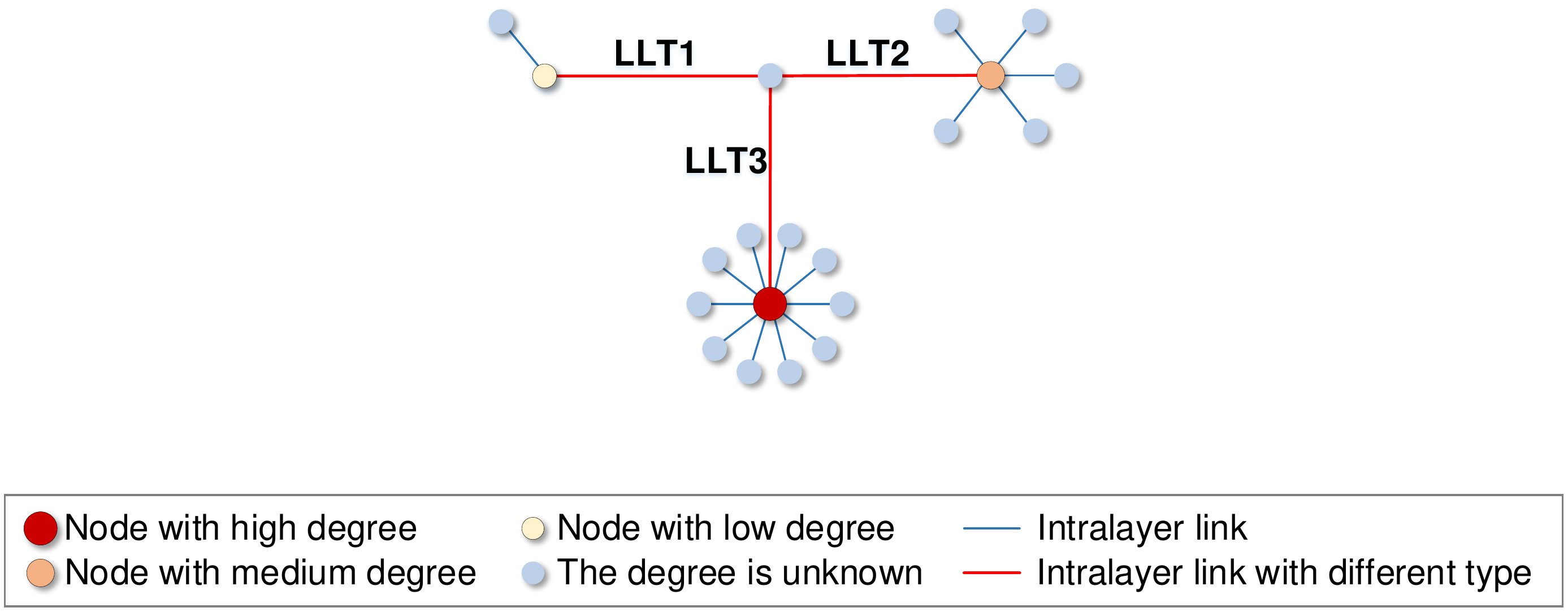}
    \caption{Example of different types of interlayer links linked to a node.}
    \label{pic:intralayer_link_types_local}
\end{figure}

For any node $v_i$ in network $G$, the intralayer links connected to it can be simply classified as (i) intralayer \textbf{l}ink \textbf{t}ype in \textbf{l}ocal strategy \textbf{(LLT) 1}: intralayer link between a node with any degree and a node with low degree; (ii) \textbf{LLT2}: intralayer link between a node with any degree and a node with medium degree; (iii) \textbf{LLT3}: intralayer link between a node with any degree and a node with high degree, as shown in Figure ~\ref{pic:intralayer_link_types_local}.

Similar with the global strategy, we choose a layer to be perturbed, leaving the other layer unchanged. Starting from any node $v_i$ in the network $G$, we can calculate the perturbation weights of all the intralayer links linked to the current node by the following equation:
\begin{equation}
w_{ij}=\frac{\bm {\mathrm{A}}(i,j) \cdot ({d_i}^\eta + {d_j}^\eta)}{\sum\limits_{a=1}^{n} {\bm {\mathrm{A}}(i,a) \cdot ({d_i}^\eta + {d_a}^\eta)}}.
\label{eq_wight_local}
\end{equation}

Eq.~(\ref{eq_wight_local}) describes how the perturbation weight of a single intralayer link is calculated. For any node $v_i$, we can obtain the perturbation weights of all intralayer links connected to it by the following method.

Denoting $\bm{\mathrm{b}}_i=[\bm{\mathrm{B}}(i,1),\cdots,\bm{\mathrm{B}}(i,a),\cdots,\bm{\mathrm{B}}(i,n)]^T$, i.e. the vector $\bm{\mathrm{b }}_i$ is a column vector formed by transposing the $i$th row of the matrix $\bm{\mathrm{B}}$, the vector formed by the weights of all the intralayer links connected to the node $v_i$ can be expressed as
\begin{equation}
\bm{\mathrm{w}}_i=\frac{\bm{\mathrm{b}}_i}{\sum\limits_{a=1}^{n}{\bm{\mathrm{B}}(i,a)}}.
\label{eq_wight_simple_local}
\end{equation}

The perturbation weights could be seen as the transition probabilities hence guiding the walk. The walker will walk to the next-hop node according to the perturbation weights. After arriving at the next-hop node, the local perturbation strategy will remove the intralayer link between the current node $v_j$ and the last node. Then, the local perturbation strategy will perform the same steps until a proportion of $\xi$ of the intralayer links are removed. It is noteworthy that if the local perturbation strategy walks to a node $v_j$ with degree 1, it will have no intralayer link connected to $v_j$ after removing the walked intralayer link. The walker could not continue. The local perturbation strategy can be re-executed by randomly selecting a node in the network as the starting node to tackle this problem. Similar to the global strategy, the backbone structure can be detected by changing $\eta$. The pseudo-code for the local perturbation strategy is shown in Algorithm 2.
\begin{algorithm}
\caption{Local perturbation strategy}
\KwIn{adjacent matrix $\bm {\mathrm{A}}$, parameters $\eta$ and $\xi$, number of nodes $n$}
\KwOut{perturbed adjacent matrix $\bm {\mathrm{A}}$}
{
    $\bm {\mathrm{D}}[n][n] \gets \{0\}$;\\
    \For{$i = 1 \to n$}{
        $\bm {\mathrm{D}}[i][i] \gets \sum\limits_{a=1}^{n}{\bm {\mathrm{A}}[i][a]}$;\\
    }
    $\bm {\mathrm{I}}[n][n]\gets \{1\}$;\\

    $i=randm(1,n)$;\\
    \For{$index = 1 \to n\xi$}{
        $\bm{\mathrm{B}}=\bm{\mathrm{A}}\odot((\bm{\mathrm{D}}\bm{\mathrm{I}})^\eta+((\bm{\mathrm{D}}\bm{\mathrm{I}})^\eta)^T)$;\\
        \While{$\bm{\mathrm{D}}[i][i]~==0$}{
            $i=randm(1,n)$;\\
        }
        $\bm {\mathrm{b}} \gets []$;\\
        $sumb\gets 0$;\\
        \For{$k = 1 \to n$}{
            $\bm {\mathrm{b}}[k] \gets \bm {\mathrm{B}}[i][k]$;\\
            $sumb\gets sumb + \bm {\mathrm{b}}[k]$;\\
        }
        $\bm {\mathrm{w}} \gets \bm {\mathrm{b}}/sumb$;\\
        Choosing a position $j$ according to the probability represented by the value of the perturbation weight vector $\bm {\mathrm{w}}$;\\
        $\bm {\mathrm{A}}[i][j] \gets 0$;\\
        $\bm {\mathrm{A}}[j][i] \gets 0$;\\
        $\bm {\mathrm{D}}[i][i]\gets\bm {\mathrm{D}}[i][i]-1$;\\
        $\bm {\mathrm{D}}[j][j]\gets\bm {\mathrm{D}}[j][j]-1$;\\
        $i \gets j$;\\
    }
    return $\bm {\mathrm{A}}$\\
}
\end{algorithm}

\subsection{Discussion}
\subsubsection{Time Complexity}
For the algorithm of global perturbation strategy, lines 2-4 calculate the degree for all nodes, and the complexity costs $O(n^2)$. Lines 6-7 obtain the perturbation weights for all intralayer links by matrix multiplication with time complexity of $O(n^3/q)$, where $q$ is the number of computed nodes~\cite{lee1997IO}. Lines 8-12 select and remove a proportion of $\xi$ intralayer links based on the perturbation weights, and the complexity costs $n^3\xi$. In summary, the complexity of Algorithm 1 is $O(n^2+n^3/q+n^3\xi)\approx O(n^3\xi)$.

For the algorithm of local perturbation strategy, the complexity is the same with algorithm 1 which costs $O(n^2)$. Lines 7-24 remove a proportion of $\xi$ intralayer links based on the weighted random walk with a time complexity of $O(n\xi \cdot(n^3/q+n+n))$. In summary, the complexity of Algorithm 2 is $O(n^2+n\xi \cdot(n^3/q+n+n))\approx O(n^4\xi/q)$.

To reduce the time complexity, we can optimize these two algorithms. For the algorithm of the global perturbation strategy, we can execute lines 8-12 using a parallel implementation. Suppose there are $n_g$ parallel units, the time complexity of algorithm 1 will become to $O(n^2+n^3/q+n^3\xi/n_g)\approx O(max(n^3\xi/n_g,n^3/q))$. For the algorithm of local perturbation strategy, several biased random walkers can simultaneously work from lines 7 to 24. Suppose there are $n_l$ walkers working simultaneously, the time complexity of algorithm 2 will become to $O(n^2+n\xi \cdot(n^3/q+n+n)/n_l)\approx O(n^4\xi/q n_l)$.

\subsubsection{Parameter $\eta$}
\begin{figure} [!t]
    \centering
    \includegraphics[width=0.8\textwidth]{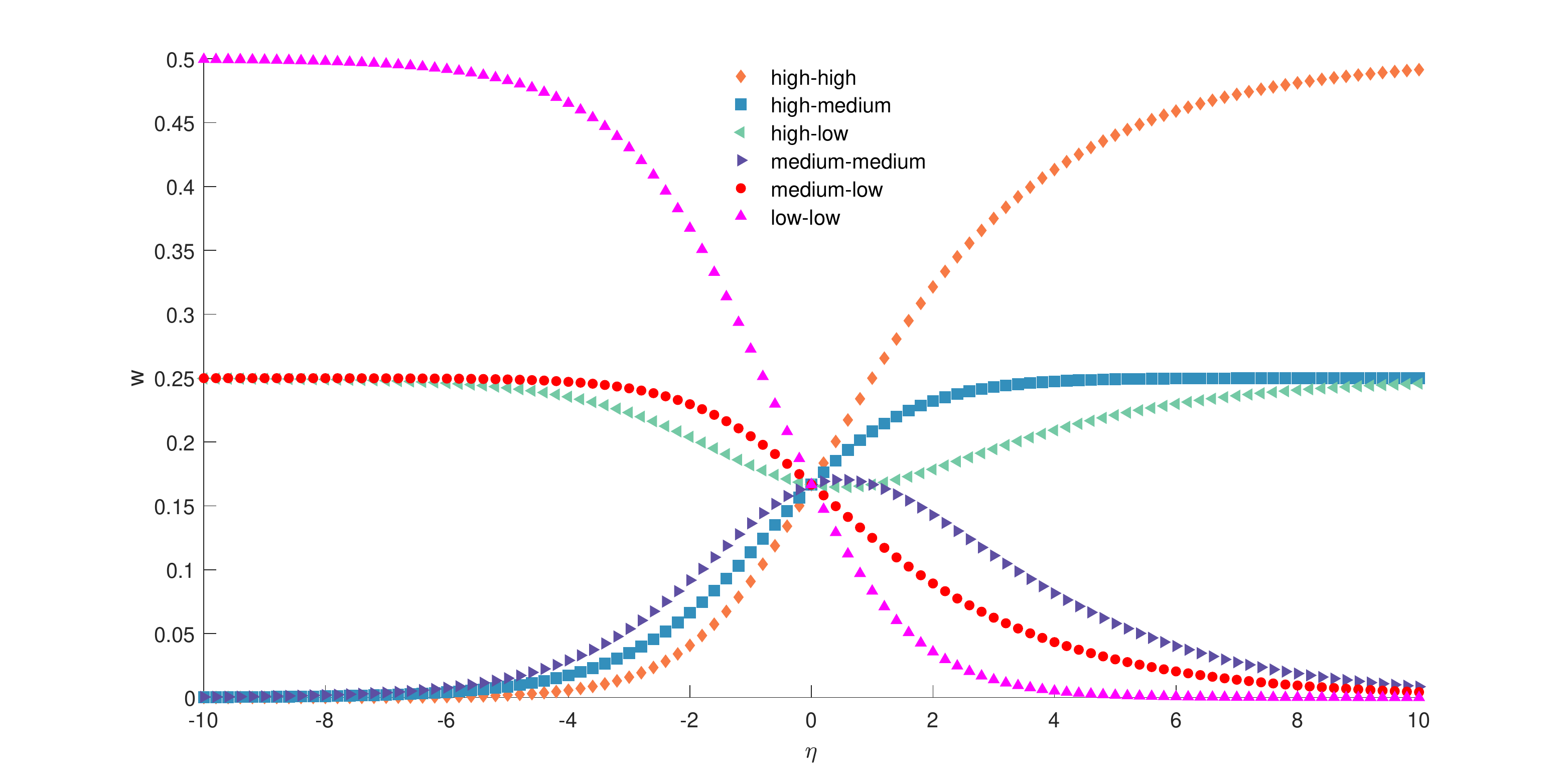}
    \caption{Example of the wights for different types of intralayer links under global perturbation strategy.}
    \label{pic:intralayer_link_types_ana}
\end{figure}

As shown in Figure ~\ref{pic:intralayer_link_types_ana}, we simulate the trend of the perturbation weights of the six types of intralayer links with the parameter $\eta$ under the global perturbation strategy. The degree of the node with the high degree is set to 30, the degree of the node with a medium degree is set to 20, and the degree of the node with a low degree is set to 10. The values of $\eta$ vary from -10 to 10 by 0.2. The perturbation weights for different types of intralayer links are calculated according to Eq.~(\ref{eq_wight_mat_simple}). We can see that (i) the weight of an intralayer link connected to a node with a high degree exhibit a trend of increasing with an increase in $\eta$; (ii) the weight of an intralayer link connected to a node with a low degree exhibit a trend of increasing with a decrease in $\eta$; and (iii) the weight of an intralayer link connected from a node with the medium degree to a node with medium degree exhibit a trend of first increasing and then decreasing with an increase in $\eta$.

\begin{figure} [!t]
    \centering
    \includegraphics[width=0.8\textwidth]{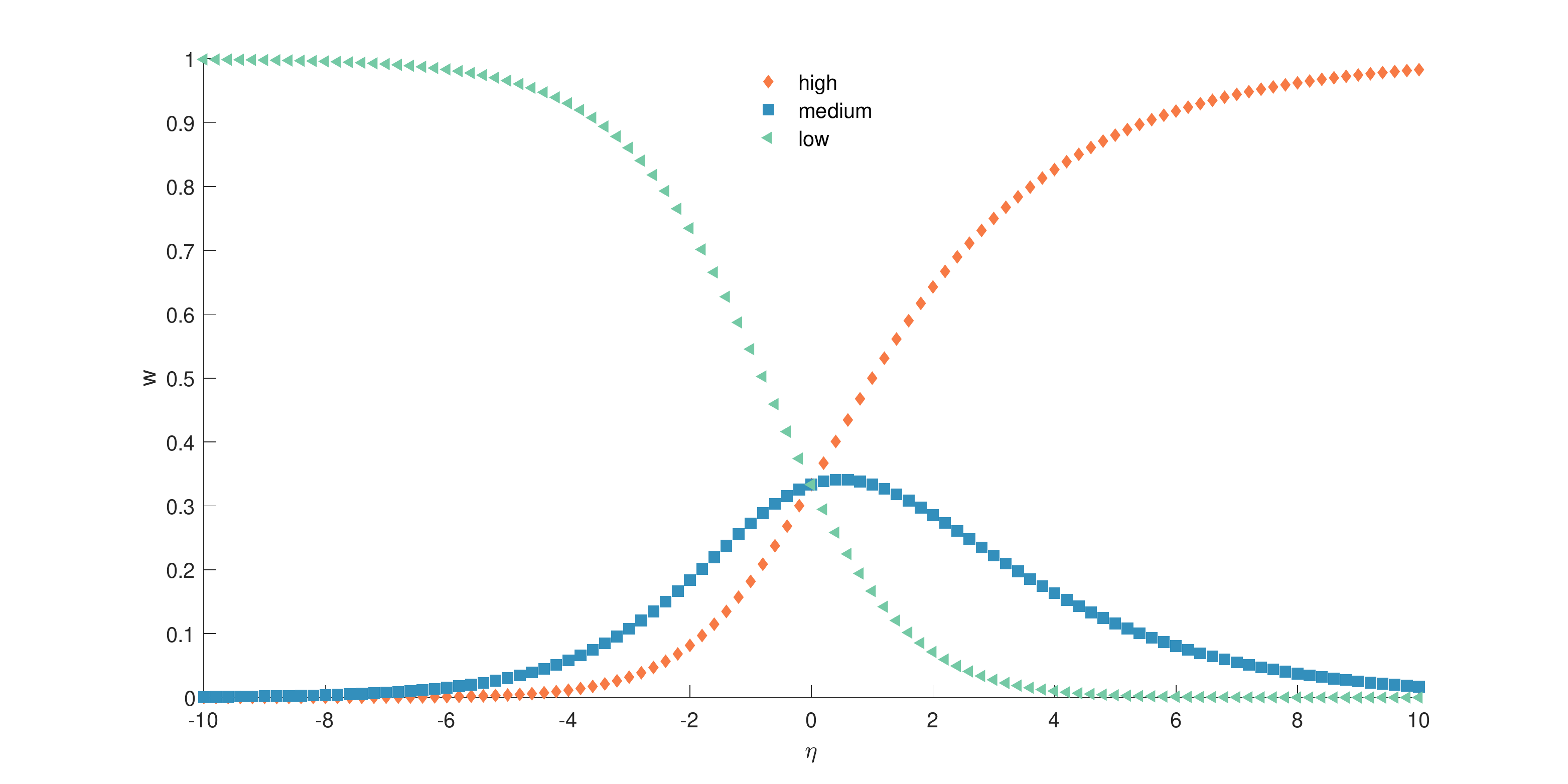}
    \caption{Example of the wights for different types of intralayer links under local perturbation strategy.}
    \label{pic:intralayer_link_types_ana_local}
\end{figure}
As shown in Figure ~\ref{pic:intralayer_link_types_ana_local}, we simulate the trend of the perturbation weights of the three types of intralayer links with the parameter $\eta$ under the local perturbation strategy. The node degree and parameter $\eta$ are the same as the global perturbation strategy above. The perturbation weights for different types of intralayer links are calculated according to Eq.~(\ref{eq_wight_simple_local}). Similarly, we can see that (i) the weight of an intralayer link connected from any node to a node with a high degree exhibit a trend of increasing with an increase in $\eta$; (ii) the weight of an intralayer link connected from any node to a node with low degree exhibit a trend of increasing with a decrease in $\eta$; and (iii) the weight of an intralayer link connected from any node to a node with medium degree exhibit a trend of first increasing and then decreasing with an increase in $\eta$.
\subsubsection{Other basic operations for perturbation}
\begin{figure} [!t]
    \centering
    \includegraphics[width=0.8\textwidth]{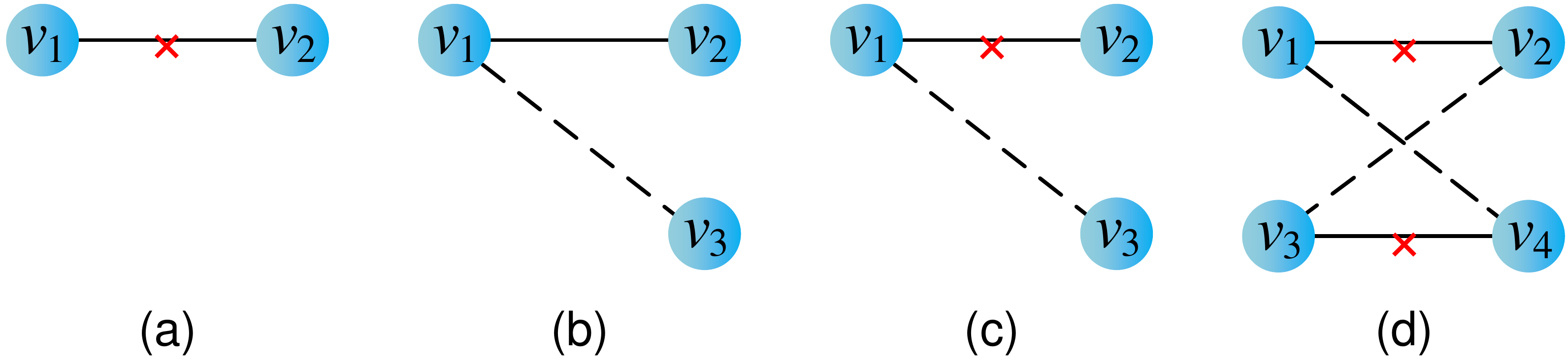}
     \caption{Example of the basic operations for perturbation. The solid lines represent the intralayer links in the original networks while the dashed links represent intralayer links will be added. The solid lines under red crosses represent that the existing interlayer links will be removed. (a) Link removal. (b) Link addition. (c) Link rotation. (d) Link switch.}
    \label{pic:operationsoflinkmod}
\end{figure}

In this paper, the global and local perturbation strategies are proposed to study the influence of different types of intralayer links for the interlayer link prediction. The intralayer links selected by these two strategies are removed directly. It is worthwhile that Removing the intralayer link is only one of the basic operations. Besides, there are several other basic operations for perturbation, such as adding a new link around the selected intralayer link, rotating the selected intralayer link to another node, and switch links between two selected intralayer links, as shown in Fig.~\ref{pic:operationsoflinkmod}.

\subsubsection{Perturbing multiple layers simultaneously}
To study the influence of different types of intralayer links for the interlayer link prediction simply, we focus on perturbing one layer of the multiplex network in this paper. In fact, multiple layers of the multiplex network can be perturbed simultaneously in the real-world scenario. Particularly, we can leverage global strategy to perturb one layer while leverage local strategy to perturb the other layer. This may be more applicable in the scenario of matching the same users from different social network platforms. Because different platforms usually belong to different companies and they would not share the relationships or interactions of the users for the concern of user privacy leakage.

\subsubsection{Influence of the observed interlayer links}
Apart from intralayer links, the observed interlayer links may also affect the interlayer link prediction. What types of interlayer links are most important for correct prediction? Are there any interlayer links whose presence leads to worse predictive performance than their absence? These questions are also worthy to answer via study. We can use the similar strategies of operating the intralayer links to explore the influence of the observed interlayer links for the interlayer link prediction. For example, we can simply classify interlayer links as six types similar with Fig.~\ref{pic:intralayer_link_types} and select them by the formula similar with Eq.(\ref{eq_wight_global}).

\subsubsection{Quasi-local perturbation strategy}
Besides of global and local strategies, quasi-local perturbation strategy can also be used which do not require global structural information but make use of more information than local strategy. For example, we can calculate the perturbation weights for the biased random walk based on one- and two-hop neighbors.

\begin{table*}[!th]
\centering
\caption{Statistical features of real-world datasets. The features include the number of nodes $|V|$, the number of intralayer links $|E|$, the maximum degree $d_{max}$, the average degree $\langle{d}\rangle$, the assortativity coefficient $r$ which is a measure of the degree-degree correlation of the network~\cite{newman2002assortative}, the clustering coefficient $c$ which quantifies how well connected are the neighbors of a node in a network~\cite{soffer2005network,watts1998collective}, the degree heterogeneity $H$ where $H=\frac{\langle{d^2}\rangle}{{\langle{d}\rangle}^2}$, and the number of interlayer links $|E^{\alpha\beta}|$.}
\label{table}
\setlength{\tabcolsep}{3pt}
\begin{tabular}{ccccccccc}
\hline
Network&$|V|$&$|E|$& $d_{max}$& $\langle{d}\rangle$ & $r$& $c$& $H$&$|E^{\alpha\beta}|$\\
\hline
Foursquare1&5,313&76,972&552&20.42&$-0.193$&0.23&3.446&\multirow{2}*{3,148}\\
Twitter1&5,120&164,920&1,725&51.01&$-0.214$&0.30&4.489&~\\
\hline
\hline
Foursquare2&1,507&18,470&282&24.51&$-0.042$&0.40&3.773&\multirow{2}*{1,507}\\
Twitter2&1,507&13,843&394&18.37&$-0.093$&0.16&3.282&~\\
\hline
\hline
Higgs\_FS&4,288&122,826&1365&57.29&$-0.140$&0.27&2.943&\multirow{2}*{3,760}\\
Higgs\_MT&3,777&13,413&1072&7.10&$-0.095$&0.22&9.619&~\\
\hline
Higgs\_FS&4,184&101,618&1086&48.57&$-0.106$&0.27&2.840&\multirow{2}*{3,219}\\
Higgs\_RT&3,238&13,571&626&8.38&$-0.090$&0.09&5.653&~\\
\hline
\end{tabular}
\label{tab:datasets}
\end{table*}
\section{Experiments}
To evaluate the effects of the two structural perturbation strategies proposed in this paper on the performance of the interlayer link prediction algorithms, we conducted experiments on four real-world datasets and nine kinds of BA artificial network Datasets. This section includes three parts: (i) datasets description; (ii) experimental settings; (iii) experimental results.

\subsection{Datasets}
\subsubsection{Real-world datasets}
 We used four real-world datasets as follows:
\begin{itemize}
\item \textbf{Foursquare-Twitter-1 (FT1)}: The FT1 dataset is collected from Foursquare, a famous location-based SMN platform, and Twitter, the hottest microblogging SMN, by Zhang et al.~\cite{ZhangJiawei2015-IJCAI}. The ground truth is obtained from the Foursquare profiles since some users may provide their Twitter account links in these profiles. The nodes in this dataset are partially aligned.
\item \textbf{Foursquare-Twitter-2 (FT2)}: The FT2 dataset is crawled from Foursquare and Twitter by Mahdi Jalili et al.~\cite{jalili2017link}. Different from FT1, the nodes in this dataset are fully aligned.
\item \textbf{Higgs\_Friendships-Higgs\_Mention (Higgs-FSMT)}: The Higgs dataset was crawled from Twitter between 1st and 7th July 2012~\cite{de2013anatomy}. It focuses on the spreading processes of the messages about discovering a new particle with the features of the elusive Higgs boson. There are four types of interactions in this dataset: FS (friendships/followers), RT (retweet), MT (mention), and RE (reply). We randomly choose the FS and MT networks (nodes with degrees greater than five) to construct the multiplex network.
\item \textbf{Higgs\_Friendships-Higgs\_Retweet (Higgs-SCRT)}: We use the same method to choose the FS and RT networks of Higgs to construct this dataset.
\end{itemize}
The statistical information of these four datasets is summarized in Table~\ref{tab:datasets}. Formally, we represent each network as an undirected network.

\subsubsection{Artificial datasets}
We generated nine groups of artificial multiplex BA networks using the method introduced in the Ref.~\cite{tang2020interlayer}. The parameter values of each group will be presented in the subsection of experimental results about these groups.

\subsection{Experimental settings}
We used the two structural perturbation strategies to perturbed any layer of the multiplex networks and then predict the unobserved interlayer links by four different interlayer link prediction algorithms hence evaluating the effects of the perturbation strategies. The four interlayer link prediction algorithms are listed as follows.
\begin{itemize}
\item \textbf{CN}: The common neighbors (CN)~\cite{lorrain1971structural} index is one of the most
well-known methods used in the link prediction problem which assumes that a link is more likely to exist between two nodes if they have many common neighbors. In interlayer link prediction problem, the matching degree between two unmatched nodes across different layers used CN index can be defined as:
\begin{equation}
r_{ij}^{CN}=\sum_{\substack{\forall(v^\alpha_a,v^\beta_b)\in \Phi, \\ v^\alpha_a \in\Gamma(u^\alpha_i),\\v^\beta_b\in\Gamma(u^\beta_j)}} |(v^\alpha_a,v^\beta_b)|,
\label{eq:CN}
\end{equation}
where $\Phi$ represents the set of matched pairs, $\Gamma(u^\alpha_i)$ and $\Gamma(u^\beta_j)$ represent the one-hop neighbor sets of nodes $u^\alpha_i$ and $u^\beta_j$, respectively.

\item \textbf{NS}: Narayanan and Shmatikov (NS)~\cite{narayanan2009anonymizing} developed this algorithm to mapping the same users across different social networks based solely on network structure. This algorithm is suitable for directed networks, which employs unmatched nodes’ in-degree and out-degree, as well as the matched node pairs, to calculate the matching degree between unmatched nodes. It can be defined as
\begin{equation}
r_{ij}^{NS}=\frac{c^{in}}{\sqrt{d^{in}(u^\beta_j)}}+\frac{c^{out}}{\sqrt{d^{out}(u^\beta_j)}},
\label{eq:NS}
\end{equation}
where $c^{in}$ and $c^{out}$ are the numbers of common incoming and outgoing neighbors of nodes $u^\alpha_i$ and $u^\beta_j$ respectively, and $d^{in}(u^\beta_j)$ and $d^{out}(u^\beta_j)$ are the in-degree and out-degree of node $u^\beta_j$ respectively.

\item \textbf{FRUI}: friend relationship-based user identification (FRUI)~\cite{ZhouXiaoping2016} algorithm aligns the same users across different social networks by the friendship structure. The main idea of FRUI is to judge whether two user accounts belong to a same individual by counting the number of shared matched pair. The matching degree of two unmatched nodes across different layers is defined as
\begin{equation}
r_{ij}^{FRUI}=|\Gamma(u^\alpha_i) \cap \Gamma(u^\beta_j)|+\frac{|\Gamma(u^\alpha_i) \cap \Gamma(u^\beta_j)|}{min(|\Gamma(u^\alpha_i)|,|\Gamma(u^\alpha_i)|)},
\label{eq:FRUI}
\end{equation}
where $(\cdot \cap \cdot)$ denotes the intersection operation between the two sets inside the parenthesis and $min(\cdot)$ is the minimum function that takes the minimum value inside the parenthesis.

\item \textbf{IDP}: iterative degree penalty (IDP) algorithm is proposed by us in our previous work in Ref.~\cite{tang2020interlayer}. It used an degree penalty principle to calculate the matching degree between two unmatched nodes, which can be represented as
\begin{equation}
r_{ij}^{IDP}=\sum_{\substack{\forall(v^\alpha_a,v^\beta_b)\in \Phi, \\ v^\alpha_a \in\Gamma(u^\alpha_i),\\v^\beta_b\in\Gamma(u^\beta_j)}} \log^{-1}(k_{v^\alpha_a}+1)\cdot\log^{-1}(k_{v^\beta_b}+1).
\label{eq:IDP}
\end{equation}
In Eq.~(\ref{eq:IDP}), $\Phi$ represents the set of PINPs, $\Gamma(u^\alpha_i)$ and $\Gamma(u^\beta_j)$ represent the neighbor sets of nodes $u^\alpha_i$ and $u^\beta_j$, respectively, $k$ represents the node degree, and the constraints in the equation indicate that the PINP $(v^\alpha_a,v^\beta_b)$ is the CMN of UNP $(u^\alpha_i,u^\beta_j)$.
\end{itemize}

The purpose of the experiments in this paper is to evaluate the effects of the perturbation strategies for the interlayer link prediction. To improve the experimental efficiency, we improved the four prediction algorithms by removing the iterative process of these algorithms.

To evaluate the performance of the interlayer link prediction algorithms, the set of interlayer links, $c$, is usually partitioned into two subsets: one is the training set, $c^T$, which is treated as the set of observed interlayer links, while the other is the validation set, $c^V$, which is used for testing and can be considered as the set of unobserved interlayer links. Clearly, $c^T \cup c^V =c$ and $c^T \cap c^V =\emptyset$. Meanwhile, two traditional metrics to measure the performance of different interlayer link prediction algorithms are adopted: $Precision@N$ ($P@N$)~\cite{LiuLi2016,ShuKai2017} and MAP (mean average precision) ~\cite{ShuKai2017}.
\begin{itemize}
\item \textbf{$P@N$}: Given an unmatched node $v^\alpha_i$ in one layer $\alpha$ of the multiplex network $\mathcal{M}$, the interlayer link prediction algorithm can provide a candidate list of unmatched interlayer node pair consisted by node $v^\alpha_i$ and any unmatched node in another layer $\beta$. In this list, the unmatched interlayer node pairs are ranked by the descending order based on the probability whether they are the true interlayer links. For a top-$N$ list, if the correct interlayer link exists in it, its indicator function ${\mathds{1}_i\{success@N\}}$ equaling to 1, and 0 otherwise. The $success@N$ measures whether the correct interlayer link will occur in top-$N$ candidate list or not. Averaging the sum of the indicator function of all unmatched nodes' lists in layer $\alpha$, the $P@N$ is defined as
    \begin{equation}
    P@N=\sum_{i=1}^{n_u}{\mathds{1}_i\{success@N\}/n_u},
    \end{equation}
    where $n_u$ is the number of unobserved interlayer links.
\item \textbf{MAP}: Mean average precision (MAP) is used to evaluate the ranked performance of different interlayer link prediction algorithms and is calculated by the average performance of average precision (AP) over all unmatched nodes that need to be matching. For the unmatched node $v^\alpha_i$ in layer $\alpha$, the (AP) is defined as
    \begin{equation}
    AP=\frac{1}{r_i},
    \end{equation}
    where $r_i$ is the rank of correct interlayer link in unmatched node $v^\alpha_i$'s candidate list. For all unobserved interlayer links, the MAP is defined as
\begin{equation}
MAP=(\sum_{i=1}^{n_u}{\frac{1}{r_i}})/n_u.
\end{equation}
\end{itemize}

There are two parameters, the parameter to adjust the weights for different types of intralayer links $\eta$ and the proportion of removed intralayer links $\xi$, in both of the two perturbation strategies. The effects of the two parameters on the prediction performance of the interlayer link prediction algorithms are investigated by experiments. The metrics to evaluate the interlayer prediction performance are $P@30$ and MAP introduced in section 3. Each of the experiments
was repeated ten times, and the average values were taken as the final results. All the experiments were performed on a personal computer with 16G memory and 3.20GHz Intel i7-8700 CPU.

\subsection{Results on real-world datasets}
In this subsection, we analyze the effects of the proposed two perturbation algorithms on the performance of the interlayer link prediction algorithms under different $\eta$ and $\xi$ in the four real-world datasets.

\subsubsection{Effects of $\eta$}
We first set $\xi=0.1, 0.2, 0.3,$ and $0.4$, $\eta$ increasing from $-20$ to $20$ by 1 to analyse the effects of $\eta$ of the two perturbation strategies.
\begin{figure*}
        \centering    
        \includegraphics[width=12cm,height=10.5cm]{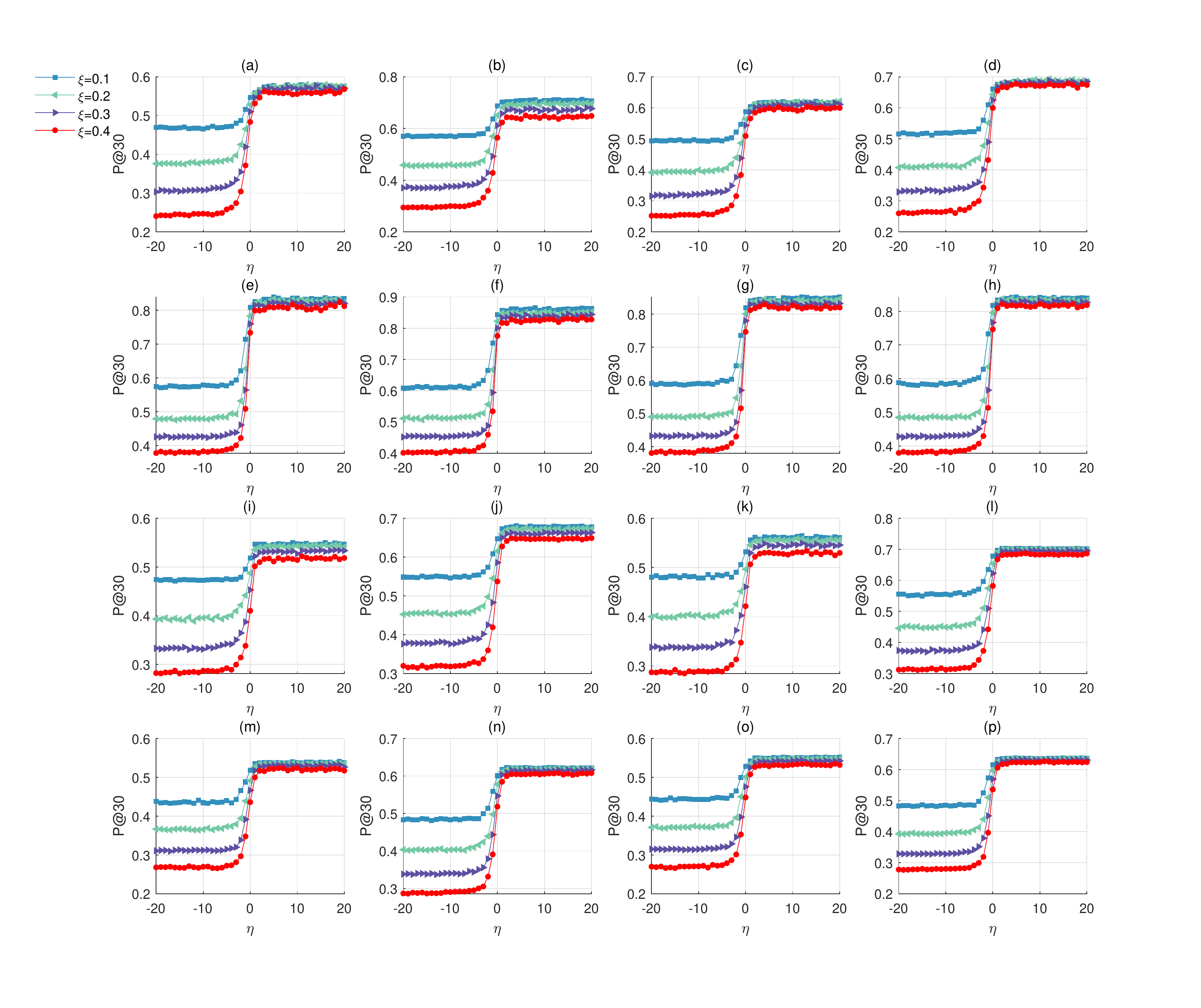}
        \caption{$P@30$ of different interlayer link prediction algorithms on the multiplex networks perturbed by the global perturbation strategy with different $\eta$. Each row uses the same dataset, and from top to bottom are datasets FT1, FT2, Higgs-SCMT, and Higgs-SCRT. Each column uses the same prediction algorithms, and from left to right are algorithms CN, NS, FRUI, and IDP.}
        \label{pic:result_real_diffeta_dmeth1}  
\end{figure*}

\textbf{(1) Result on global perturbation strategy.}
\begin{figure*}
        \centering    
        \includegraphics[width=12cm,height=10.5cm]{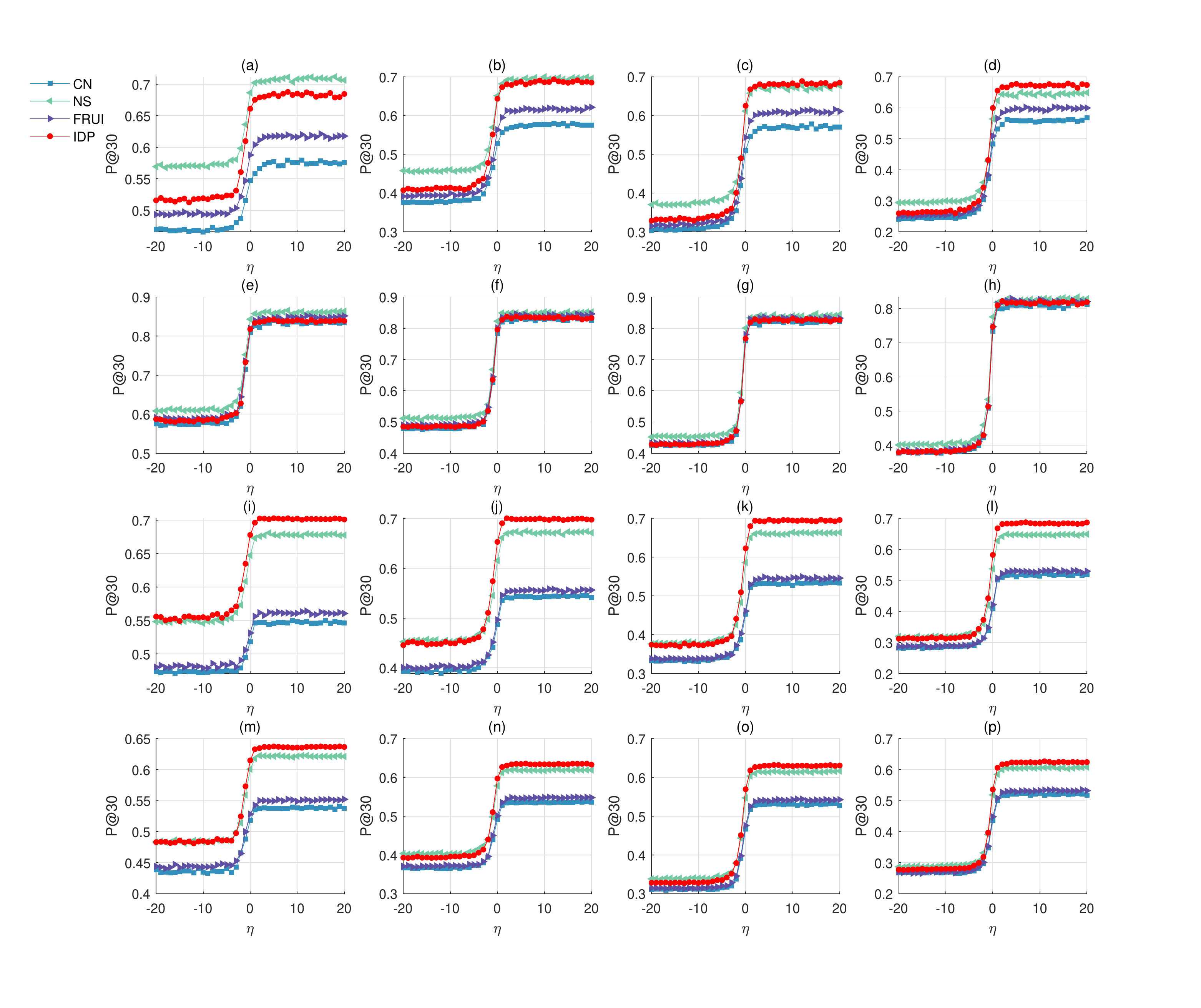}
        \caption{Comparison of different interlayer link prediction algorithms on the multiplex networks perturbed by the global perturbation strategy with different $\eta$. Each row uses the same dataset, and from top to bottom are datasets FT1, FT2, Higgs-SCMT, and Higgs-SCRT. Each column uses the same $\xi$, and from left to right are 0.1, 0.2, 0.3 and 0.4.}
        \label{pic:result_real_diffeta_p30_dmeth1_comppmethod}  
\end{figure*}

\begin{figure*}
        \centering    
        \label{pic:result_real_diffeta_p30_dmeth1_comppmethod}  
        \includegraphics[width=12cm,height=10.5cm]{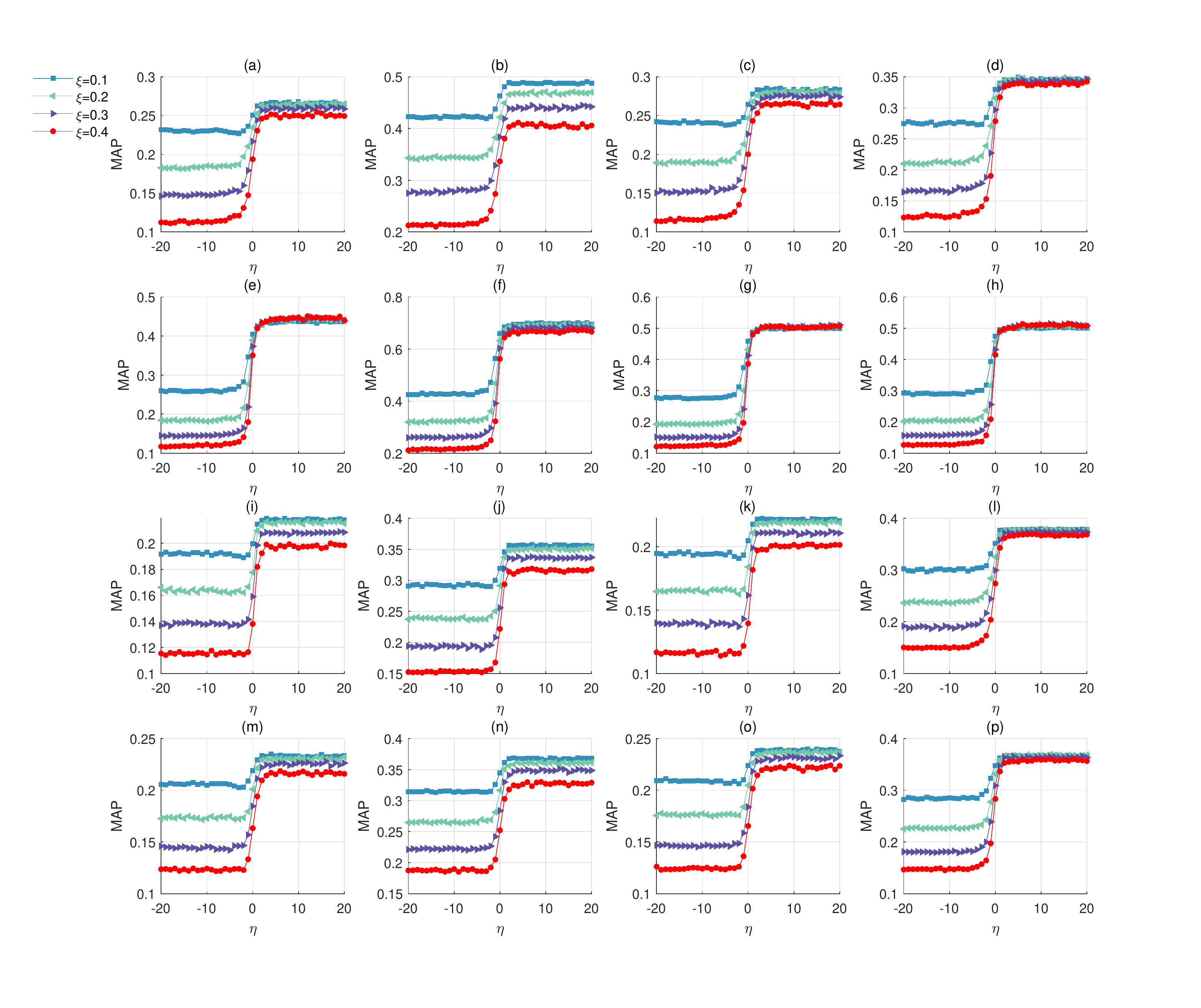}
        \caption{MAP of different interlayer link prediction algorithms on the multiplex networks perturbed by the global perturbation strategy with different $\eta$. The datasets and prediction algorithms in different subfigures are the same as Fig.~\ref{pic:result_real_diffeta_dmeth1}.}
        \label{pic:result_real_diffeta_map_dmeth1}  
\end{figure*}

Figure~\ref{pic:result_real_diffeta_dmeth1} is the $P@30$ of different prediction algorithms on the four kinds of real-world multiplex networks perturbed by the global perturbation strategy under the above settings. From the figure, we can see that:
(i) For a given $\xi$, $P@30$ of different prediction algorithms exhibit the same trend that it has almost no change firstly, then increases rapidly, and finally becomes stable with an increase in $\eta$. The differences in datasets, prediction algorithms, and $\xi$ did not change this trend.
(ii) When $-20\leq \eta \leq -5$, $P@30$ of the four prediction algorithms are generally stable and $P@30$ hardly varies with $\eta$ on different datasets. For example, when $\xi=0.1, \eta=-20$, the $P@30$ of CN is 0.4697 and when $\xi=0.1, \eta=-10$, the $P@30$ of CN is 0.4656 on the FT1 dataset. The difference of these two $P@30$ is 0.0041, which is very small.
(iii) When $-5<\eta\leq3$, $P@30$ of different prediction algorithms shows a trend of increasing with an increase in $\eta$. The increasing trend was first slowly growing, then rapidly growing, and finally slowly increasing again.
(iv) When $3<\eta\leq20$, $P@30$ of the four prediction algorithms is stable again. In a same $\xi$, $P@30$ hardly varies with $\eta$.
(v) The differences of $P@30$ for different $\xi$ in a same $\eta$ are very small when $3<\eta\leq20$. In contrast, the differences of $P@30$ for different $\xi$ in a same $\eta$ are relatively large when $\eta<-5$.
\begin{figure*}
\centering
\subfigure[$\eta=-20$]{
\begin{minipage}[t]{0.33\linewidth}
\centering
\includegraphics[width=4.5cm]{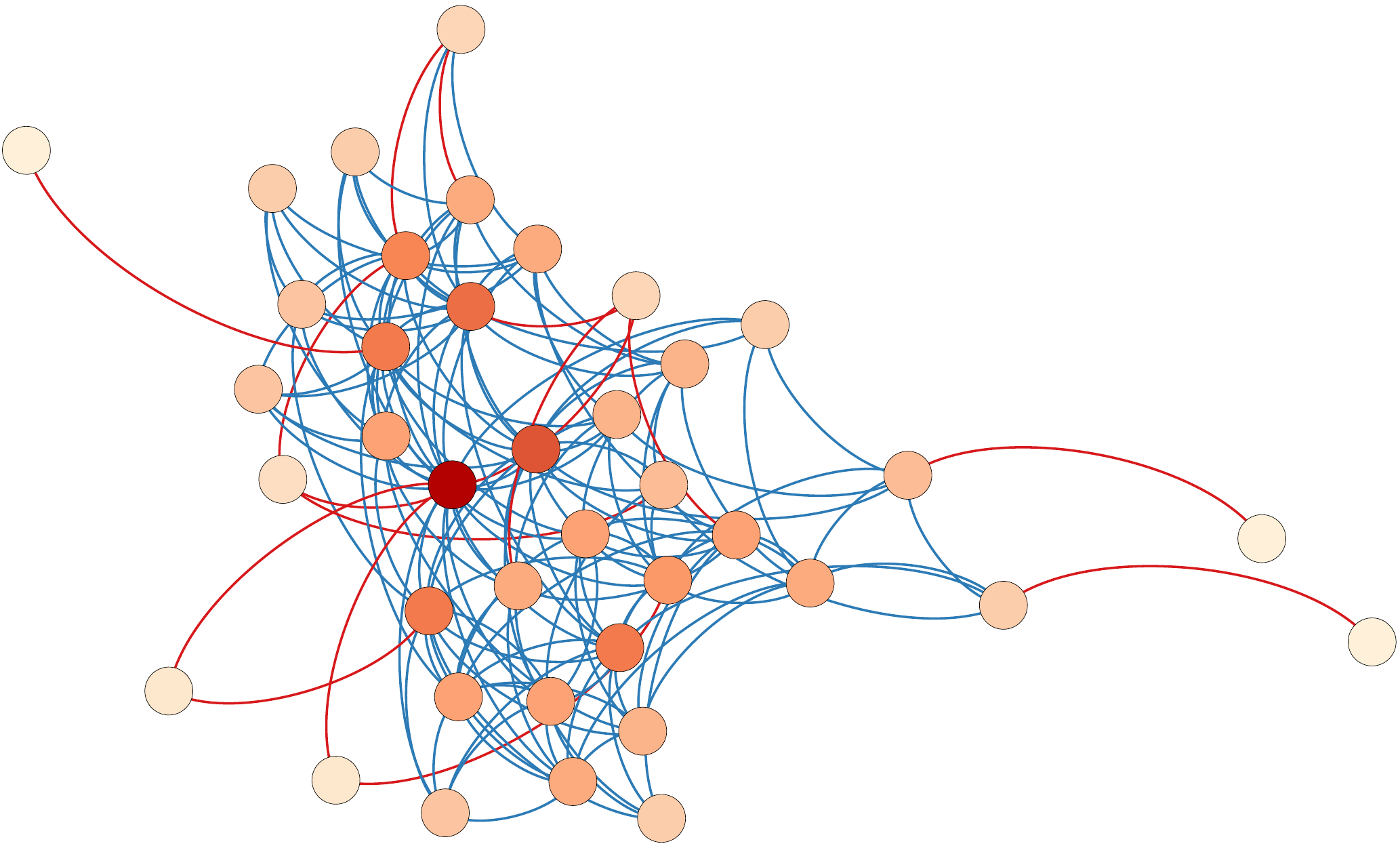}
\end{minipage}%
}%
\subfigure[$\eta=-10$]{
\begin{minipage}[t]{0.33\linewidth}
\centering
\includegraphics[width=4.5cm]{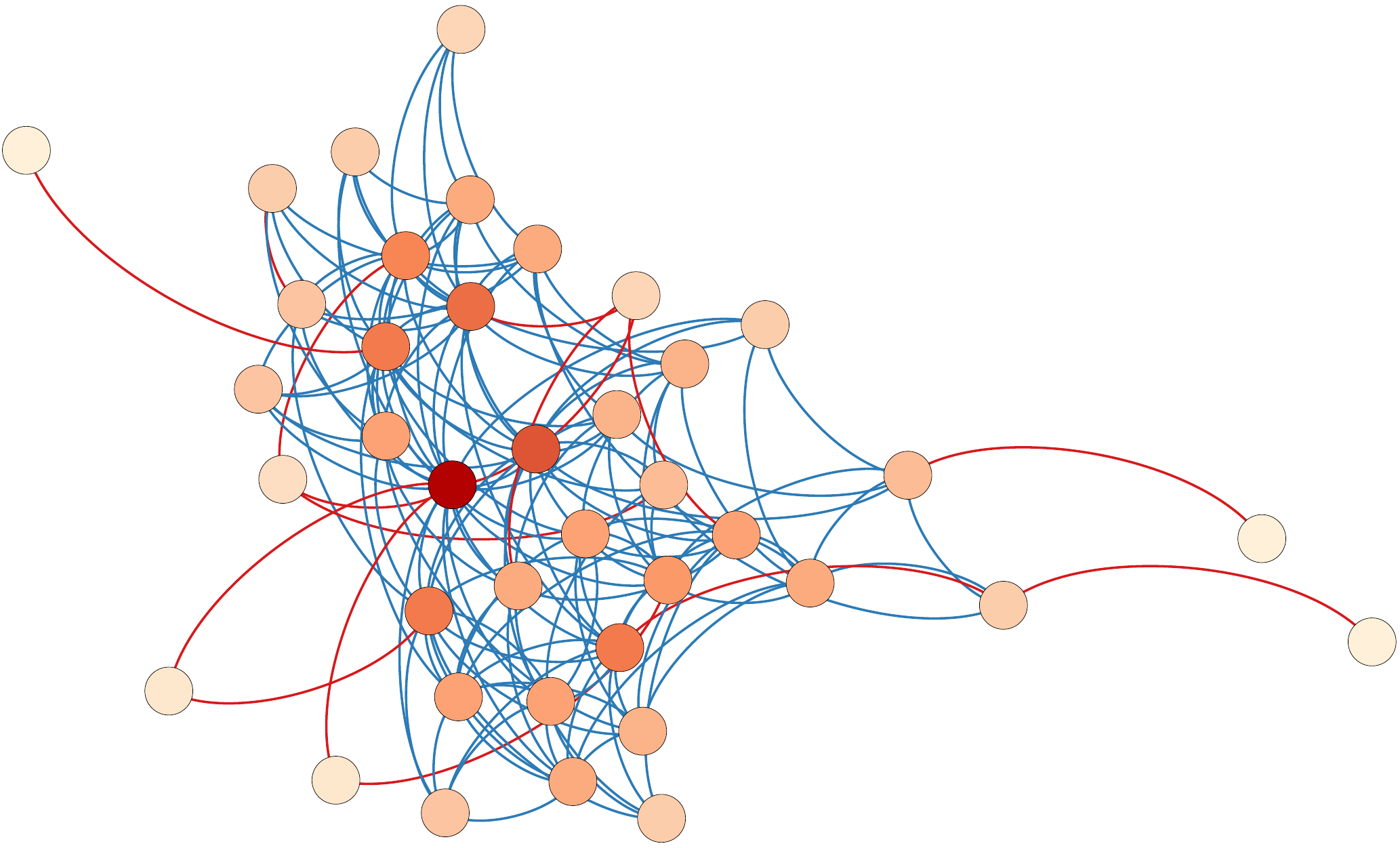}
\end{minipage}%
}%
\subfigure[$\eta=-5$]{
\begin{minipage}[t]{0.33\linewidth}
\centering
\includegraphics[width=4.5cm]{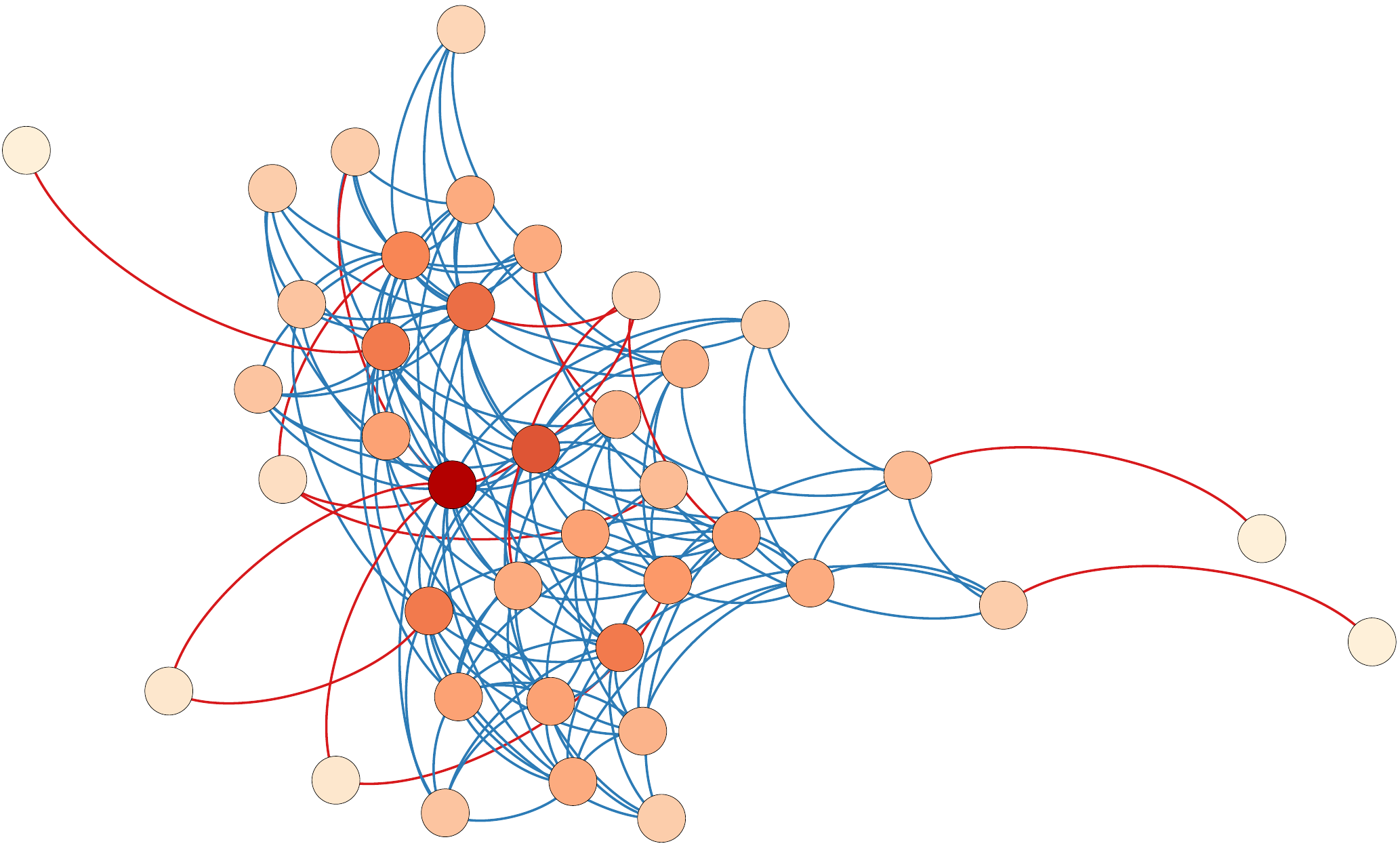}
\end{minipage}%
}%
\quad
\subfigure[$\eta=-1$]{
\begin{minipage}[t]{0.33\linewidth}
\centering
\includegraphics[width=4.5cm]{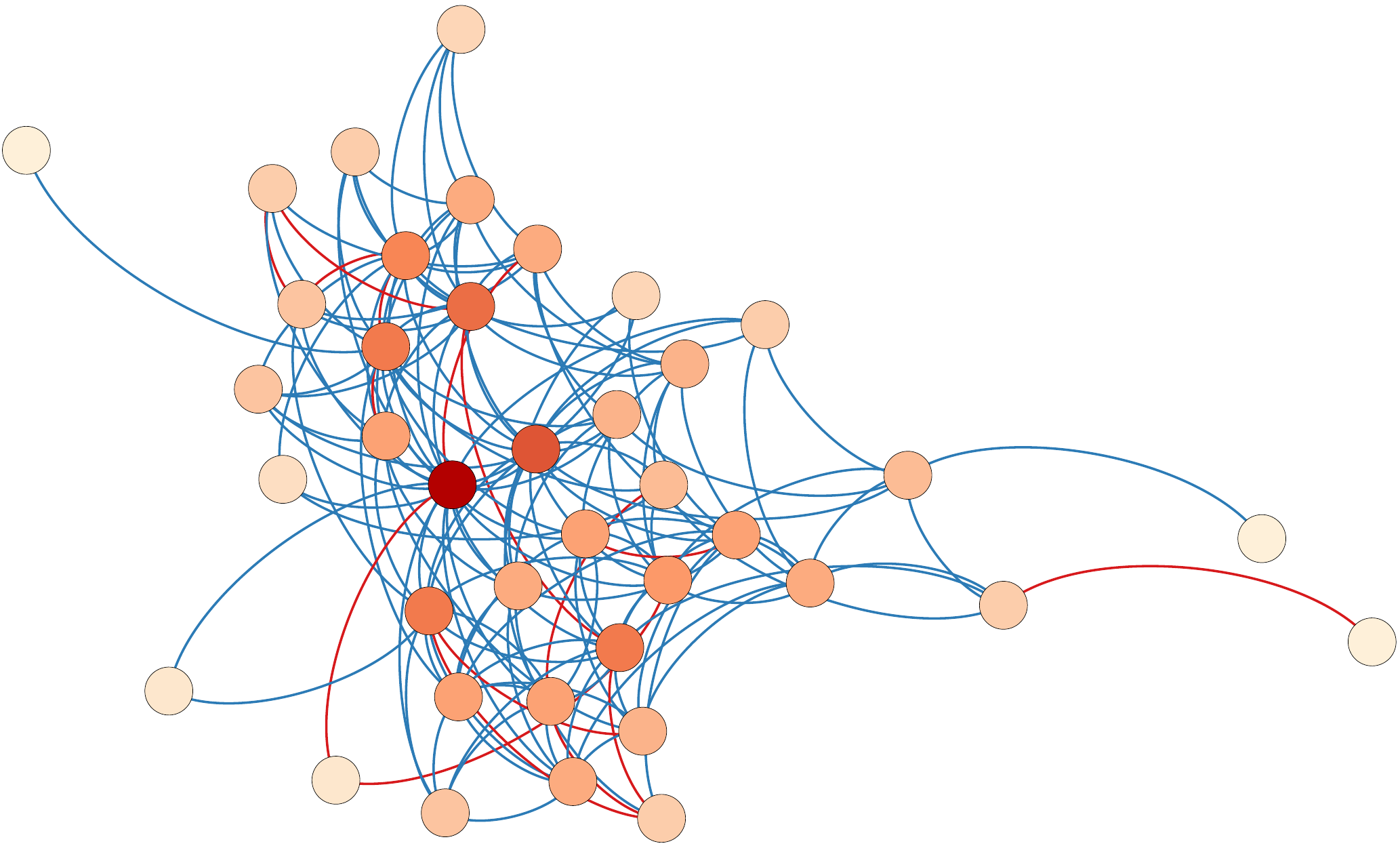}
\end{minipage}%
}%
\subfigure[$\eta=0$]{
\begin{minipage}[t]{0.33\linewidth}
\centering
\includegraphics[width=4.5cm]{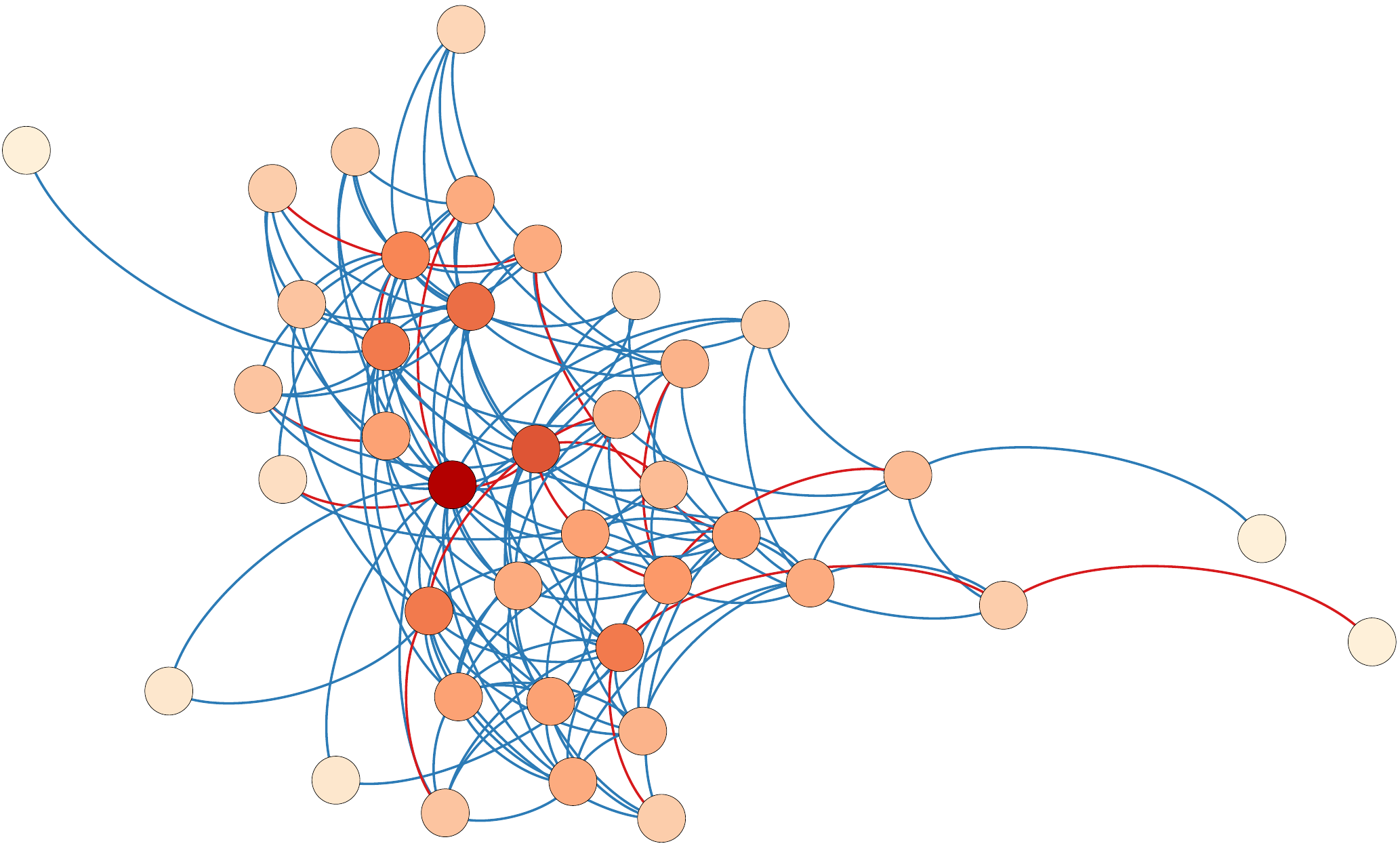}
\end{minipage}%
}%
\subfigure[$\eta=1$]{
\begin{minipage}[t]{0.33\linewidth}
\centering
\includegraphics[width=4.5cm]{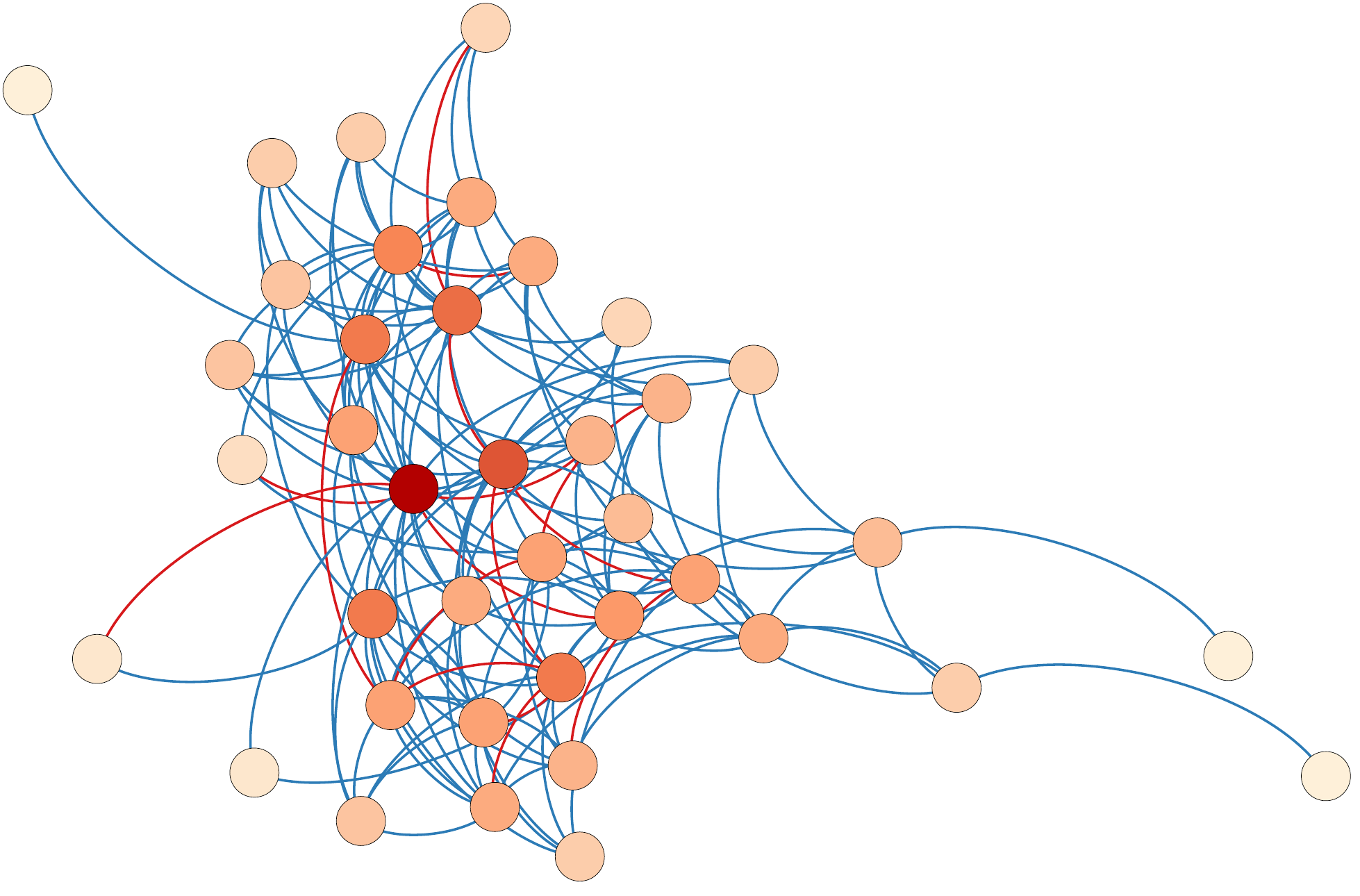}
\end{minipage}%
}%
\quad
\subfigure[$\eta=5$]{
\begin{minipage}[t]{0.33\linewidth}
\centering
\includegraphics[width=4.5cm]{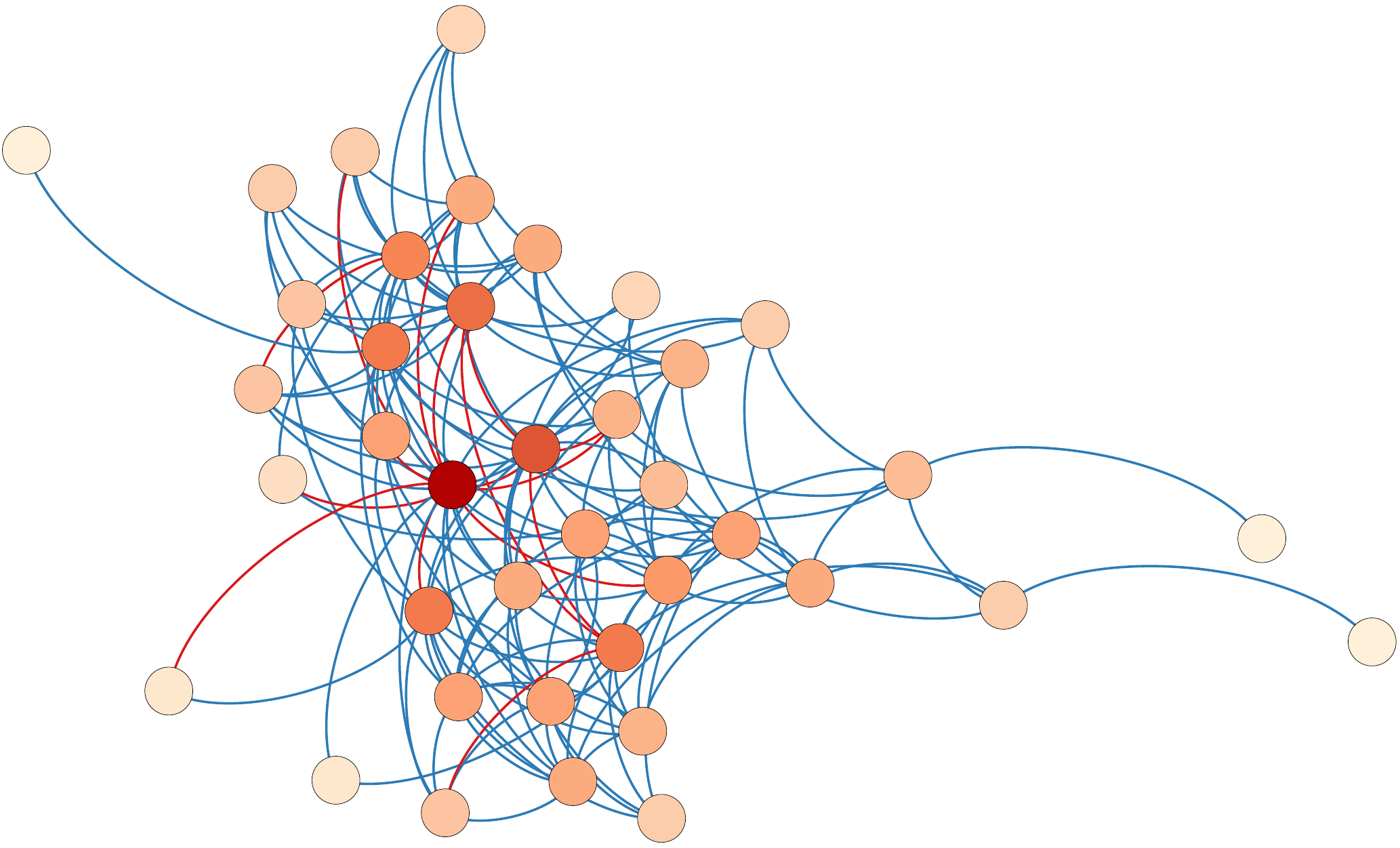}
\end{minipage}%
}%
\subfigure[$\eta=10$]{
\begin{minipage}[t]{0.33\linewidth}
\centering
\includegraphics[width=4.5cm]{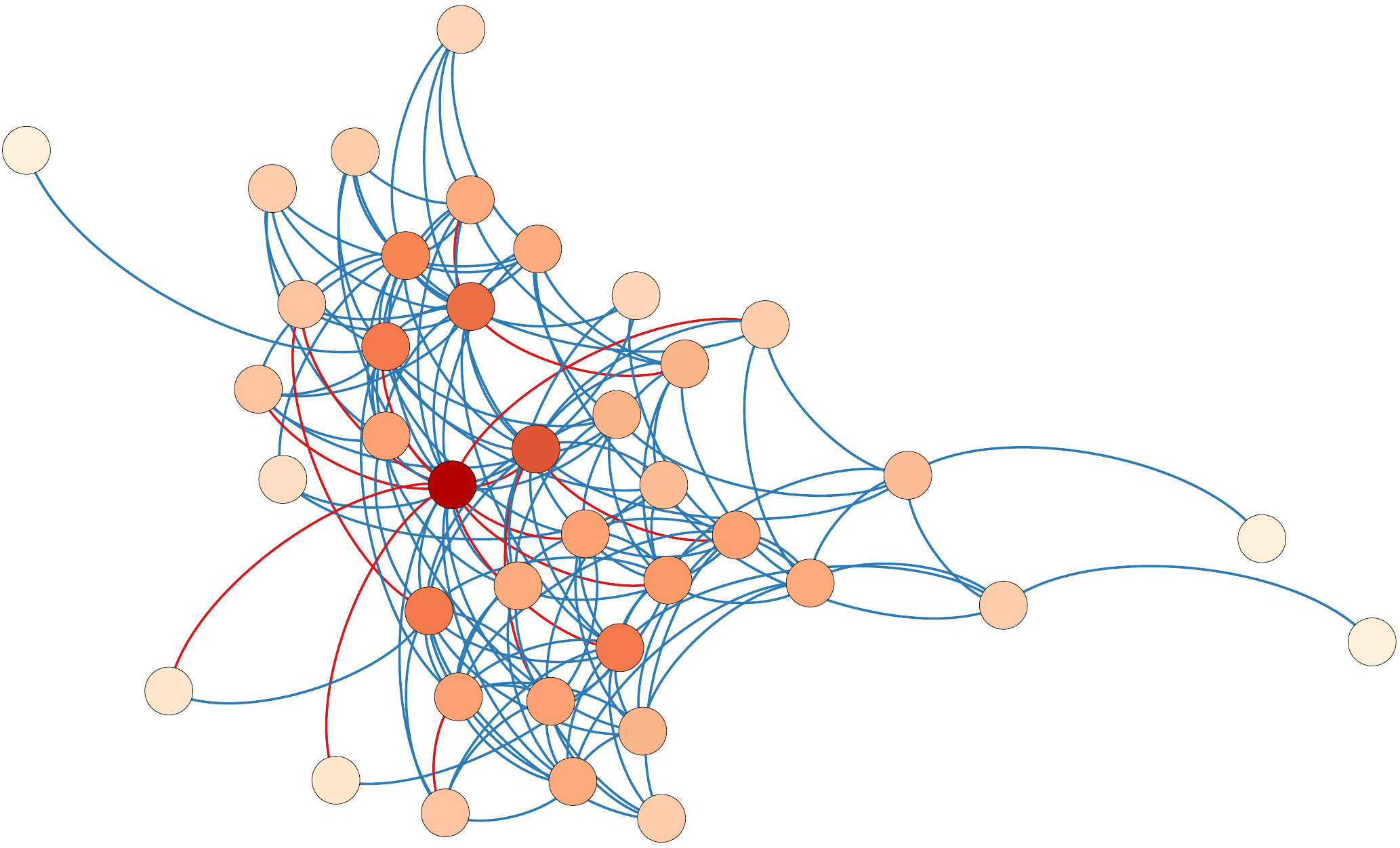}
\end{minipage}%
}%
\subfigure[$\eta=20$]{
\begin{minipage}[t]{0.33\linewidth}
\centering
\includegraphics[width=4.5cm]{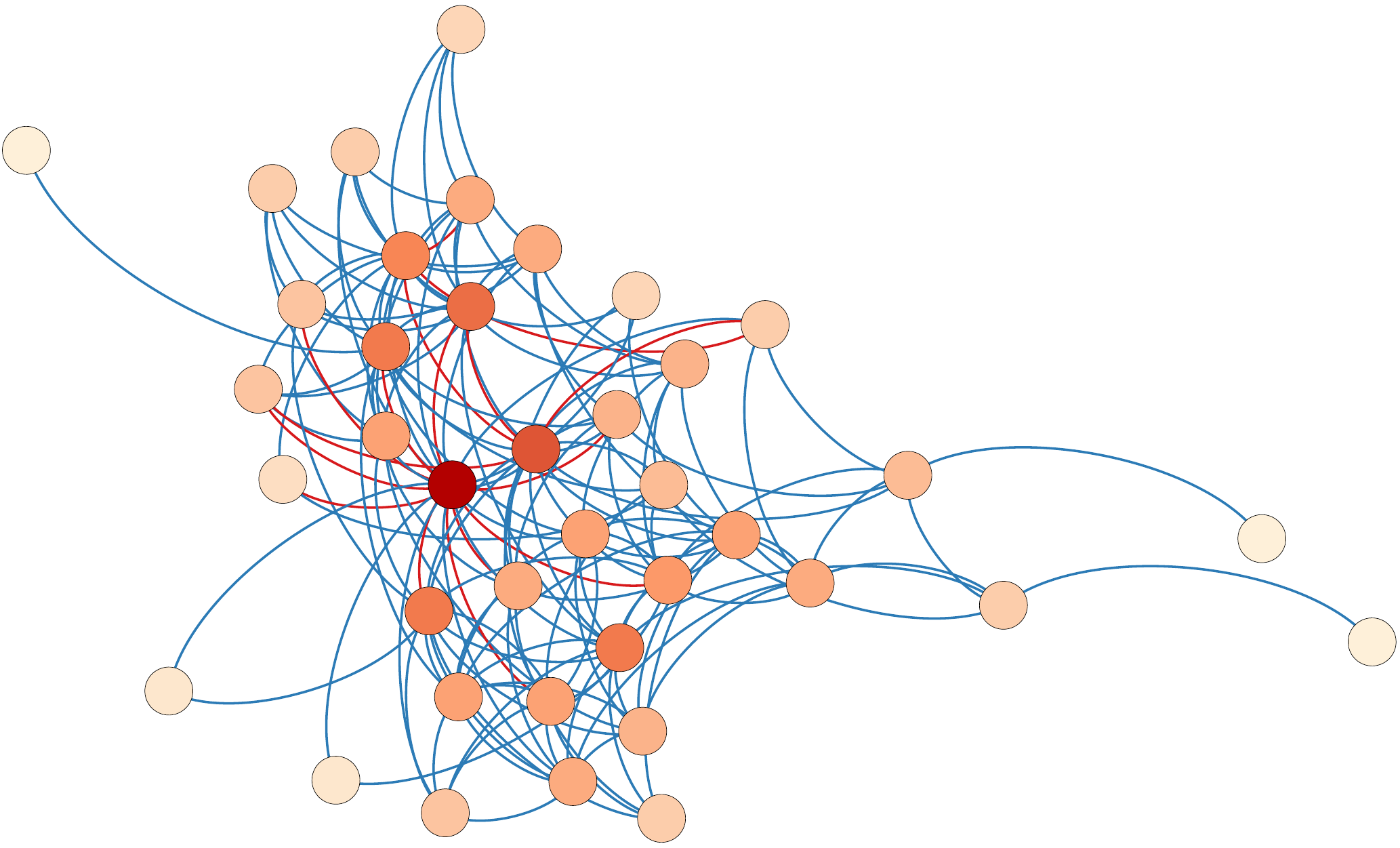}
\end{minipage}%
}%
\centering
\caption{Example of intralayer links removed by the global perturbation strategy. The node with darker color represents that it has a greater degree. The red line represents the intralayer links removed by the global perturbation strategy, and the blue line represents the remaining intralayer links. In all subfigures, $\xi=0.1$.}
\label{pics_dellinks_d1_exa}
\end{figure*}

To compare the different interlayer link prediction algorithms obviously on the same perturbation settings, we put the results of four interlayer link prediction algorithms with fixed parameter $\xi$ in each subfigure in Fig.~\ref{pic:result_real_diffeta_p30_dmeth1_comppmethod}. As can be seen from the figure, the rankings of the $P@30$ of these four interlayer link prediction algorithms are almost identical under different $\xi$ in each dataset. NS and IDP have better predictive performance than FRUI and CN.

MAP reflects the ranking ability of different prediction algorithms. We also investigate the effects of the global perturbation strategy on the MAP. Fig.~\ref{pic:result_real_diffeta_map_dmeth1} is the experimental results. As can be seen from the figure, the effects of the global perturbation strategy on MAP is almost the same as that of $P@30$.
\begin{figure*}
\centering
\subfigure[$\eta=-20$]{
\begin{minipage}[t]{0.33\linewidth}
\centering
\includegraphics[width=4.5cm]{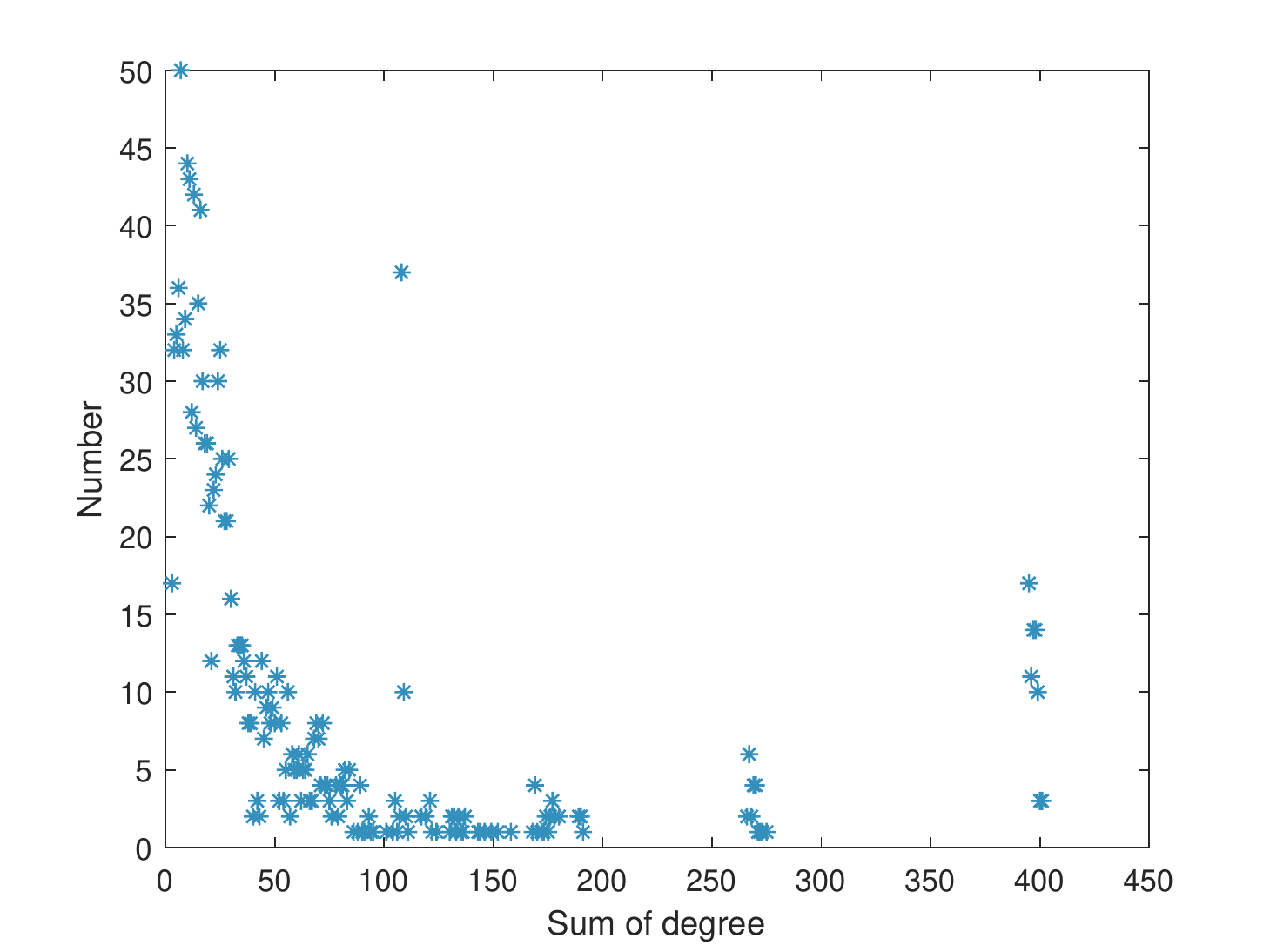}
\end{minipage}%
}%
\subfigure[$\eta=-10$]{
\begin{minipage}[t]{0.33\linewidth}
\centering
\includegraphics[width=4.5cm]{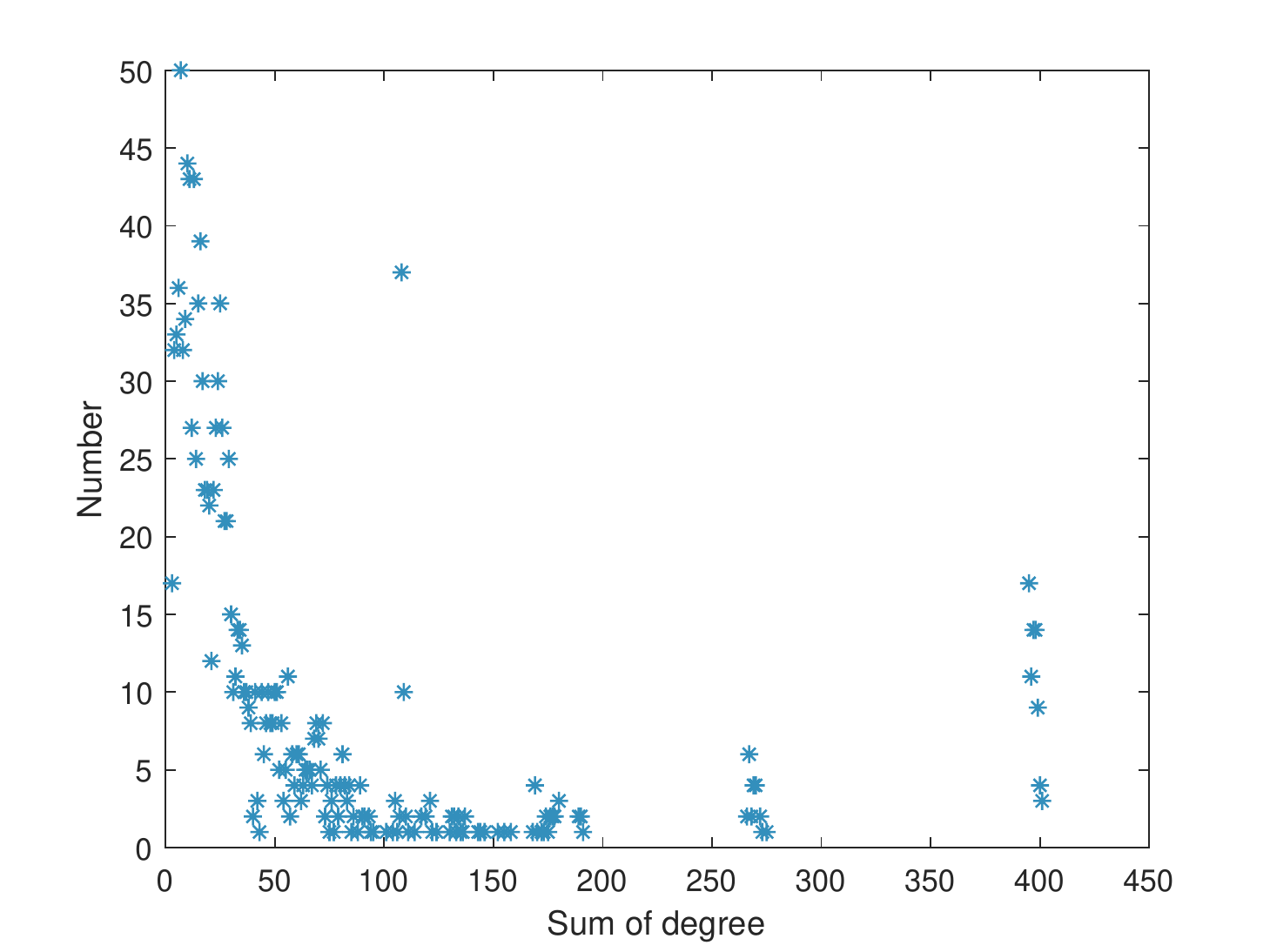}
\end{minipage}%
}%
\subfigure[$\eta=-5$]{
\begin{minipage}[t]{0.33\linewidth}
\centering
\includegraphics[width=4.5cm]{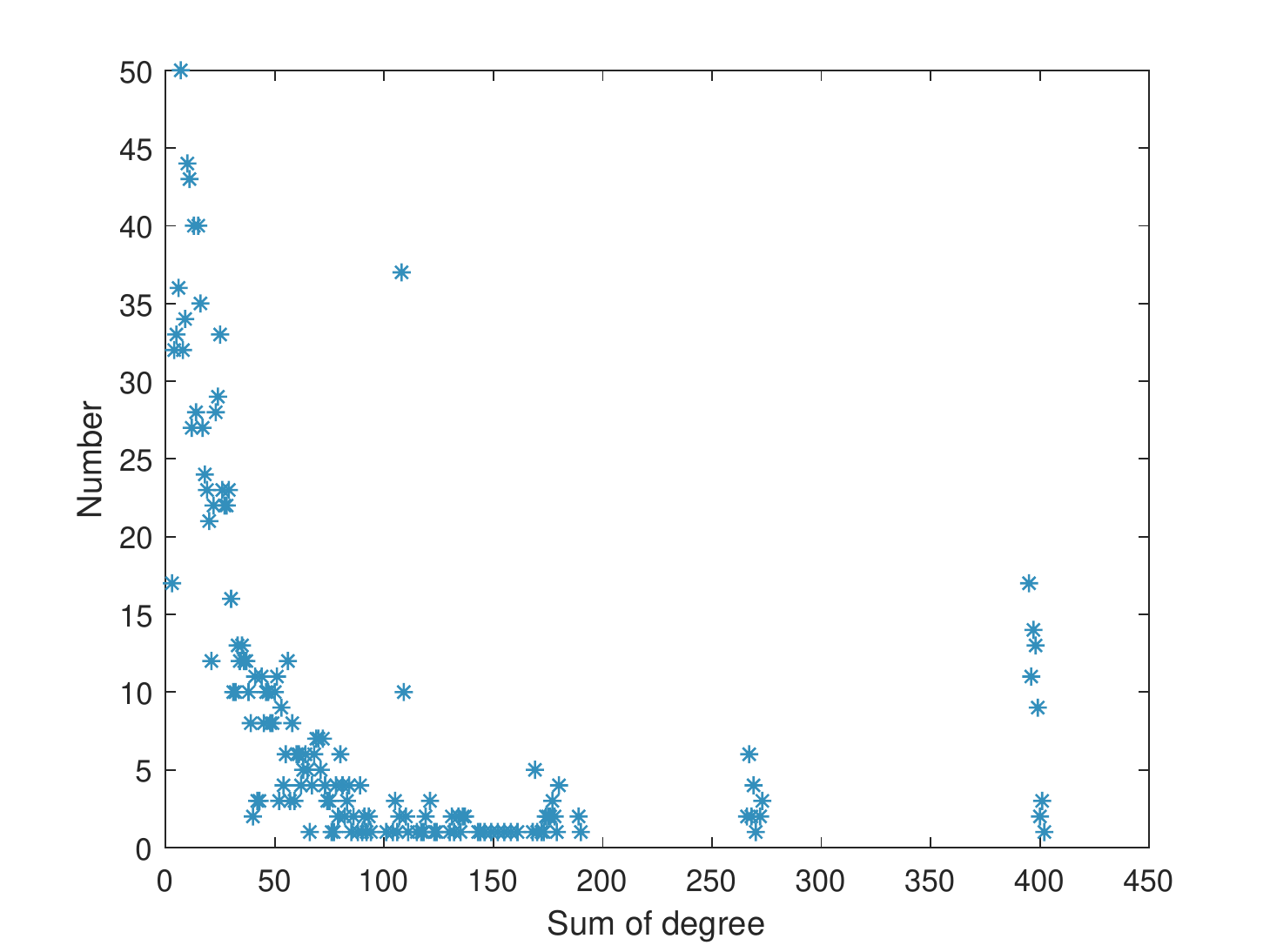}
\end{minipage}%
}%
\quad
\subfigure[$\eta=-1$]{
\begin{minipage}[t]{0.33\linewidth}
\centering
\includegraphics[width=4.5cm]{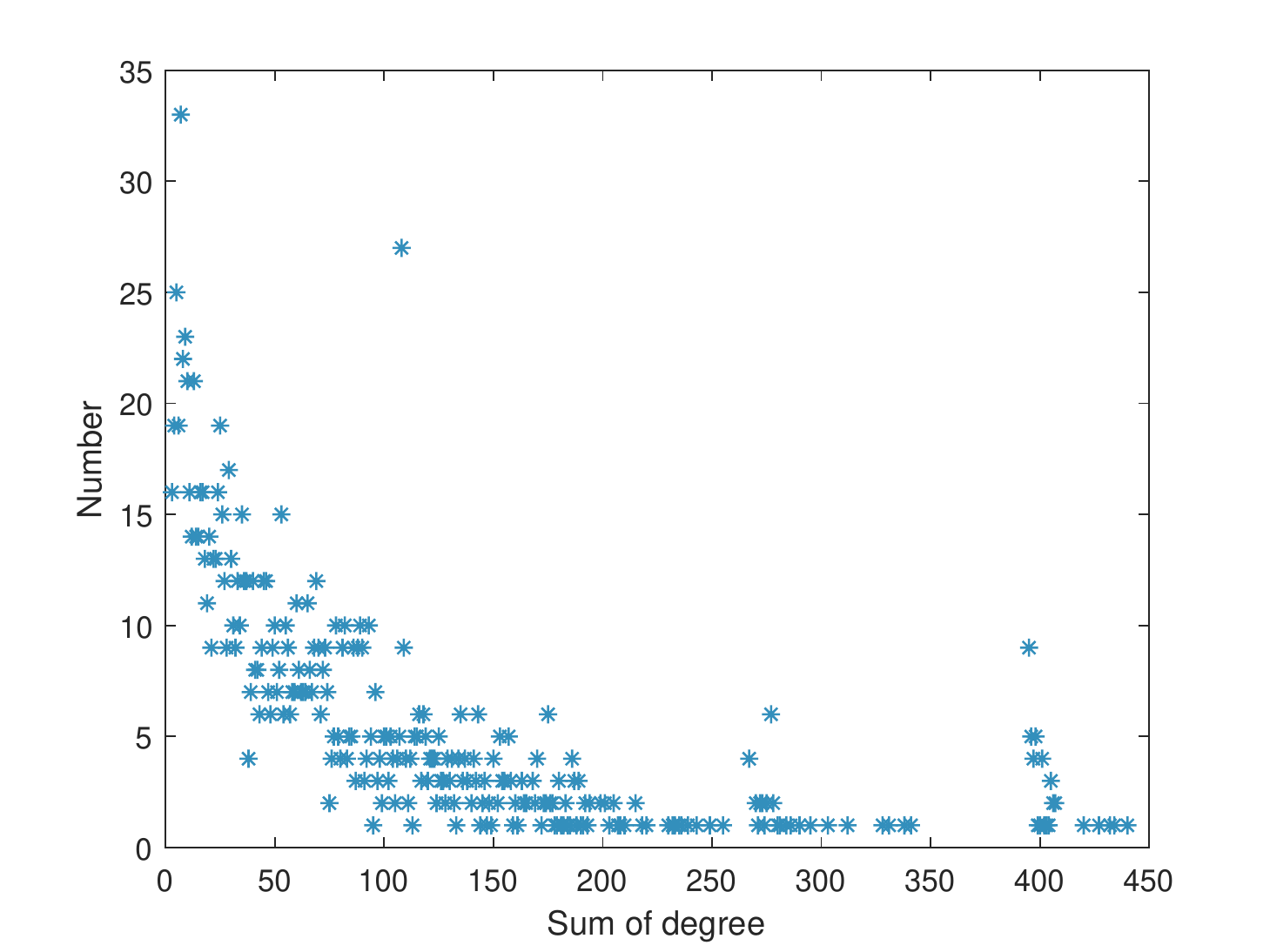}
\end{minipage}%
}%
\subfigure[$\eta=0$]{
\begin{minipage}[t]{0.33\linewidth}
\centering
\includegraphics[width=4.5cm]{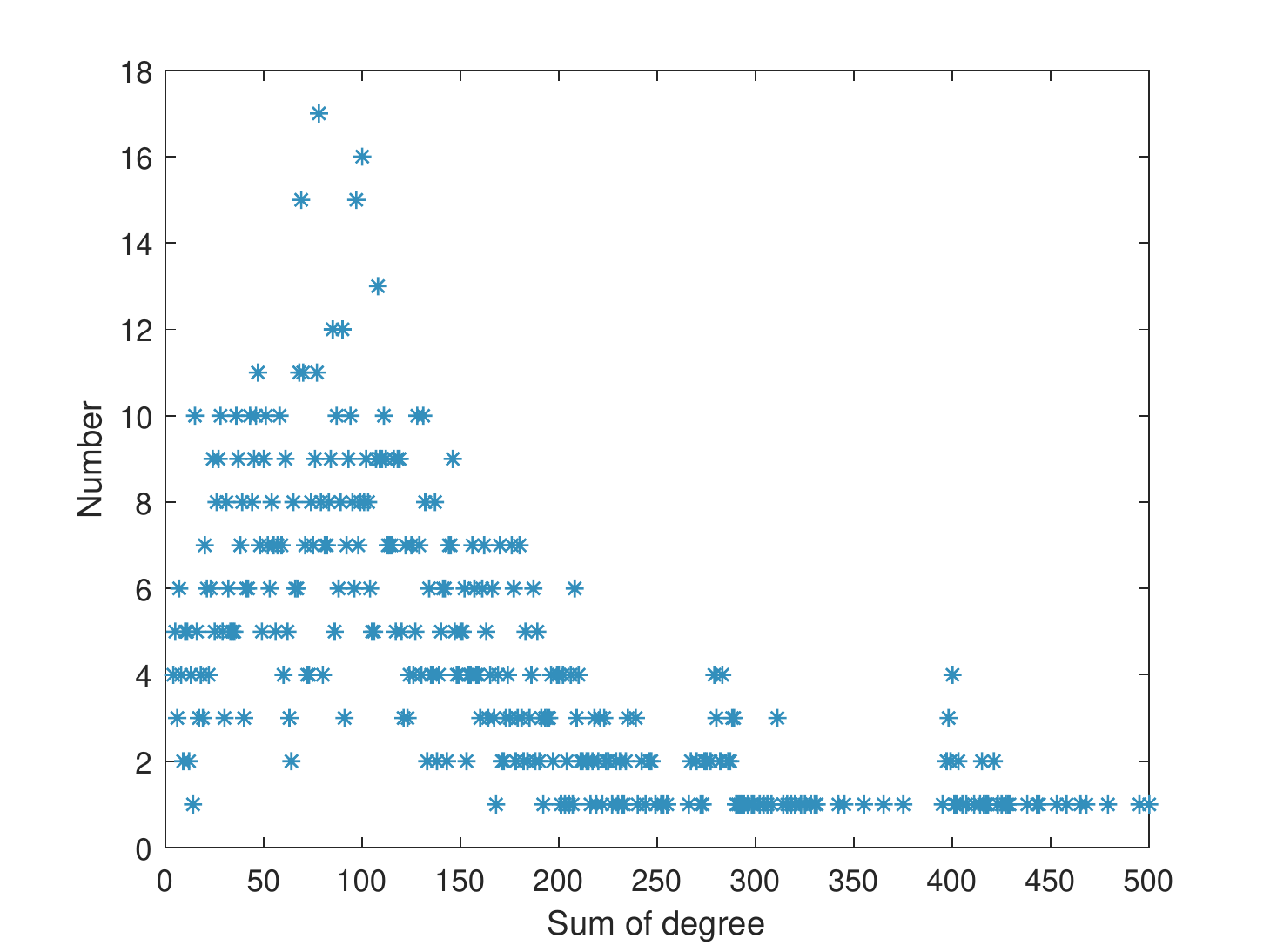}
\end{minipage}%
}%
\subfigure[$\eta=1$]{
\begin{minipage}[t]{0.33\linewidth}
\centering
\includegraphics[width=4.5cm]{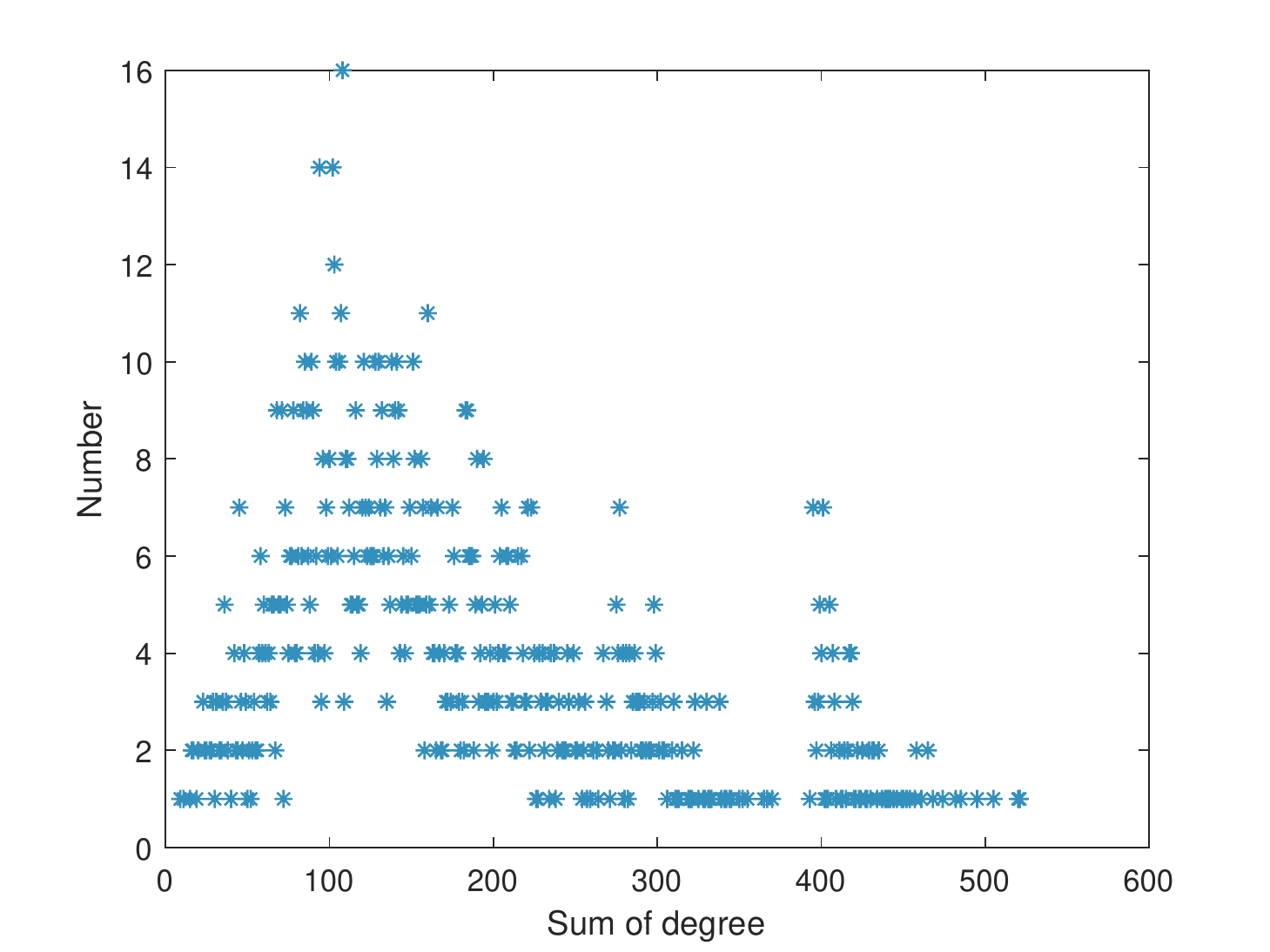}
\end{minipage}%
}%
\quad
\subfigure[$\eta=5$]{
\begin{minipage}[t]{0.33\linewidth}
\centering
\includegraphics[width=4.5cm]{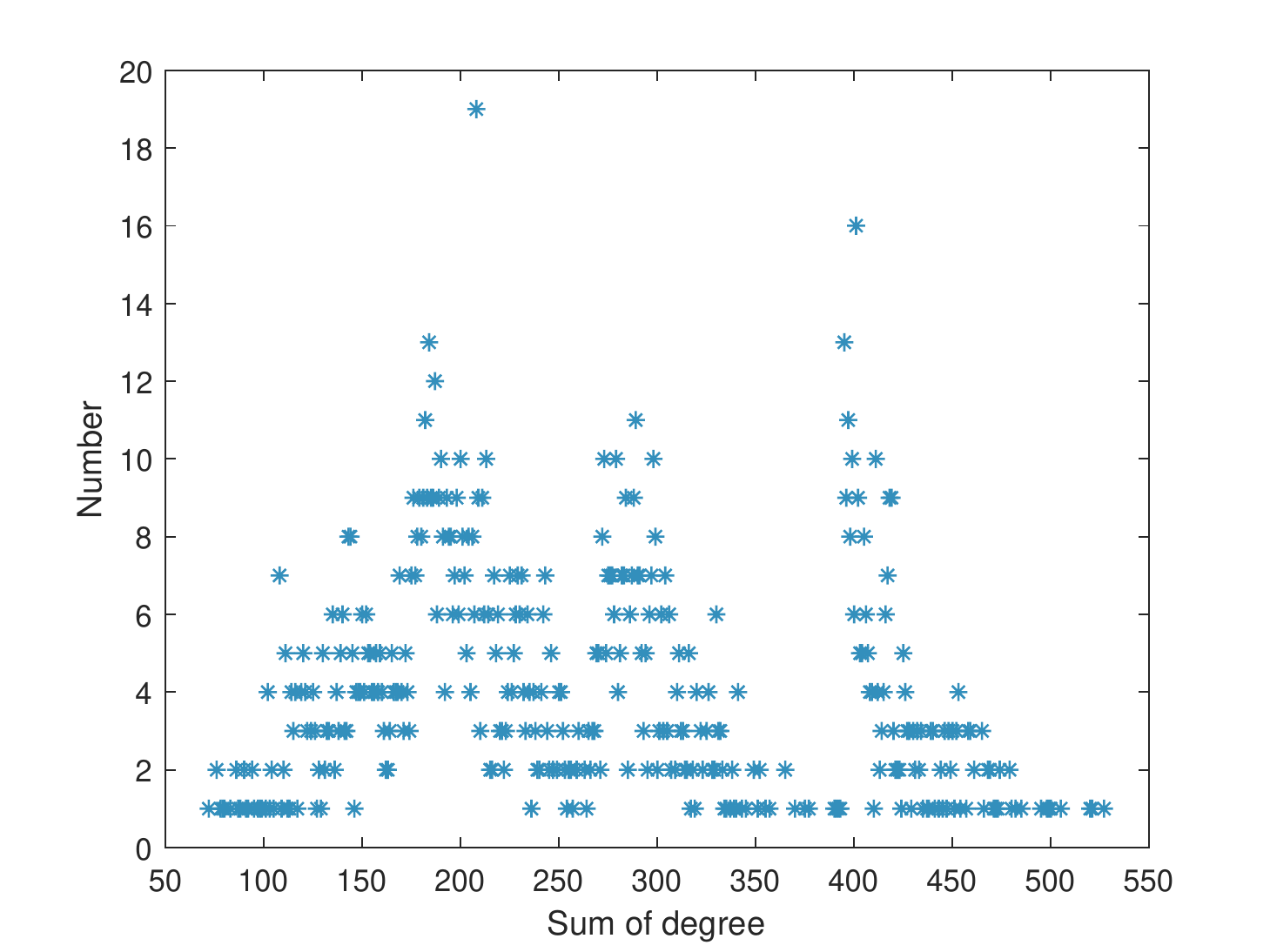}
\end{minipage}%
}%
\subfigure[$\eta=10$]{
\begin{minipage}[t]{0.33\linewidth}
\centering
\includegraphics[width=4.5cm]{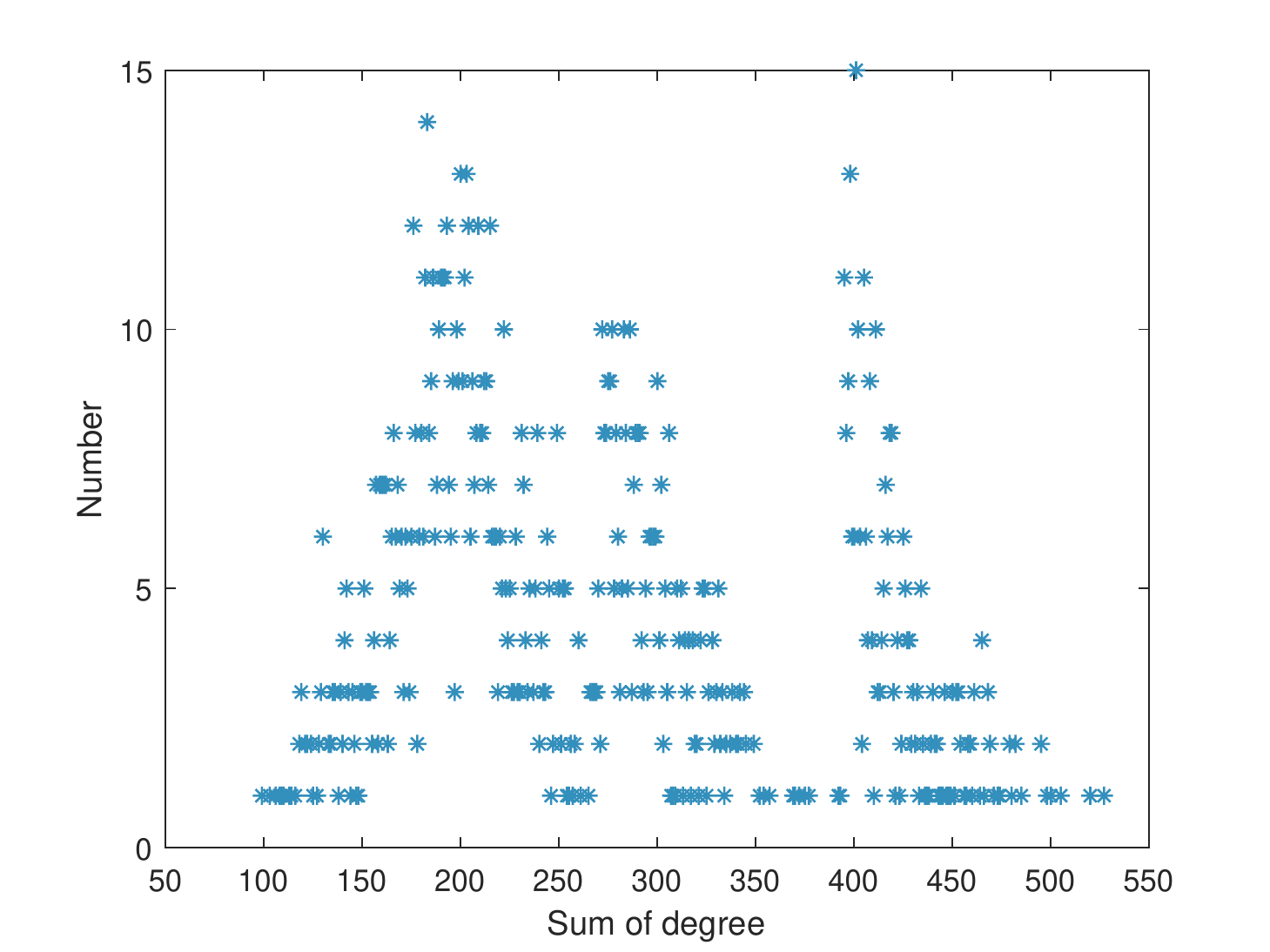}
\end{minipage}%
}%
\subfigure[$\eta=20$]{
\begin{minipage}[t]{0.33\linewidth}
\centering
\includegraphics[width=4.5cm]{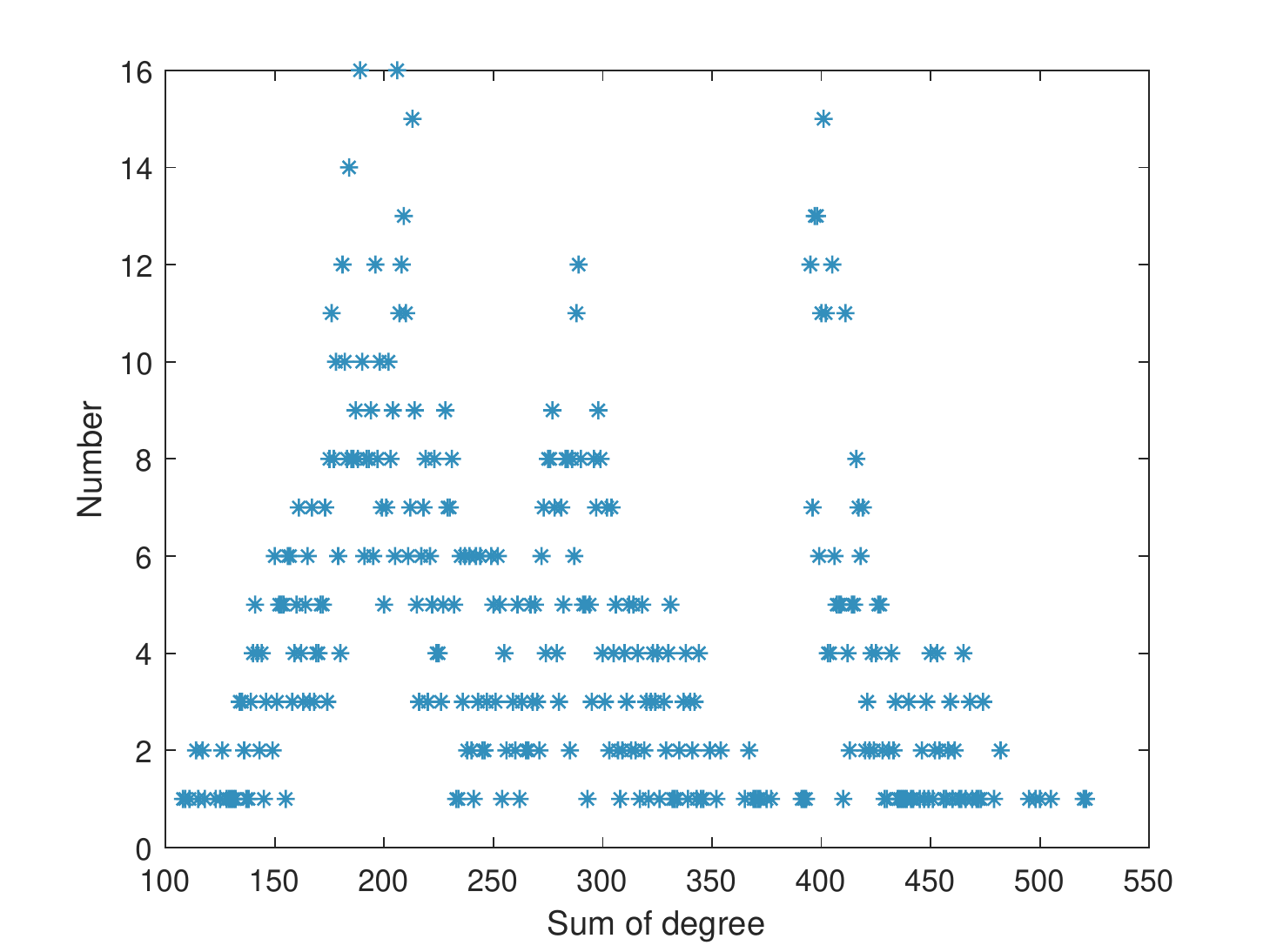}
\end{minipage}%
}%
\centering
\caption{Sums of degree for pairs of nodes connected by an intralayer link removed by the global perturbation strategy on dataset FT1. The horizontal ordinate is the sum of degrees, while the vertical ordinate is the number of times the value of a sum occurs.}
\label{pics_deg_dellinks_d1}
\end{figure*}

The phenomena observed in Figs.~\ref{pic:result_real_diffeta_dmeth1} and ~\ref{pic:result_real_diffeta_map_dmeth1} illustrate that the effects of different types of intralayer links selected by the parameter $\eta$ of the global strategy for the interlayer link prediction are different. There is a certain regularity of the effects on the prediction.
When $\eta$ is very small, almost all of the removed intralayer links are the links connected to nodes with low degree, including types of \textbf{GLT1}, \textbf{GLT4}, and \textbf{GLT6} in Fig.~\ref{pic:intralayer_link_types}. The perturbation weight for each intralayer link does not change much when $-20\leq\eta<-5$ since $\eta$ is an exponent. Therefore, the $P@30$ and MAP under the same $\xi$ hardly varies with $\eta$. When $\eta\geq-5$, the proportion of other types of intralayer links that are removed begins to increase. The $P@30$ and MAP start to increase with the increase of $\eta$. This indicates that the intralayer links connected to nodes with a low degree have a greater impact on the interlayer link prediction algorithms; other types of intralayer links have relatively less impact on it. When $\eta=0$, the perturbation weights for all intralayer links are the same, all intralayer links will be removed with the same probability. The $P@30$ and MAP do not get the maximum or minimum value when $\eta=0$, which again illustrates that different types of intralayer links selected by the parameter $\eta$ of the global strategy have different effects on interlayer link prediction, when $\eta>0$, intralayer links connected to nodes with high degree start to obtain larger perturbation weights. When $\eta=3$, almost all the curves of $P@30$ and MAP for different $\xi$ in each subfigure achieve the maximum value. When $3<\eta\leq20$, the $P@30$ and MAP under the same $\xi$ hardly varies with $\eta$. This is because $\eta$ is an exponent, the perturbation weight for each intralayer link does not change much when $3<\eta\leq20$. In this interval, almost all of the removed intralayer links are the links connected to nodes with high degree, including types of \textbf{GLT2}, \textbf{GLT3}, and \textbf{GLT5} in Fig.~\ref{pic:intralayer_link_types}. This indicates that the intralayer links connected to nodes with a high degree have a smaller impact on the interlayer link prediction algorithms; other types of intralayer links have a greater impact.
\begin{figure*}
    \centering
    \includegraphics[width=12cm,height=10.5cm]{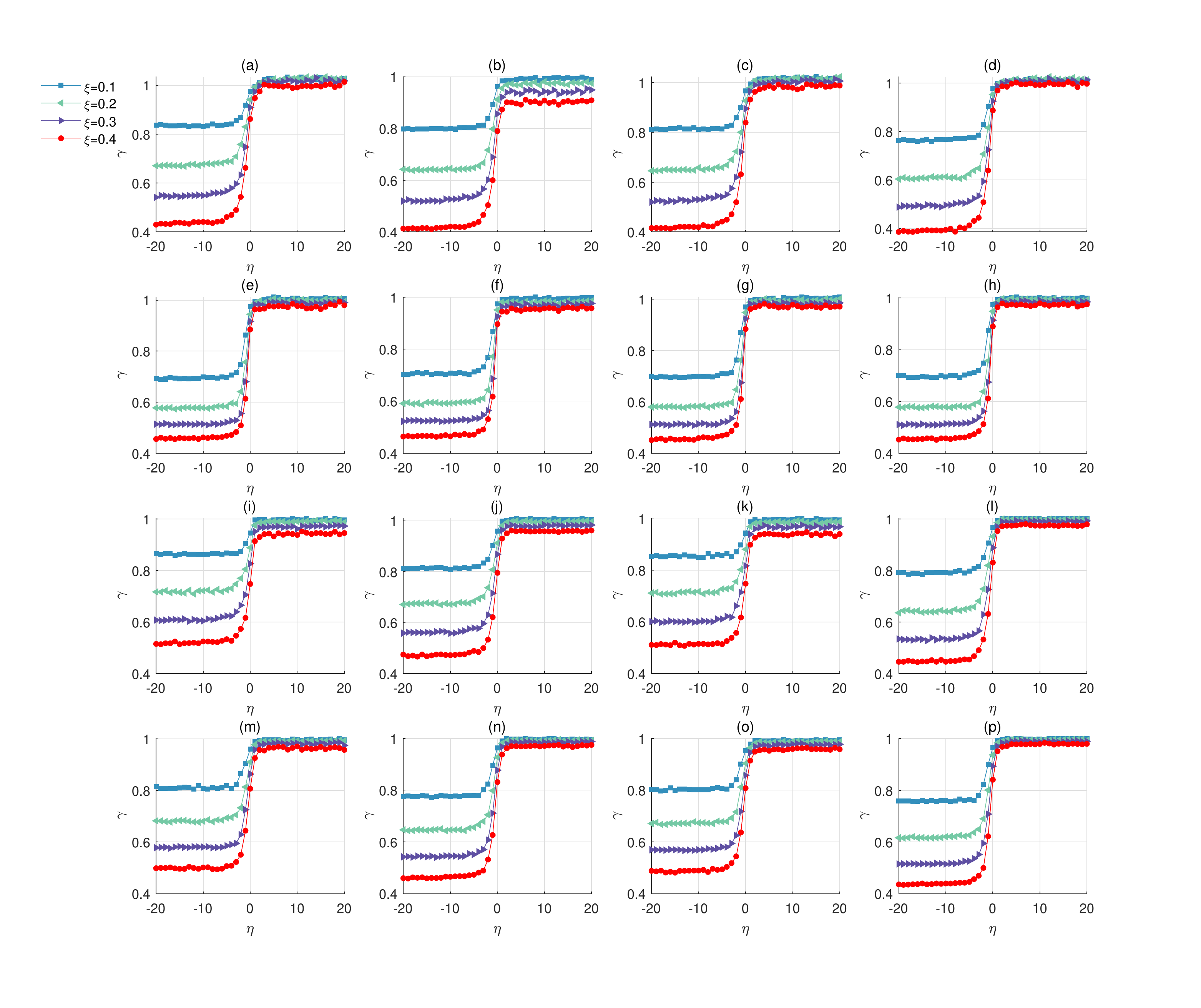}
    \caption{$\gamma$ of different interlayer link prediction algorithms on the multiplex networks perturbed by the global perturbation strategy with different $\eta$. $\gamma$ is the ratio between the $P@30$ on the perturbed networks and $P@30$ on the original networks. The datasets and prediction algorithms in different subfigures are the same as Fig.~\ref{pic:result_real_diffeta_dmeth1}.}
    \label{pic:result_real_diffeta_dmeth1_gamma}
\end{figure*}

\begin{figure*}
    \centering
    \includegraphics[width=12cm,height=10.5cm]{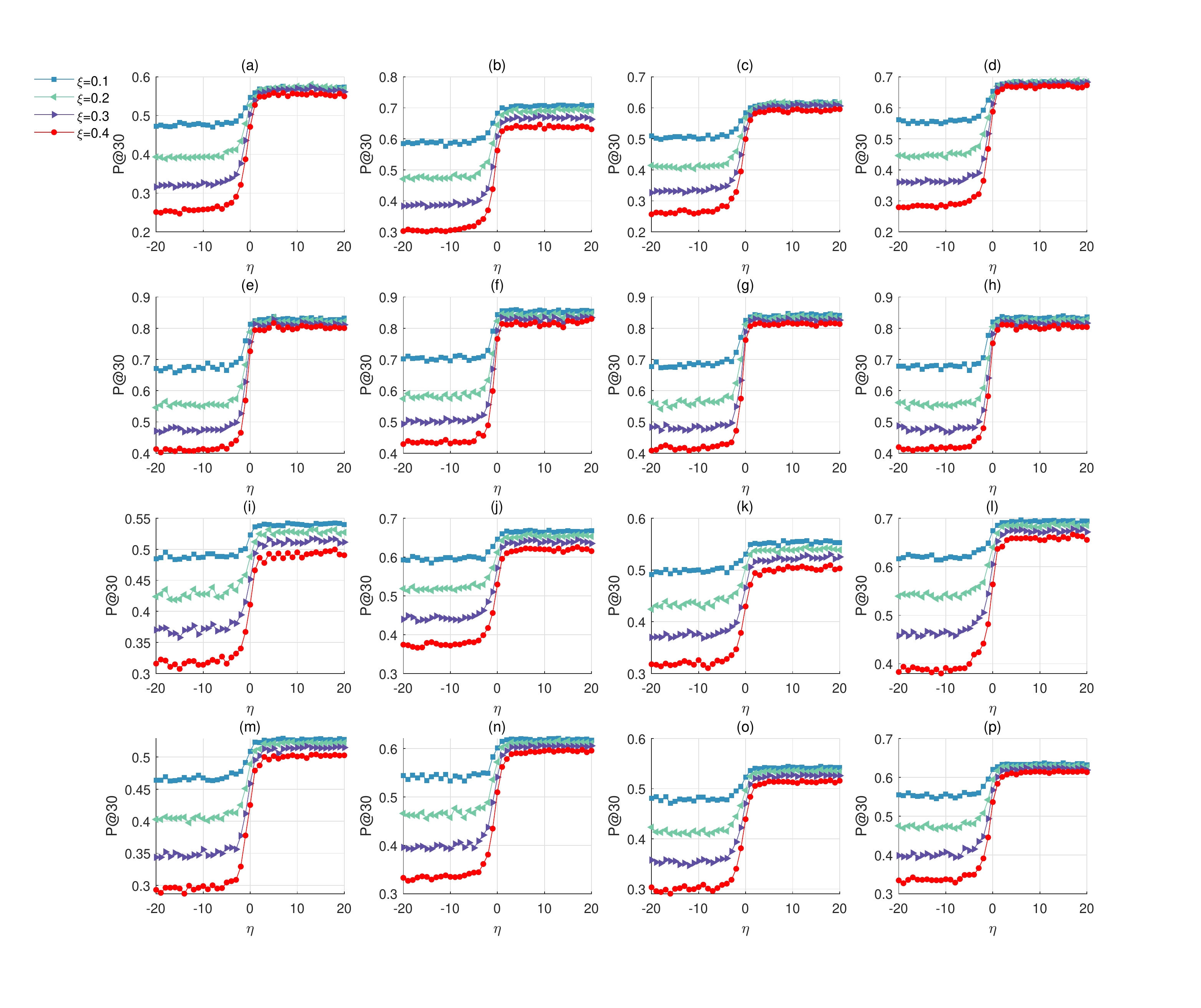}
    \caption{$P@30$ of different interlayer link prediction algorithms on the multiplex networks perturbed by the local perturbation strategy with different $\eta$. The datasets and prediction algorithms in different subfigures are the same as Fig.~\ref{pic:result_real_diffeta_dmeth1}.}
    \label{pic:result_real_diffeta_dmeth2}
\end{figure*}

To understand the mechanism of the global perturbation strategy deeply, we apply it to the network of Kapferer friendship interactions~\cite{kapferer1972strategy}, which has 39 nodes and 158 intralayer links; hence easy to visualize and analyze the influence in detail. Fig.~\ref{pics_dellinks_d1_exa} presents the perturbed network under different $\eta$. When $\eta=-20$, $\eta=-10$, and $\eta=-5$, the removed intralayer links are similar, connected to at least one node with low degree. All the intralayer links connected to the nodes with the lowest degree (equaling to 1 or 2) are removed. When $\eta=-1$, the removed intralayer links are less biased to nodes with a low degree than the subfigures of $\eta\leq-5$. Some intralayer links without connecting to nodes with a low degree are also removed. When $\eta=0$, the perturbation weights for all intralayer links are the same, all intralayer links have the same probability of being removed. The red line is not very regular. When $\eta=5$, $\eta=10$, and $\eta=20$, the removed intralayer links are also similar, connected to at least one node with a high degree. The observations in Fig.~\ref{pics_dellinks_d1_exa} are consistent with our analysis of the results of $P@30$ and MAP of different interlayer link prediction algorithms on the multiplex networks perturbed by the global perturbation strategy.

We further take dataset FT1 as an example to count the sum of degrees for pairs of nodes connected by an intralayer link removed by the global perturbation strategy. Fig.~\ref{pics_deg_dellinks_d1} presents the results. When $\eta$ is small, most of the sums of degrees are small. As $\eta$ increases, the sums of a degree increase. The observations in Fig.~\ref{pics_deg_dellinks_d1} are also consistent with our analysis of the results of $P@30$ and MAP of different interlayer link prediction algorithms on the multiplex networks perturbed by the global perturbation strategy.

Based on the above analysis, we can draw the following conclusion that the intralayer links of type \textbf{GLT4} and \textbf{GLT6} in Fig.~\ref{pic:intralayer_link_types} have a greater effects on interlayer link prediction than type \textbf{GLT5}, intralayer links of type \textbf{GLT5} has a greater effects than type \textbf{GLT1}, and intralayer links of type \textbf{GLT1} has a greater effects than type \textbf{GLT2} and \textbf{GLT3}.

Moreover, we analyzed the ratio between the $P@30$ on the perturbed networks and $P@30$ on the original networks, as shown in Fig.~\ref{pic:result_real_diffeta_dmeth1_gamma}. From the figure, we can see that the trend of $\gamma$ varying with $\eta$ is the same as that of $P@30$ and MAP varying with $\eta$. The value of $\gamma$ in most conditions is less than 1. This is because some of the intralayer links are removed by the global perturbation strategy. Less information on the multiplex networks leads to worse interlayer link prediction performance. This is consistent with our intuition. However, in some of the subfigures, when $\eta>3$, the value of $\gamma$ is greater than 1. For example, when $\xi=0.2$ and $\eta=10$, the CN algorithm's value of $\gamma$ is 1.03 on dataset FT1 while IDP is 1.02. The performance of the interlayer link prediction algorithms is unexpectedly improved by removing the intralayer link. When $\eta>3$, almost all of the removed intralayer links selected by the global perturbation strategy are connected to nodes with a high degree. This phenomenon indicates that the intralayer links connected to nodes with a high degree are not necessarily helpful for the interlayer link prediction. It is not that the more structural information, the more conducive to interlayer link prediction.

\textbf{(2) Result on local perturbation strategy.}
\begin{figure*}
    \centering
    \includegraphics[width=12cm,height=10.5cm]{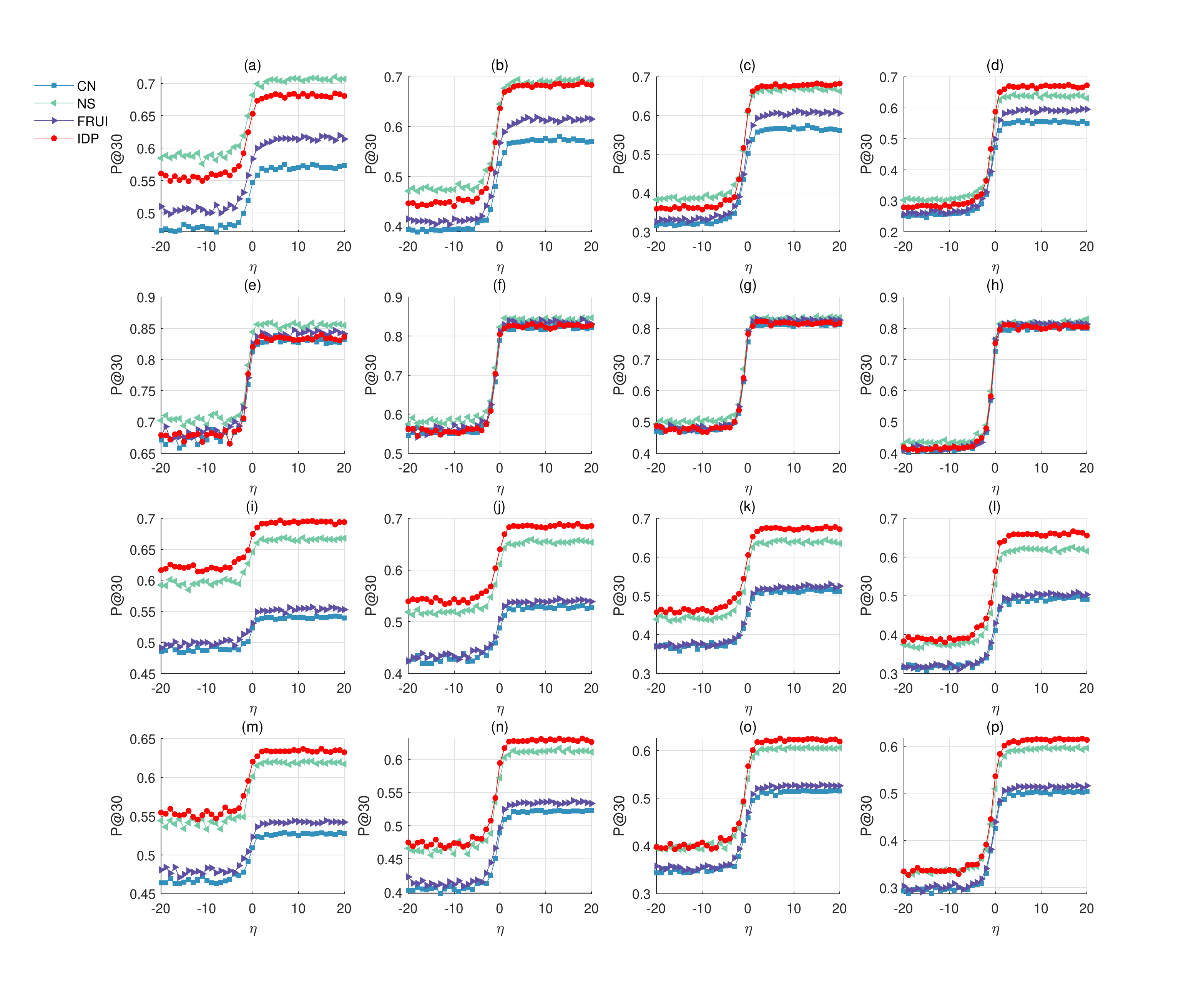}
        \caption{Comparison of different interlayer link prediction algorithms on the multiplex networks perturbed by the local perturbation strategy with different $\eta$. Each row uses the same dataset, and from top to bottom are datasets FT1, FT2, Higgs-SCMT, and Higgs-SCRT. Each column uses the same $\xi$, and from left to right are 0.1, 0.2, 0.3 and 0.4.}
        \label{pic:result_real_diffeta_p30_dmeth2_comppmethod}
\end{figure*}
\begin{figure*}
    \centering
    \includegraphics[width=12cm,height=10.5cm]{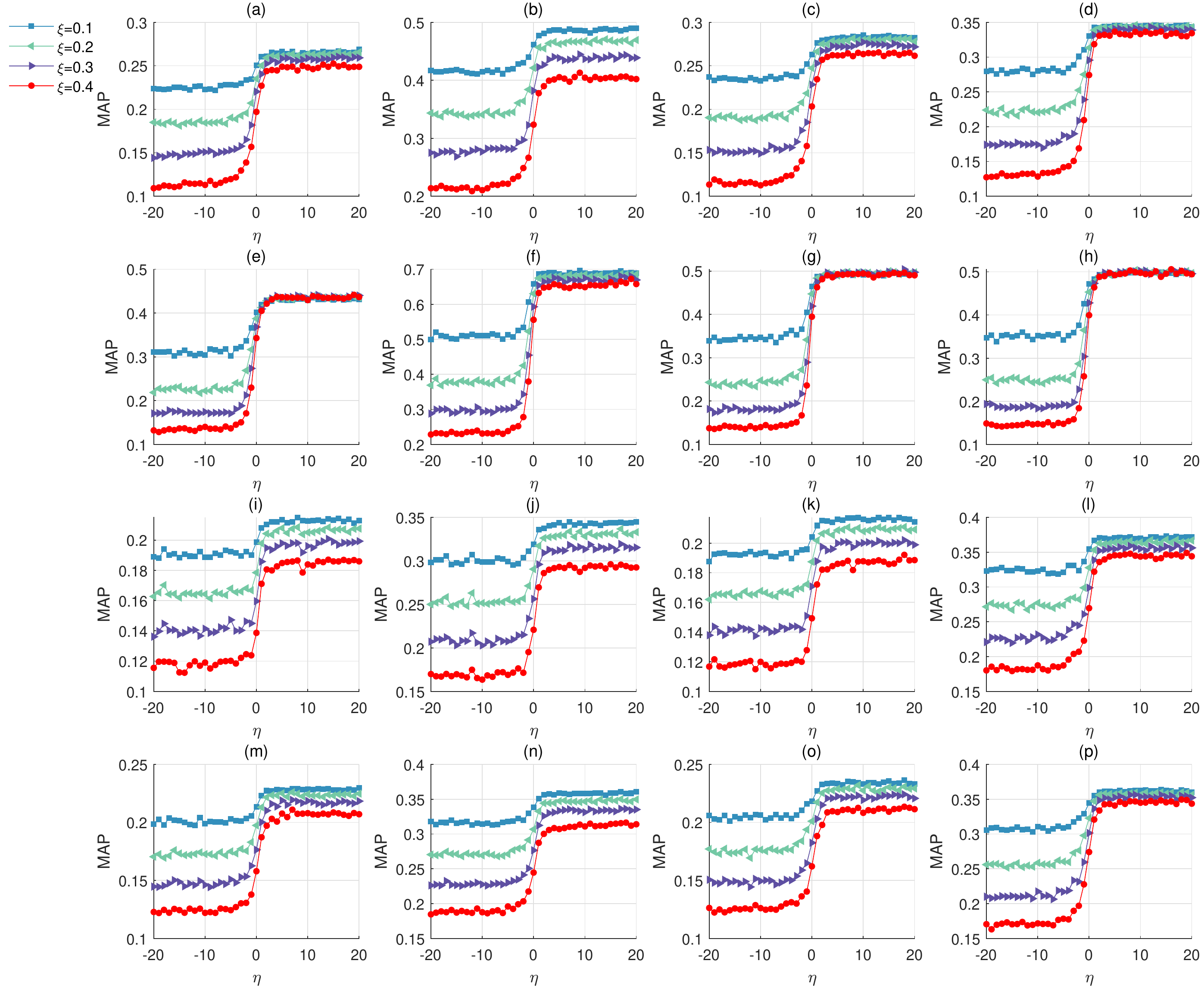}
    \caption{MAP of different interlayer link prediction algorithms on the multiplex networks perturbed by the local perturbation strategy with different $\eta$. The datasets and prediction algorithms in different subfigures are the same as Fig.~\ref{pic:result_real_diffeta_dmeth1}.}
    \label{pic:result_real_diffeta_map_dmeth2}
\end{figure*}

\begin{figure*}
\centering
\subfigure[$\eta=-20$]{
\begin{minipage}[t]{0.33\linewidth}
\centering
\includegraphics[width=4.5cm]{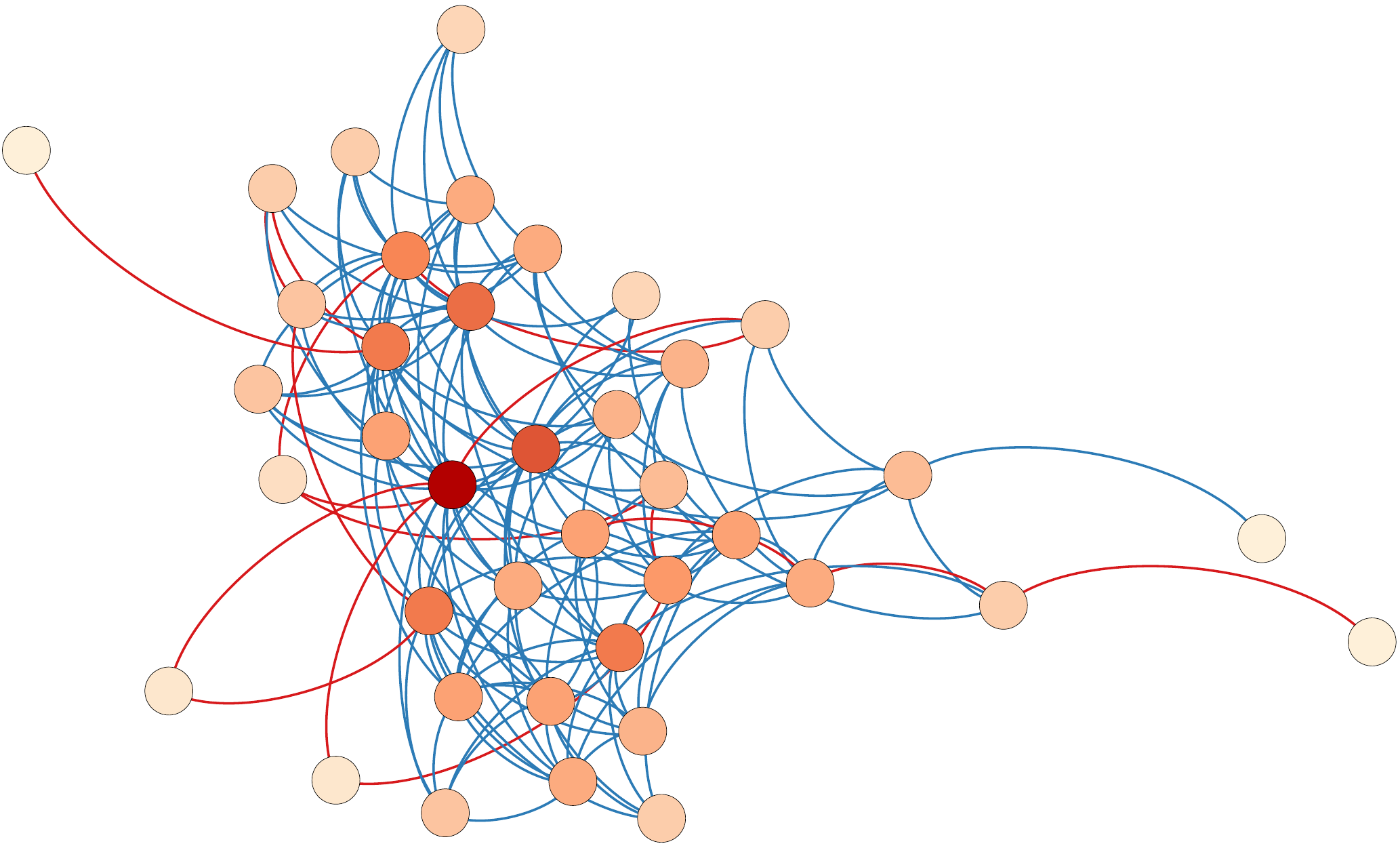}
\end{minipage}%
}%
\subfigure[$\eta=-10$]{
\begin{minipage}[t]{0.33\linewidth}
\centering
\includegraphics[width=4.5cm]{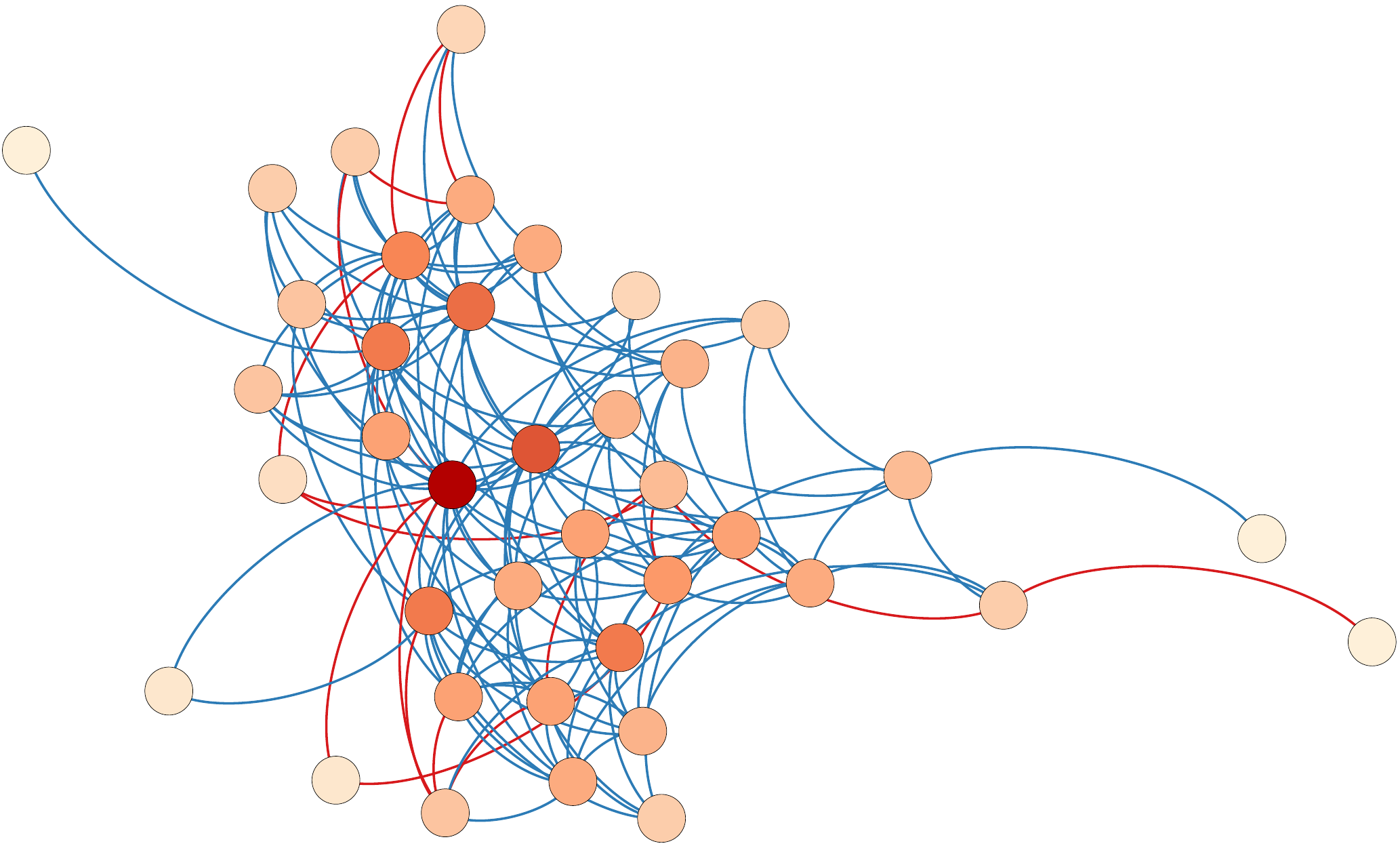}
\end{minipage}%
}%
\subfigure[$\eta=-5$]{
\begin{minipage}[t]{0.33\linewidth}
\centering
\includegraphics[width=4.5cm]{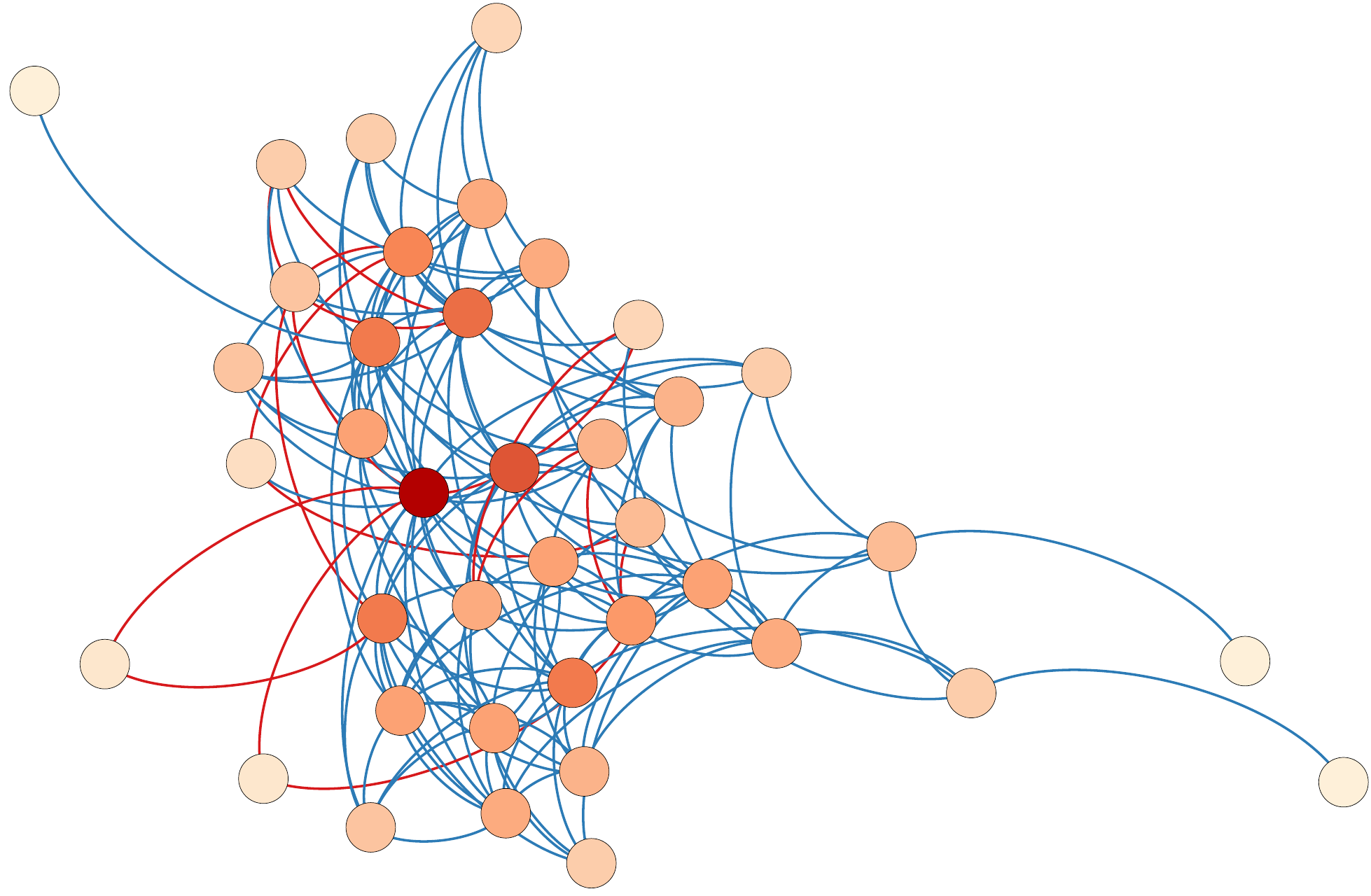}
\end{minipage}%
}%
\quad
\subfigure[$\eta=-1$]{
\begin{minipage}[t]{0.33\linewidth}
\centering
\includegraphics[width=4.5cm]{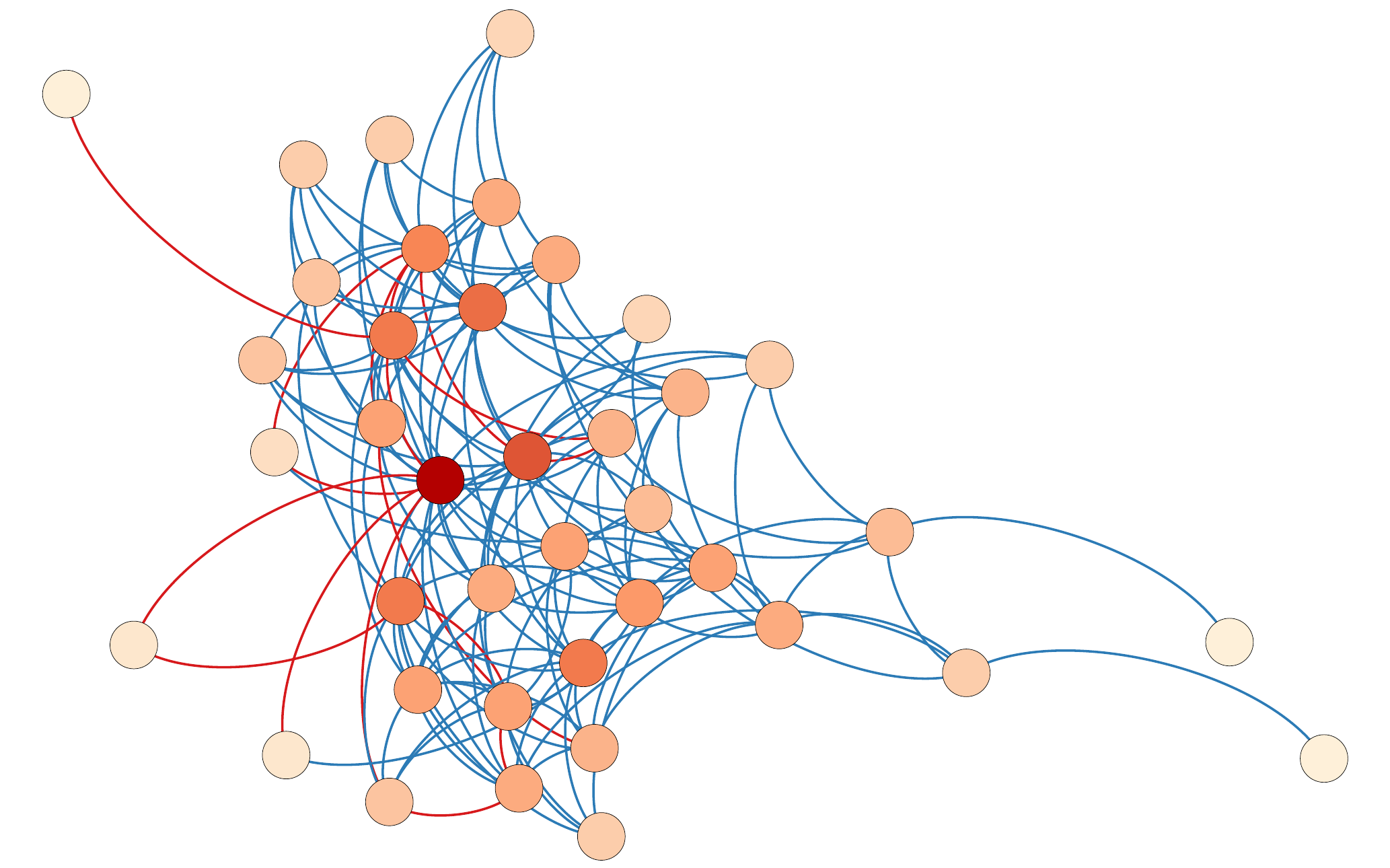}
\end{minipage}%
}%
\subfigure[$\eta=0$]{
\begin{minipage}[t]{0.33\linewidth}
\centering
\includegraphics[width=4.5cm]{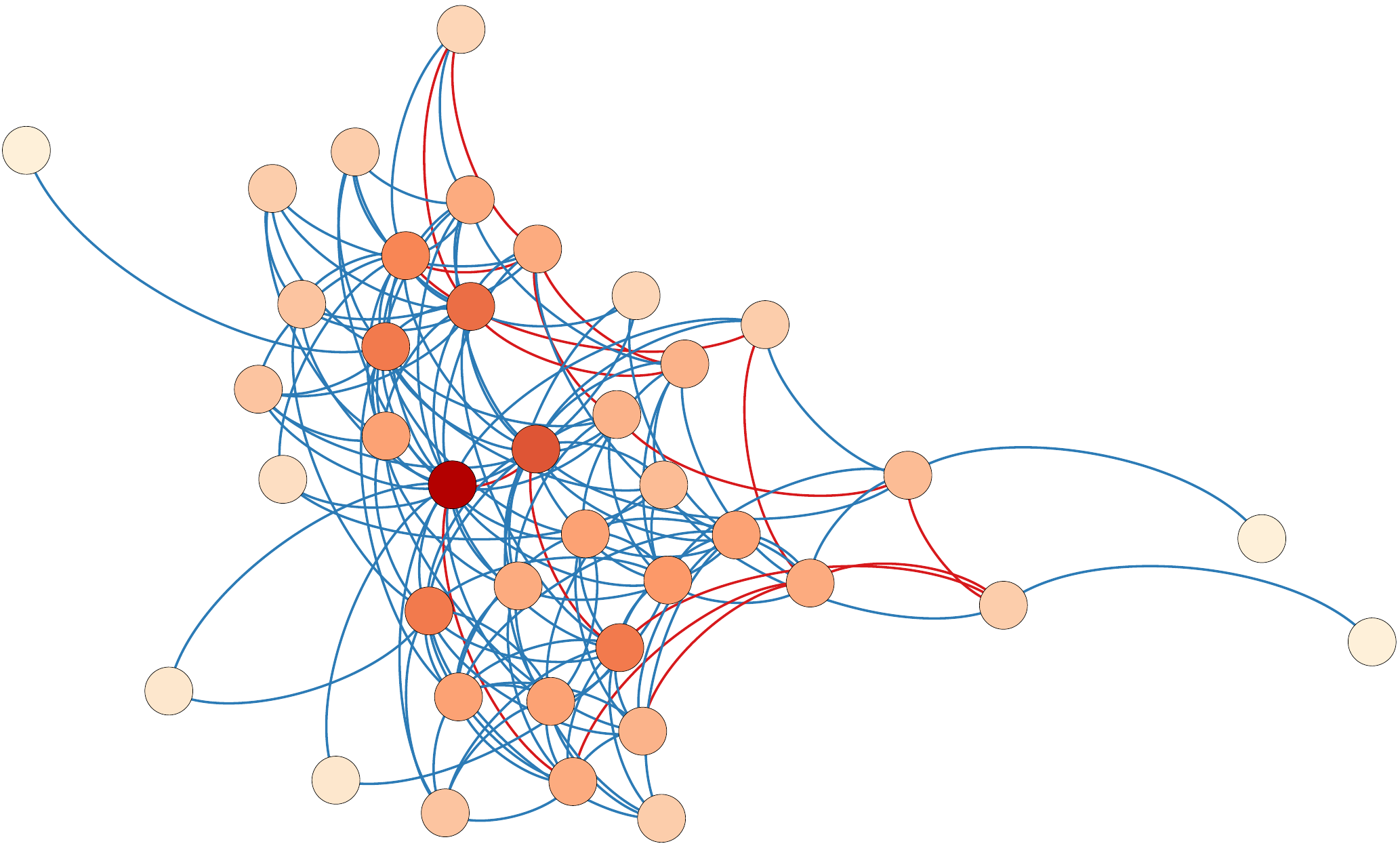}
\end{minipage}%
}%
\subfigure[$\eta=1$]{
\begin{minipage}[t]{0.33\linewidth}
\centering
\includegraphics[width=4.5cm]{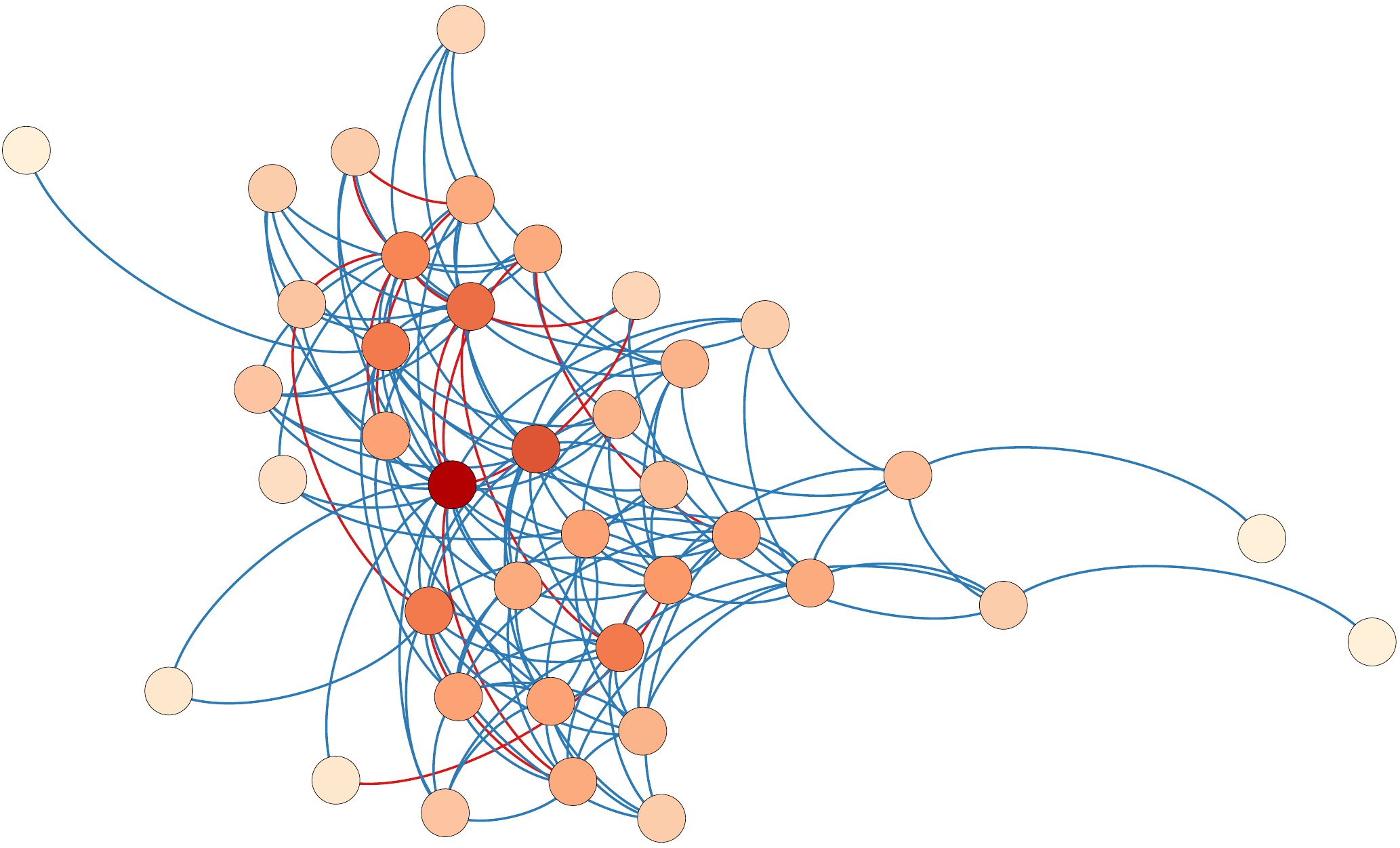}
\end{minipage}%
}%
\quad
\subfigure[$\eta=5$]{
\begin{minipage}[t]{0.33\linewidth}
\centering
\includegraphics[width=4.5cm]{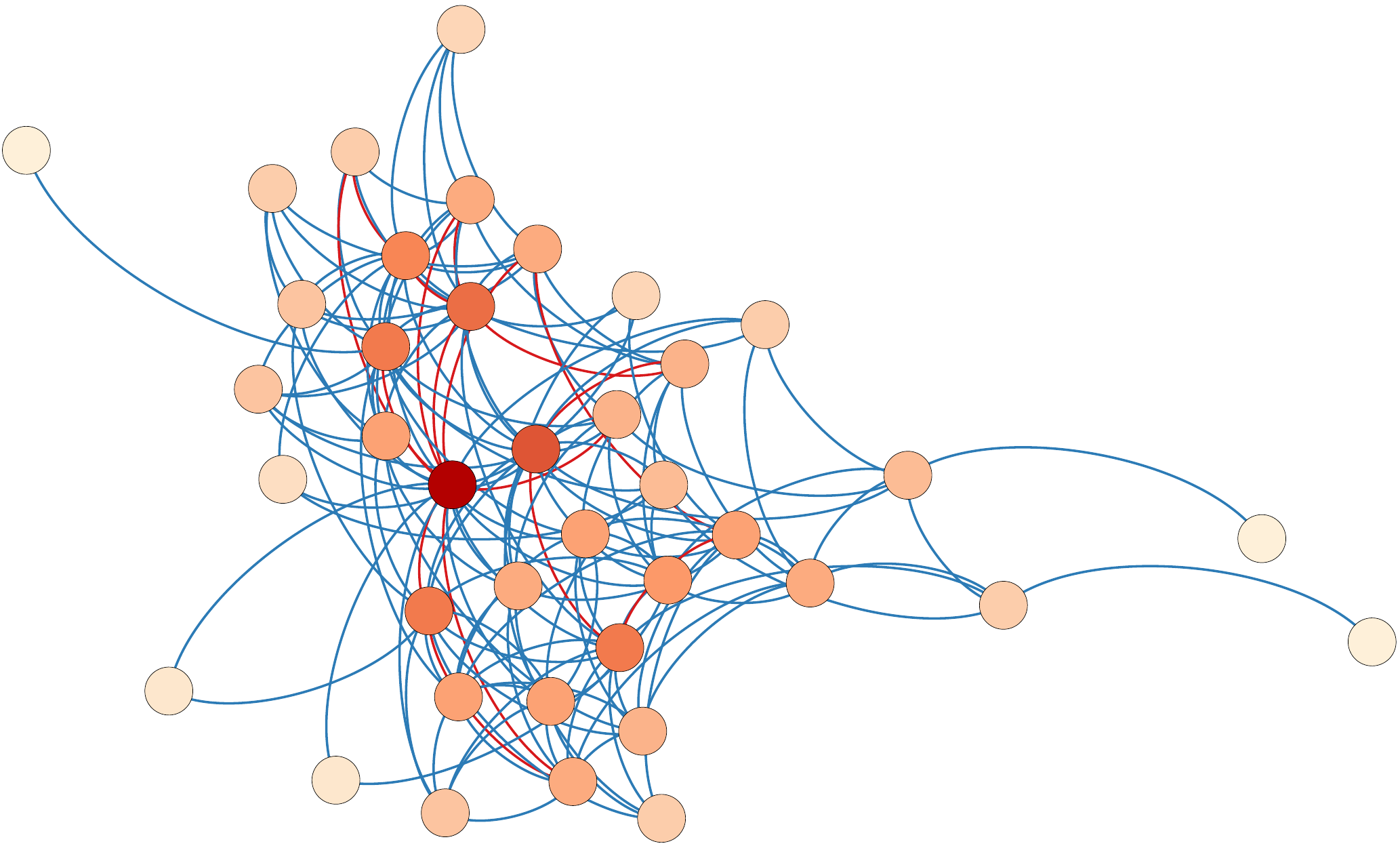}
\end{minipage}%
}%
\subfigure[$\eta=10$]{
\begin{minipage}[t]{0.33\linewidth}
\centering
\includegraphics[width=4.5cm]{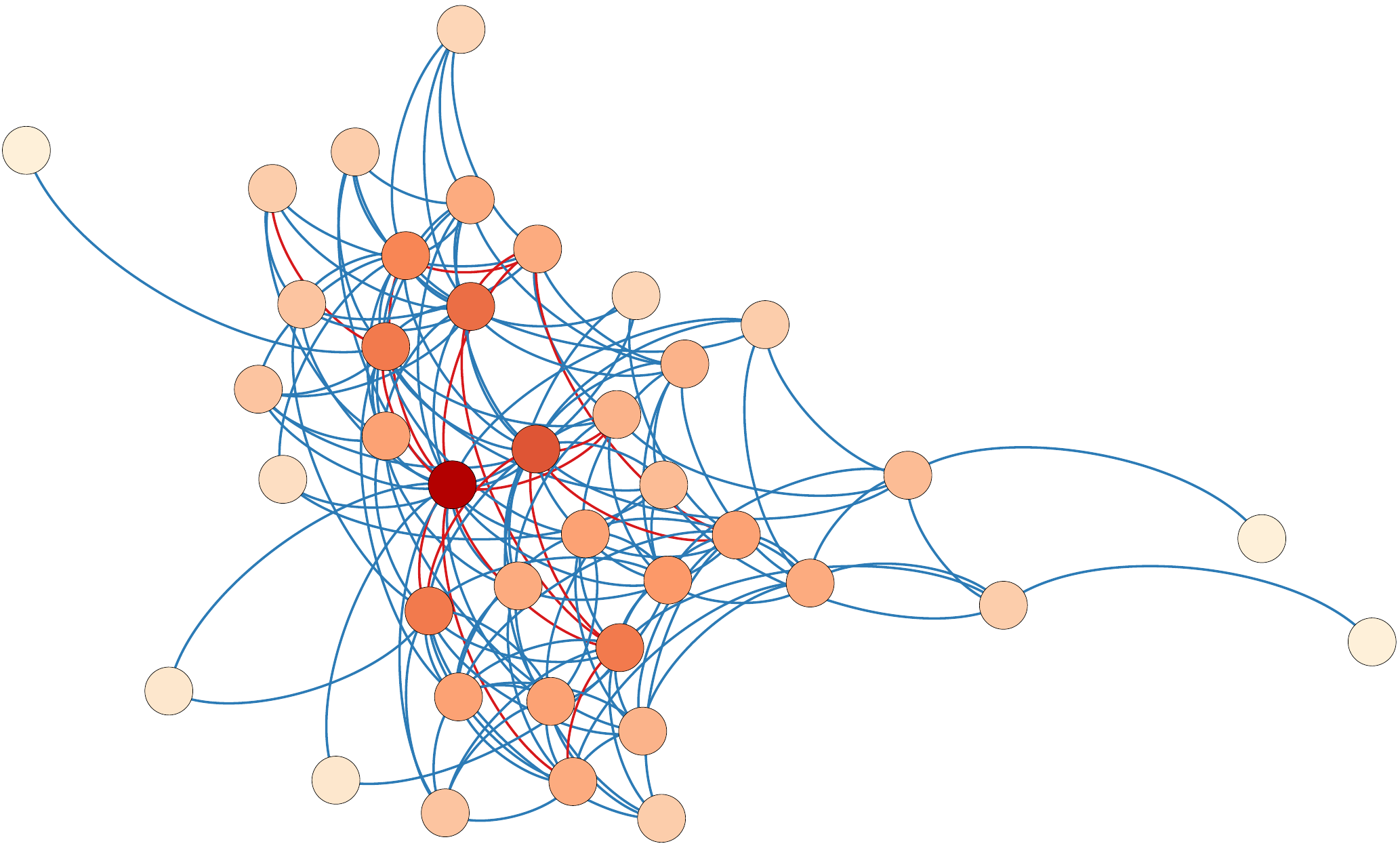}
\end{minipage}%
}%
\subfigure[$\eta=20$]{
\begin{minipage}[t]{0.33\linewidth}
\centering
\includegraphics[width=4.5cm]{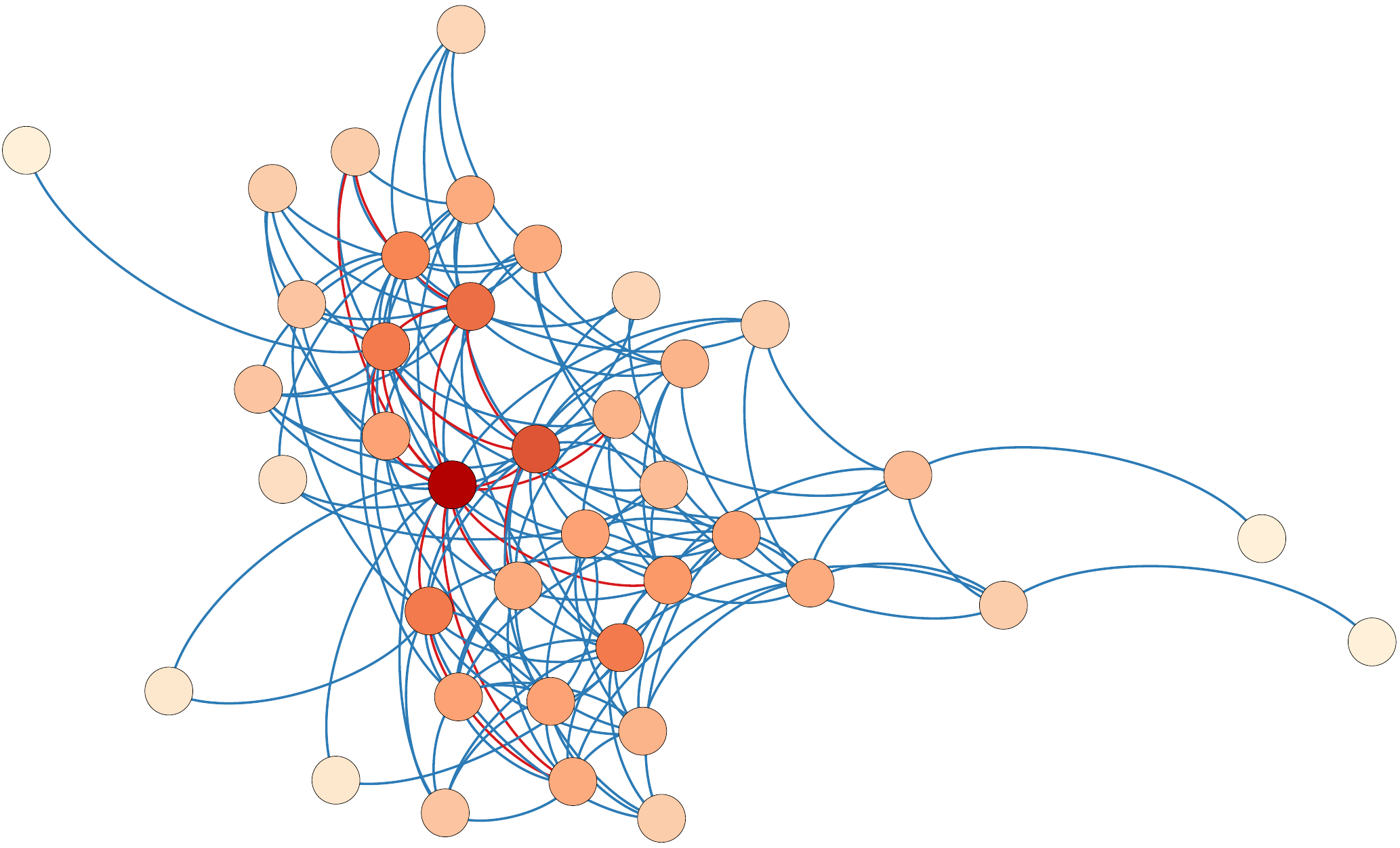}
\end{minipage}%
}%
\centering
\caption{Example of intralayer links removed by the local perturbation strategy. The value of $\xi$ and the meaning of nodes and links in each subfigure is the same as Fig~\ref{pics_dellinks_d1_exa}.}
\label{pics_dellinks_d2_exa}
\end{figure*}

Figure~\ref{pic:result_real_diffeta_dmeth2} is the $P@30$ of different prediction algorithms on the four kinds of real-world multiplex networks perturbed by the local perturbation strategy under the same settings as the global strategy. From the figure, we can see that:
(i) For a given $\xi$, $P@30$ of different prediction algorithms exhibit the same trend that it has almost no change firstly, then increases rapidly, and finally becomes stable with an increase in $\eta$. The differences in datasets, prediction algorithms, and $\xi$ did not change this trend.
(ii) When $-20\leq \eta \leq -5$, $P@30$ of the four prediction algorithms are generally stable and $P@30$ hardly varies with $\eta$ on different datasets. For example, when $\xi=0.1, \eta=-20$, the $P@30$ of CN is 0.4723 and when $\xi=0.1, \eta=-10$, the $P@30$ of CN is 0.4764 on the FT1 dataset. The difference of these two $P@30$ is 0.0041, which is very small.
(iii) When $-5<\eta\leq3$, $P@30$ of different prediction algorithms shows a trend of increasing with an increase in $\eta$. The increasing trend was first slowly growing, then rapidly increasing, and finally slowly growing again.
(iv) When $3<\eta\leq20$, $P@30$ of the four prediction algorithms is stable again. In a same $\xi$, $P@30$ hardly varies with $\eta$.
(v) The differences of $P@30$ for different $\xi$ in a same $\eta$ are very small when $3<\eta\leq20$. In contrast, the differences of $P@30$ for different $\xi$ in a same $\eta$ are relatively large when $\eta<-5$.

Similar to experimental result analysis in global perturbation strategy, we put the results of four interlayer link prediction algorithms with fixed parameter $\xi$ in each subfigure in Fig.~\ref{pic:result_real_diffeta_p30_dmeth2_comppmethod} to compare the different interlayer link prediction algorithms. As can be seen from the figure, the rankings of the $P@30$ of these four interlayer link prediction algorithms are almost identical under different $\xi$ in each dataset. NS and IDP have better predictive performance than FRUI and CN.

We also investigate the effects of the local perturbation strategy on the MAP. Fig. ~\ref{pic:result_real_diffeta_map_dmeth2} is the experimental results. As can be seen from the figure, the effects of the local perturbation strategy on MAP is almost the same as that of $P@30$.
\begin{figure*}
\centering
\subfigure[$\eta=-20$]{
\begin{minipage}[t]{0.33\linewidth}
\centering
\includegraphics[width=4.5cm]{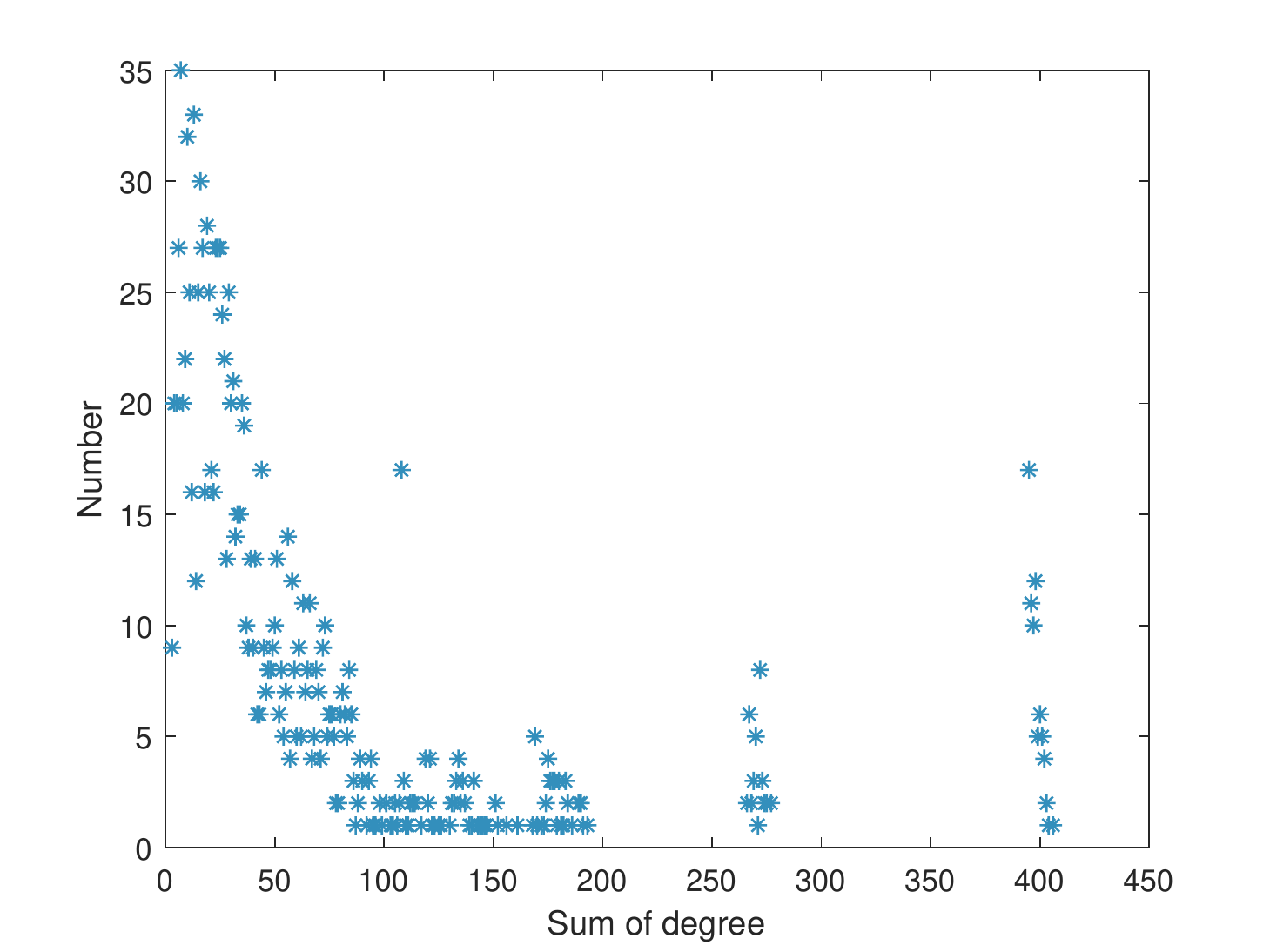}
\end{minipage}%
}%
\subfigure[$\eta=-10$]{
\begin{minipage}[t]{0.33\linewidth}
\centering
\includegraphics[width=4.5cm]{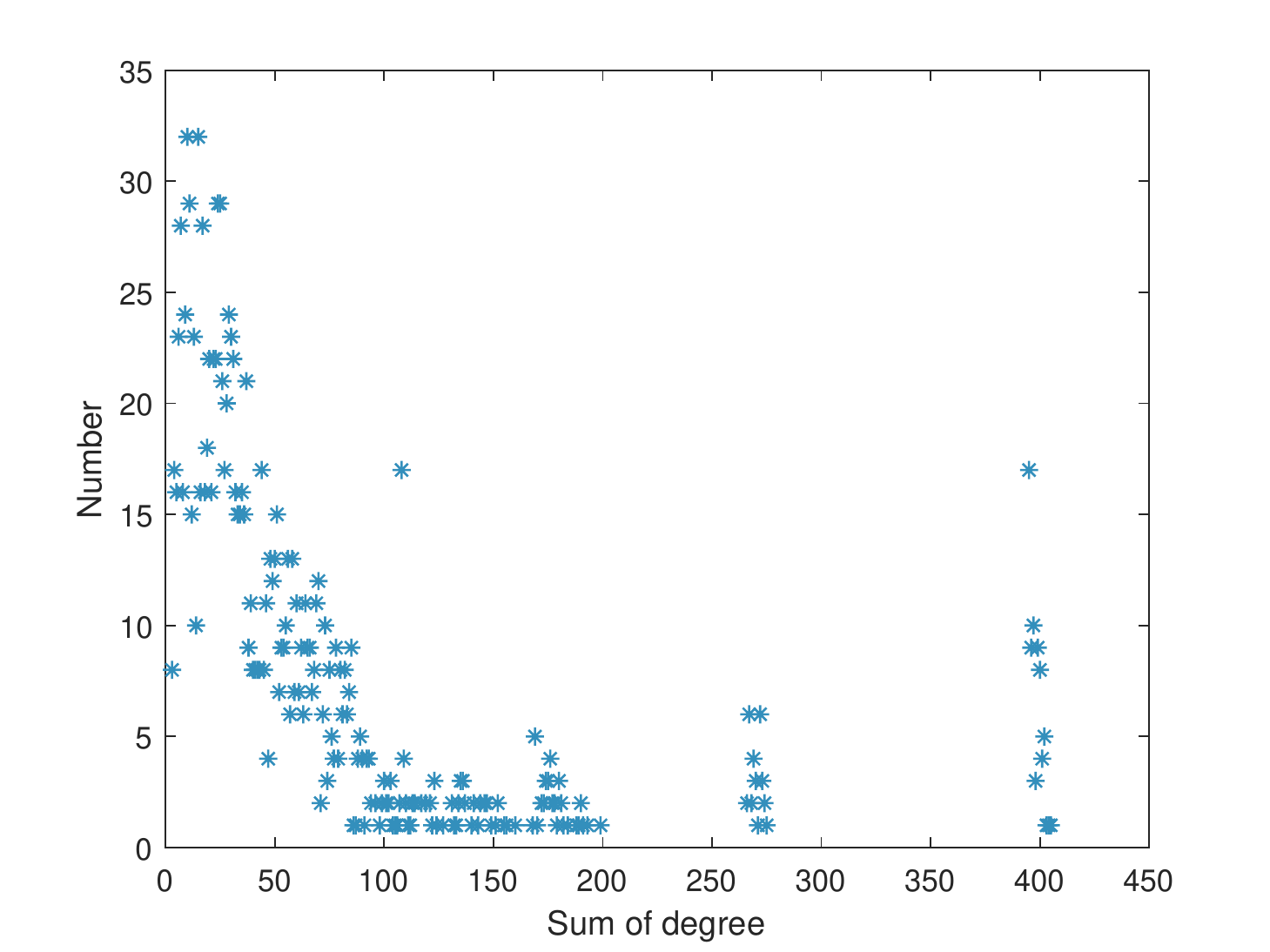}
\end{minipage}%
}%
\subfigure[$\eta=-5$]{
\begin{minipage}[t]{0.33\linewidth}
\centering
\includegraphics[width=4.5cm]{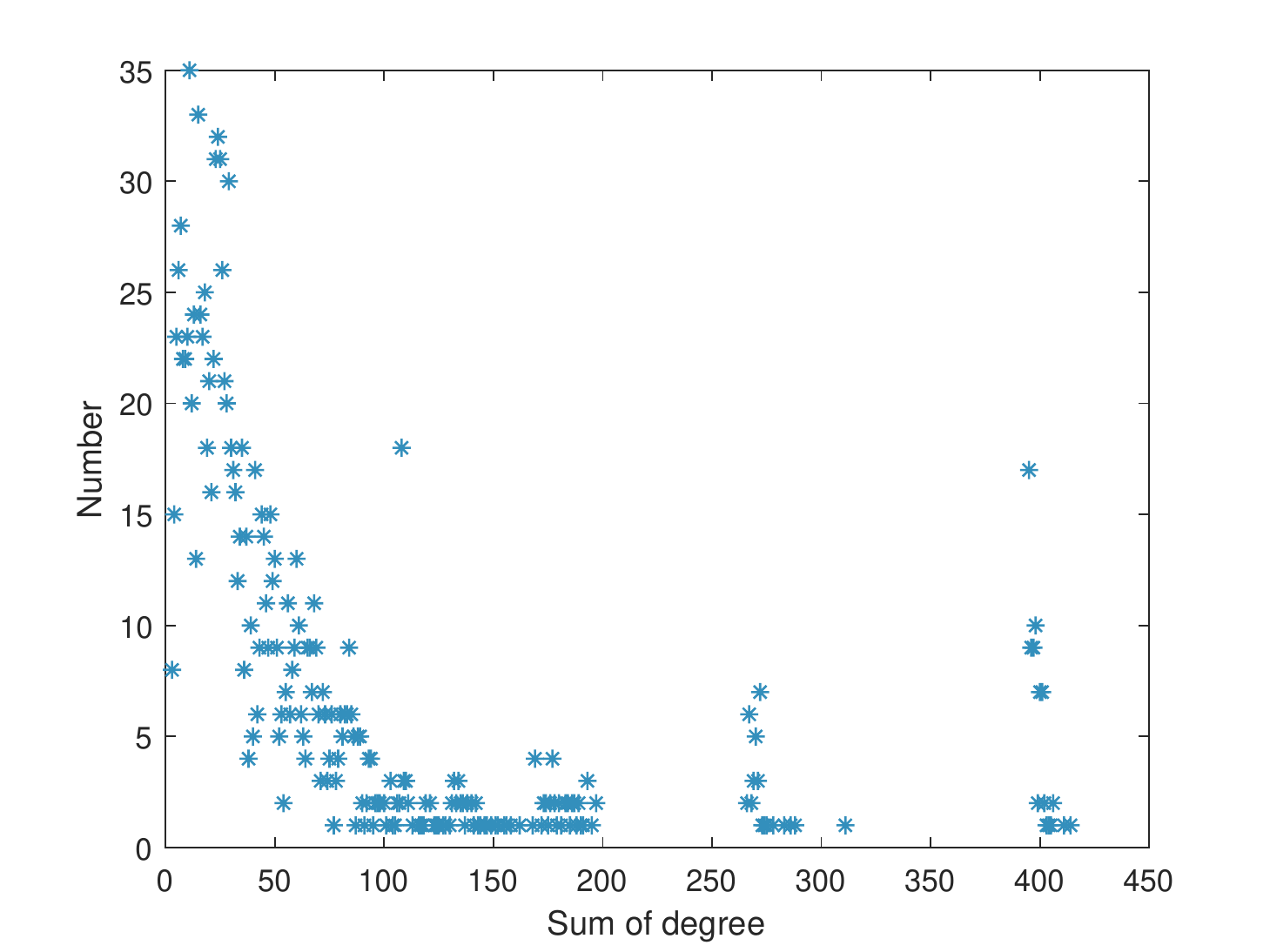}
\end{minipage}%
}%
\quad
\subfigure[$\eta=-1$]{
\begin{minipage}[t]{0.33\linewidth}
\centering
\includegraphics[width=4.5cm]{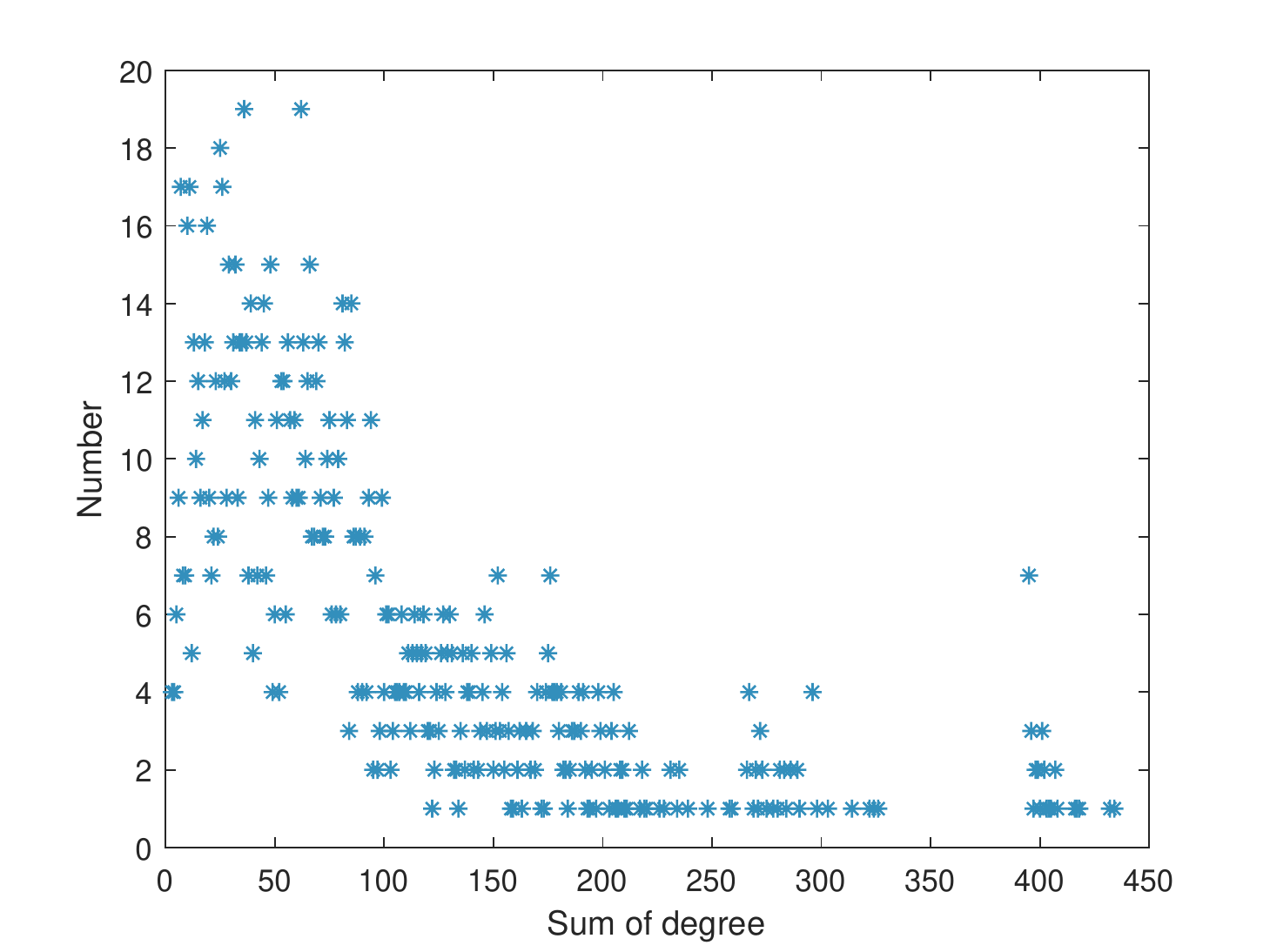}
\end{minipage}%
}%
\subfigure[$\eta=0$]{
\begin{minipage}[t]{0.33\linewidth}
\centering
\includegraphics[width=4.5cm]{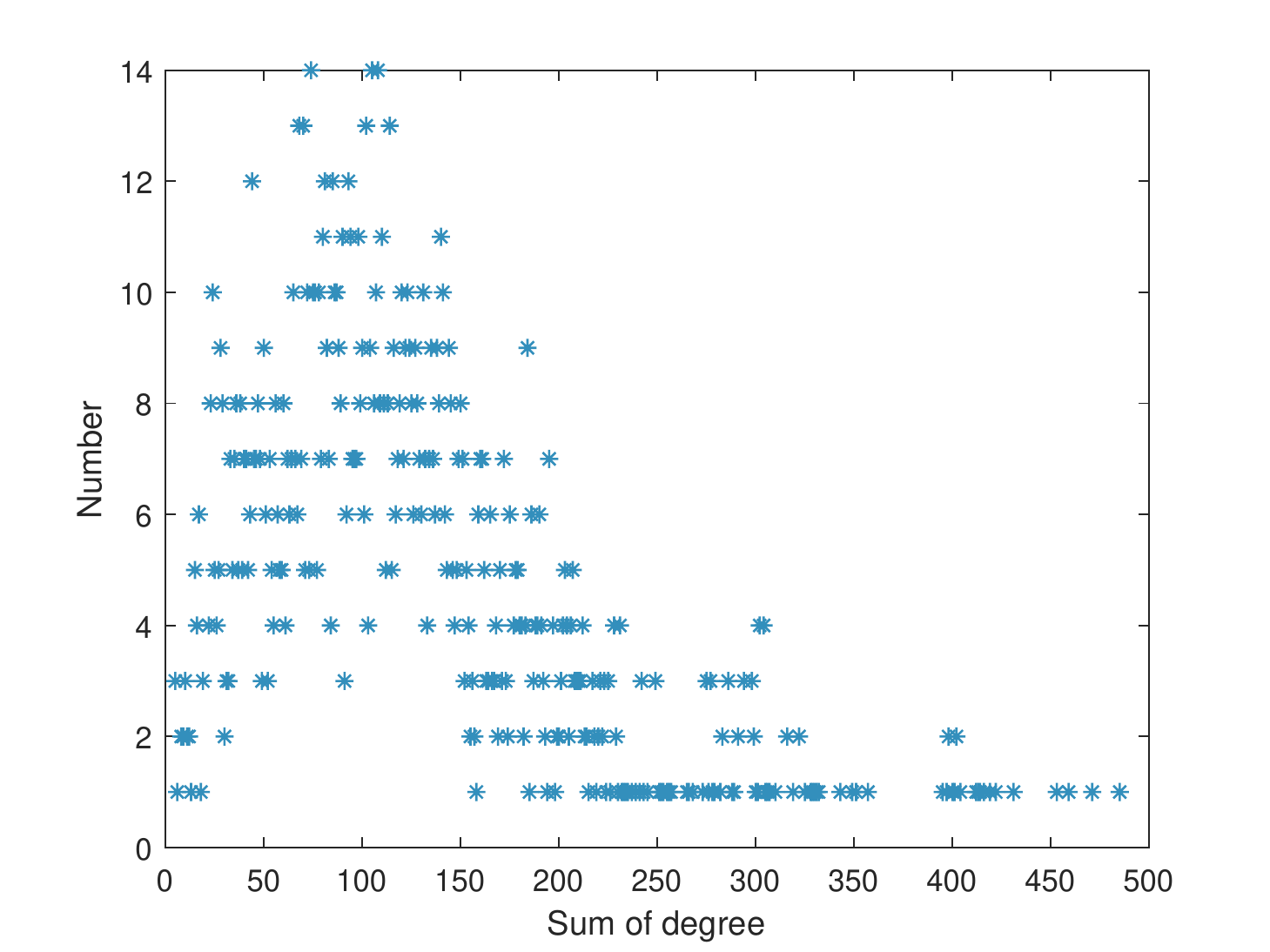}
\end{minipage}%
}%
\subfigure[$\eta=1$]{
\begin{minipage}[t]{0.33\linewidth}
\centering
\includegraphics[width=4.5cm]{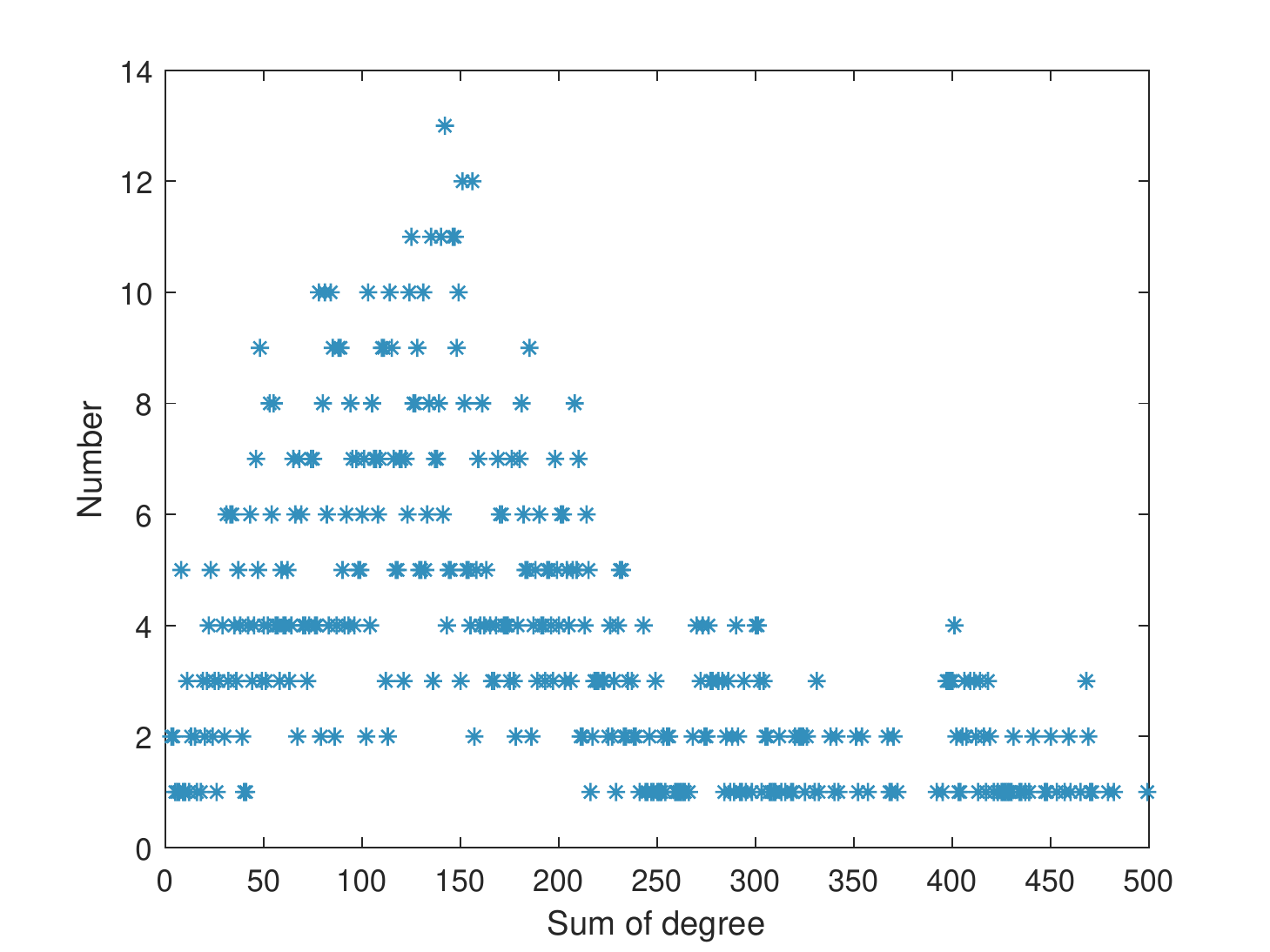}
\end{minipage}%
}%
\quad
\subfigure[$\eta=5$]{
\begin{minipage}[t]{0.33\linewidth}
\centering
\includegraphics[width=4.5cm]{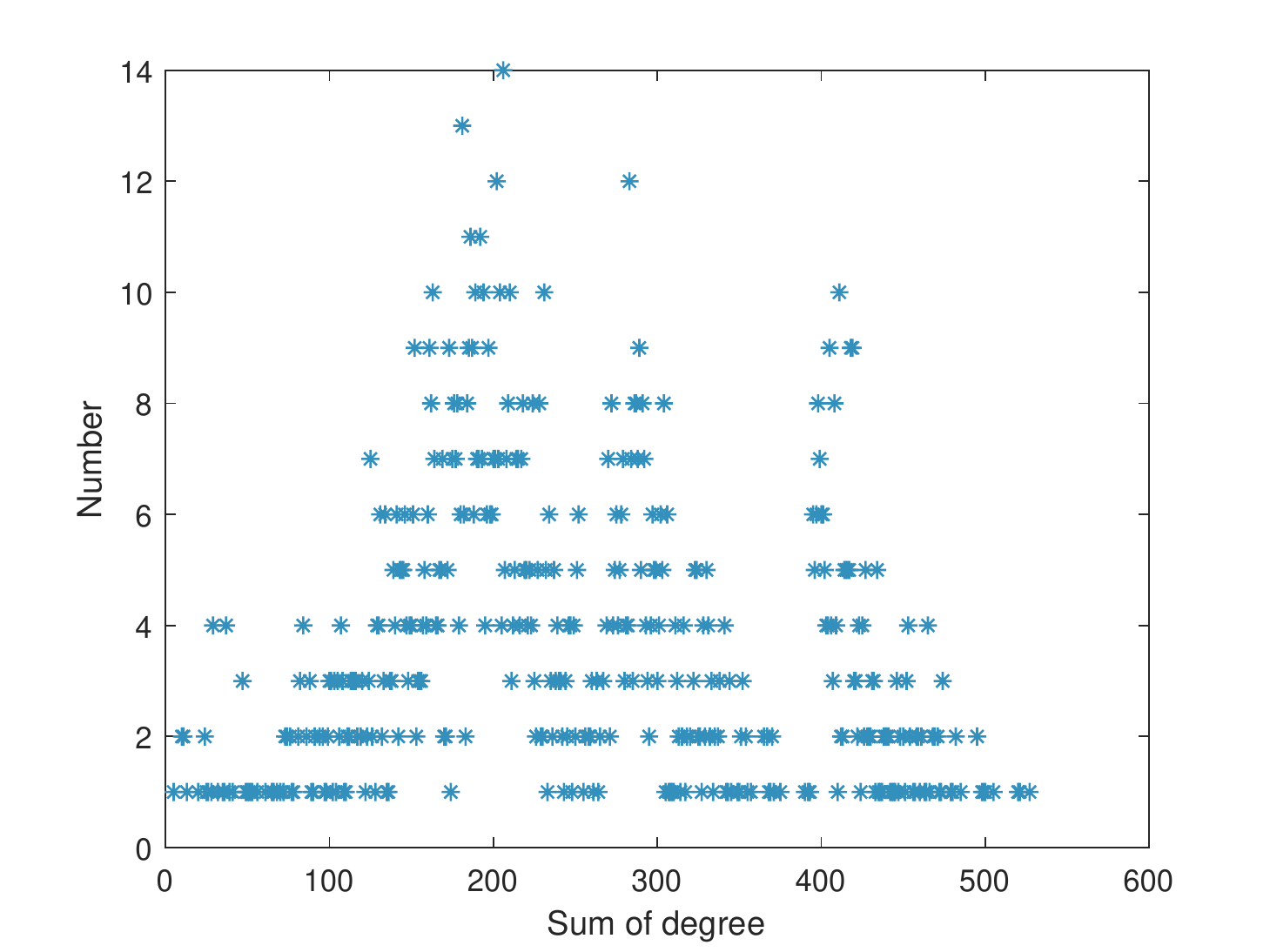}
\end{minipage}%
}%
\subfigure[$\eta=10$]{
\begin{minipage}[t]{0.33\linewidth}
\centering
\includegraphics[width=4.5cm]{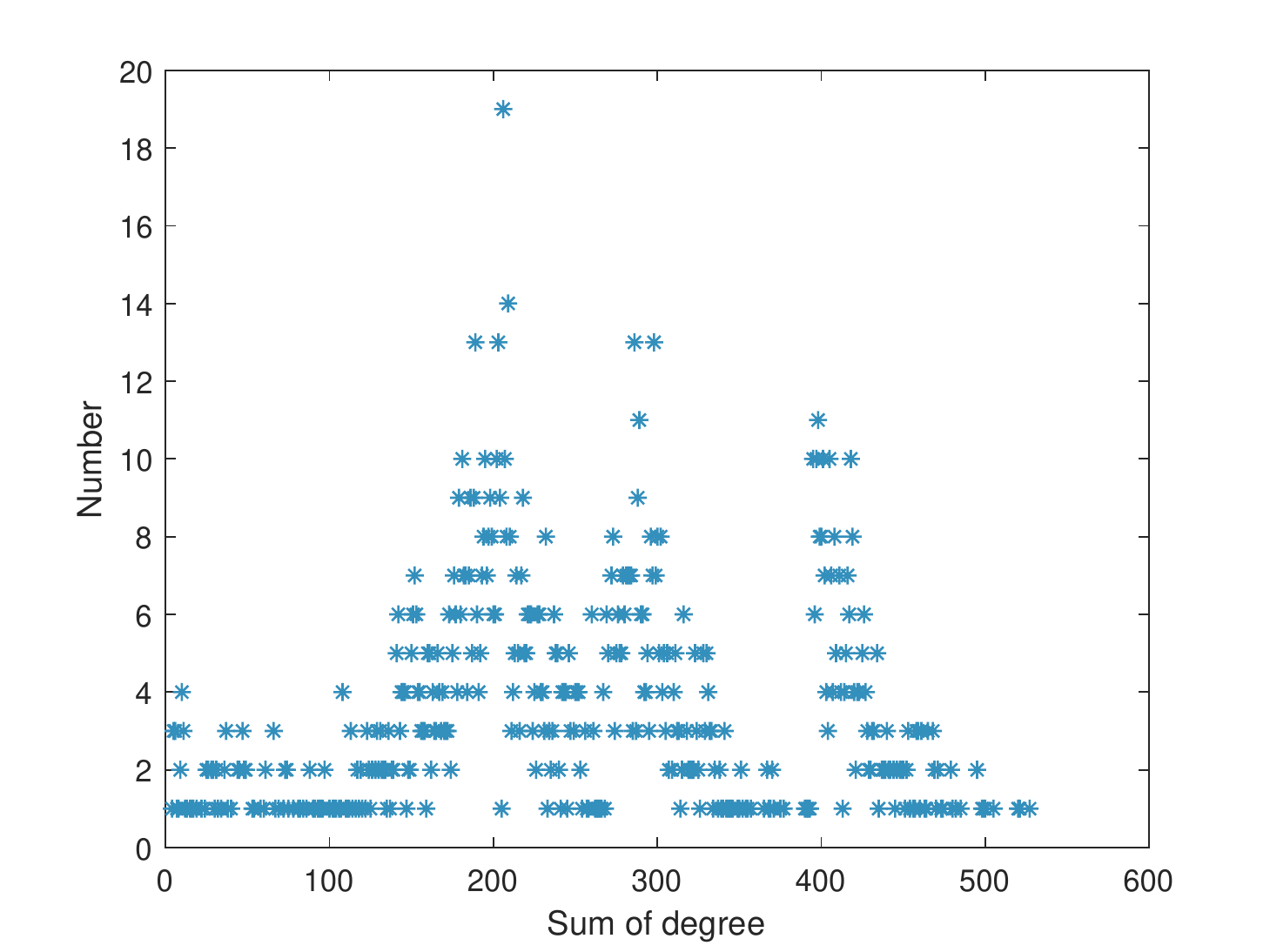}
\end{minipage}%
}%
\subfigure[$\eta=20$]{
\begin{minipage}[t]{0.33\linewidth}
\centering
\includegraphics[width=4.5cm]{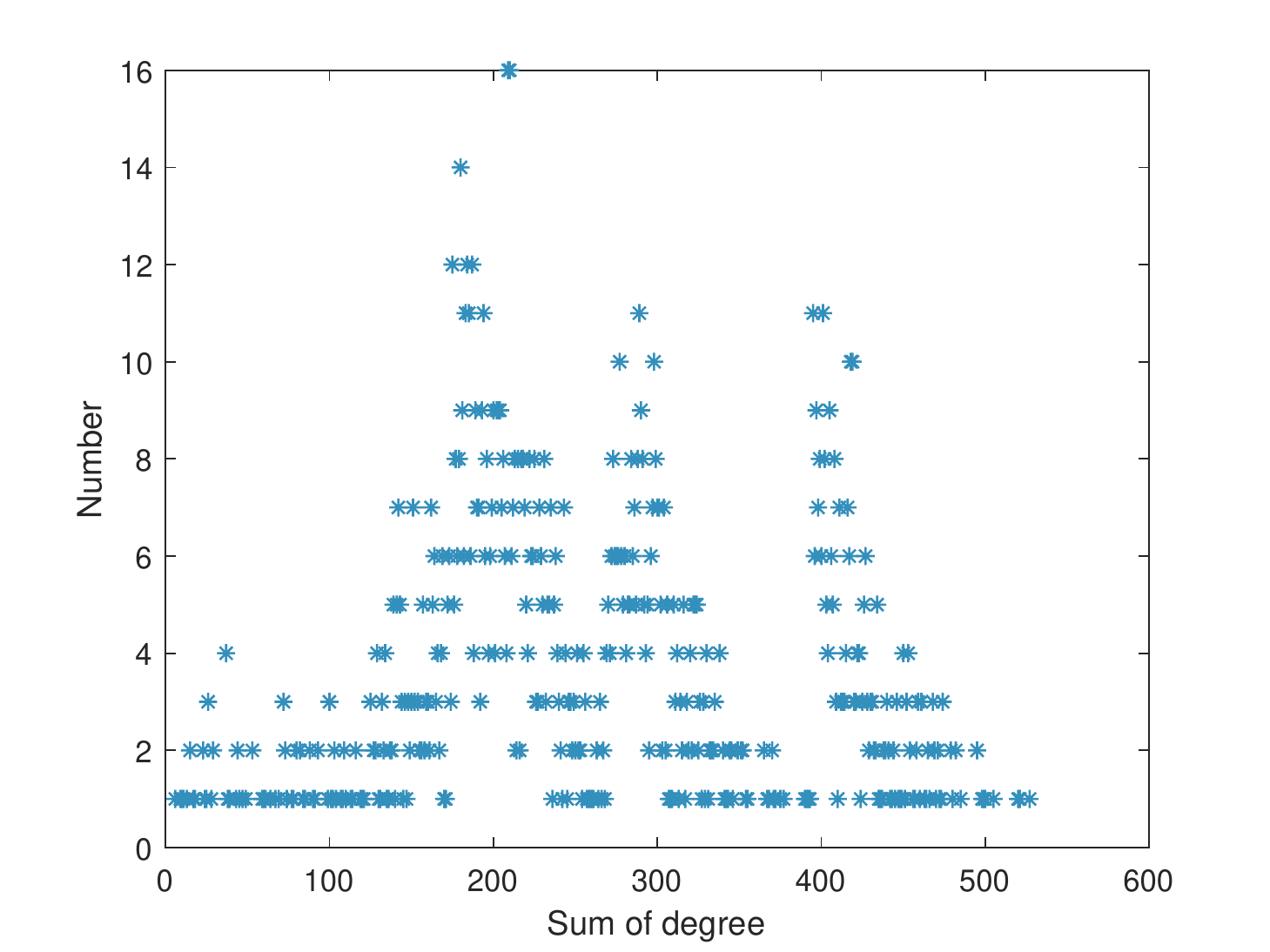}
\end{minipage}%
}%
\centering
\caption{Sums of degree for pairs of nodes connected by an intralayer link removed by the local perturbation strategy on dataset FT1. The meaning of horizontal ordinate and vertical ordinate are the same as Fig.~\ref{pics_deg_dellinks_d1}.}
\label{pics_deg_dellinks_d2}
\end{figure*}

The phenomena observed in Figs.~\ref{pic:result_real_diffeta_dmeth2} and ~\ref{pic:result_real_diffeta_map_dmeth2} illustrate that the effects of different types of intralayer links selected by the parameter $\eta$ of the local strategy for the interlayer link prediction are different. There is a certain regularity of the effects on the prediction.
The local perturbation strategy first chooses any node as the starting node, walks to the next node, and removes the intralayer link between the two walked nodes. When $\eta$ is very small, the local perturbation strategy prefers to select the nodes with a low degree as the next node. Therefore, the intralayer links of type \textbf{LLT1} in Fig.~\ref{pic:intralayer_link_types_local} are more likely to be removed. The perturbation weight for each node to be selected as the next node does not change much when $-20\leq\eta<-5$ since $\eta$ is an exponent. Therefore, the $P@30$ and MAP under the same $\xi$ hardly varies with $\eta$. When $\eta\geq-5$, the local perturbation strategy prefers to select other types of nodes as the next node. The proportion of other types of intralayer links that are removed begins to increase. The $P@30$ and MAP start to increase with the increase of $\eta$. This indicates that the intralayer links connected to nodes with a low degree have a greater impact on the interlayer link prediction algorithms; other types of intralayer links have relatively less impact on it. When $\eta=0$, the probabilities of all nodes to be selected as the next node are the same so that the perturbation weights for all intralayer links are the same. All intralayer links will be removed with the same probability. The $P@30$ and MAP do not get the maximum or minimum value when $\eta=0$, which again illustrates that different types of intralayer links selected by the parameter $\eta$ of the global strategy have different effects on interlayer link prediction. When $\eta>0$, the local perturbation strategy prefers to select the nodes with a high degree as the next node. The intralayer links of type \textbf{LLT3} are more likely to be removed. When $\eta=3$, almost all the curves of $P@30$ and MAP for different $\xi$ in each subfigure achieve the maximum value. When $3<\eta\leq20$, the $P@30$ and MAP under the same $\xi$ hardly varies with $\eta$. This is because $\eta$ is an exponent, the perturbation weight for each node to be selected as the next node does not change much. Almost all of the removed intralayer links are the links connected to nodes with a high degree in this interval. This indicates that the intralayer links connected to nodes with a high degree have a smaller impact on the interlayer link prediction algorithms; other types of intralayer links have a greater impact on it.
\begin{figure*}
    \centering
    \includegraphics[width=12cm,height=10.5cm]{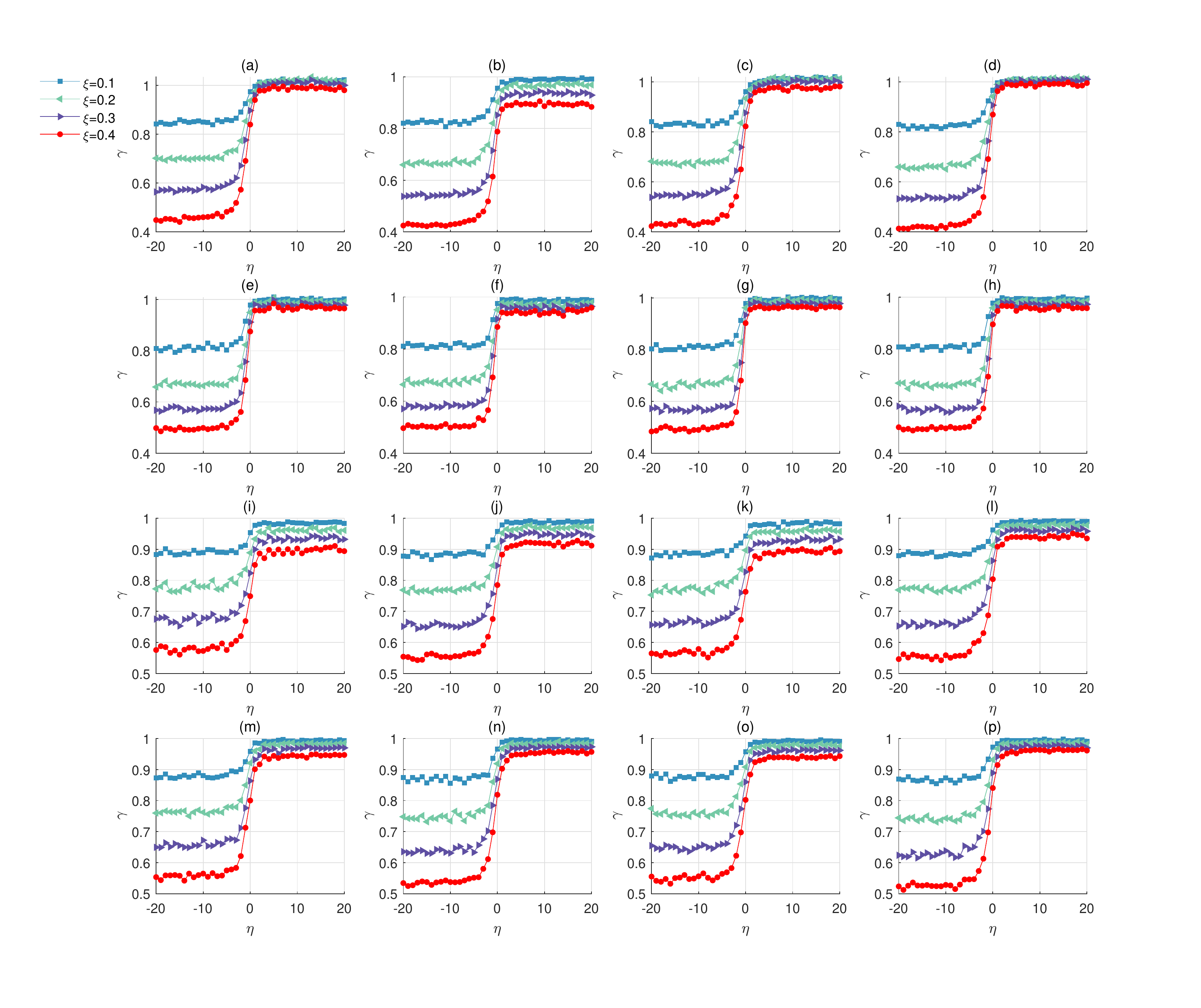}
    \caption{$\gamma$ of different interlayer link prediction algorithms on the multiplex networks perturbed by the local perturbation strategy with different $\eta$. The datasets and prediction algorithms in different subfigures are the same as Fig.~\ref{pic:result_real_diffeta_dmeth1}.}
    \label{pic:result_real_diffeta_dmeth2_gamma}
\end{figure*}

To deeply understand the local perturbation strategy mechanism, we also apply it to the network of Kapferer friendship interactions and analyze the influence in detail. Fig.~\ref{pics_dellinks_d2_exa} presents the perturbed network under different $\eta$. When $\eta=-20$, $\eta=-10$, and $\eta=-5$, the removed intralayer links are similar, prefer to connecting to node with low degree. When $\eta=-1$, the removed intralayer links are less biased to nodes with a low degree than the subfigures of $\eta\leq-5$. When $\eta=0$, the probabilities of all nodes to be selected as the next node are the same. The red line is not very regular. When $\eta=5$, $\eta=10$, and $\eta=20$, the removed intralayer links are also similar, prefer connecting to the node with a high degree. The observations in Fig.~\ref{pics_dellinks_d2_exa} are consistent with our analysis of the results of $P@30$ and MAP of different interlayer link prediction algorithms on the multiplex networks perturbed by the local perturbation strategy.

Similar to the global perturbation strategy analysis, we further take dataset FT1 as an example to count the sum of degrees for pairs of nodes connected by an intralayer link removed by the global perturbation strategy. Fig.~\ref{pics_deg_dellinks_d2} presents the results. When $\eta$ is small, most of the sums of degrees are small. As $\eta$ increases, the sums of a degree increase. The observations in Fig.~\ref{pics_deg_dellinks_d2} are also consistent with our analysis of the results of $P@30$ and MAP of different interlayer link prediction algorithms on the multiplex networks perturbed by the local perturbation strategy.

Based on the above analysis, we can draw the following conclusion that the intralayer links of type \textbf{LT1} in Fig.~\ref{pic:intralayer_link_types_local} have a greater effects on interlayer link prediction than type \textbf{LLT2}, and intralayer links of type \textbf{LLT2} have a greater effects than type \textbf{LLT3}.

Moreover, we analyzed the ratio between the $P@30$ on the perturbed networks and $P@30$ on the original networks, as shown in Fig.~\ref{pic:result_real_diffeta_dmeth2_gamma}. From the figure, we can see that the local strategy's trend of $\gamma$ varying with $\eta$ is the same as that of $P@30$ and MAP varying with $\eta$. The value of $\gamma$ in most conditions is less than 1. This is because some of the intralayer links are removed by the local perturbation strategy. Less information on the multiplex networks leads to worse interlayer link prediction performance. This is consistent with our intuition. However, in some of the subfigures, when $\eta>3$, the value of $\gamma$ is greater than 1. For example, when $\xi=0.2$ and $\eta=10$, the CN algorithm's value of $\gamma$ is 1.03 on dataset FT1 while IDP is 1.01. The performance of the interlayer link prediction algorithms is unexpectedly improved by removing the intralayer link. When $\eta>3$, most of the removed intralayer links selected by the local perturbation strategy are connected to nodes with a high degree. This phenomenon indicates that the intralayer links connected to nodes with a high degree are not necessarily helpful for the interlayer link prediction. It is not that the more structural information, the more conducive to interlayer link prediction.

\subsubsection{Effects of $\xi$}
We then set $\eta=-20, -10, 0, 10,$ and $20$, $\xi$ increasing from $0$ to $0.4$ by $0.02$ to analyse the effects of $\xi$ of the two perturbation strategies.
\begin{figure*}
    \centering
    \includegraphics[width=12cm,height=10.5cm]{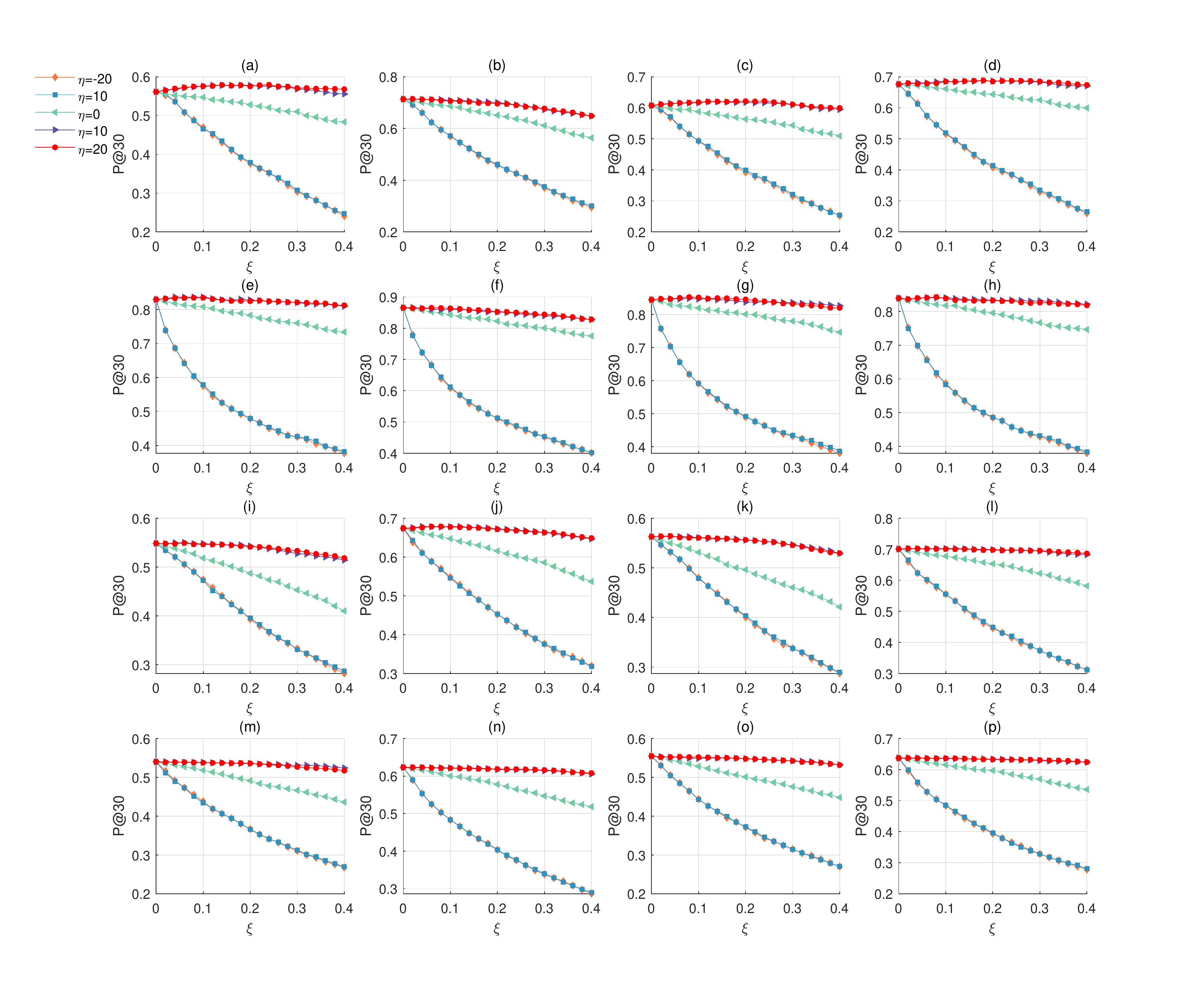}
    \caption{$P@30$ of different interlayer link prediction algorithms on the multiplex networks perturbed by the global perturbation strategy with different $\xi$. The datasets and prediction algorithms in different subfigures are the same as Fig.~\ref{pic:result_real_diffeta_dmeth1}.}
    \label{pic:result_real_diffxi_dmeth1}
\end{figure*}
\begin{figure*}
    \centering
    \includegraphics[width=12cm,height=10.5cm]{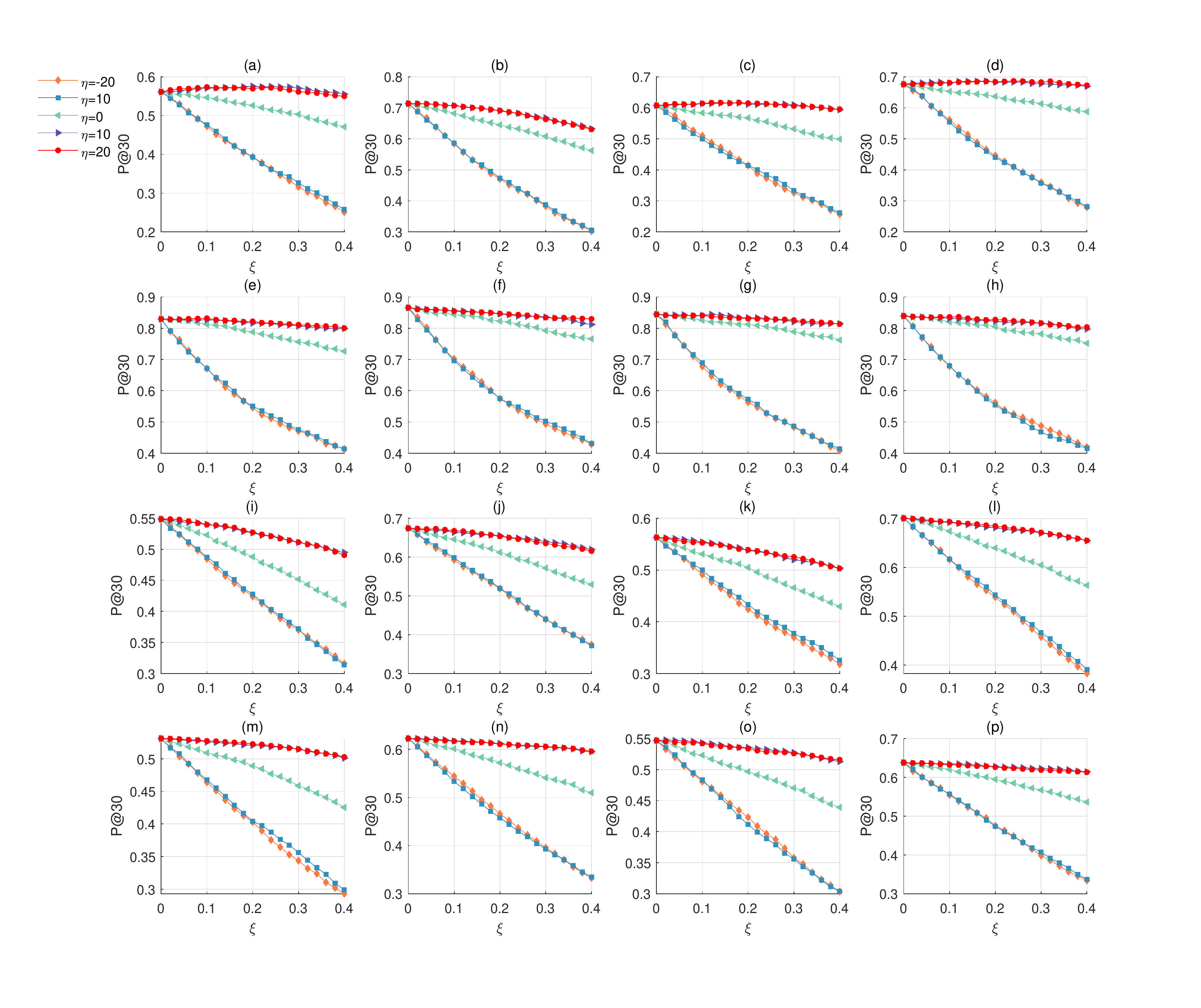}
    \caption{$P@30$ of different interlayer link prediction algorithms on the multiplex networks perturbed by the local perturbation strategy with different $\xi$. The datasets and prediction algorithms in different subfigures are the same as Fig.~\ref{pic:result_real_diffeta_dmeth1}.}
    \label{pic:result_real_diffxi_dmeth2}
\end{figure*}

\begin{figure*}
    \centering
    \includegraphics[width=12cm,height=10.5cm]{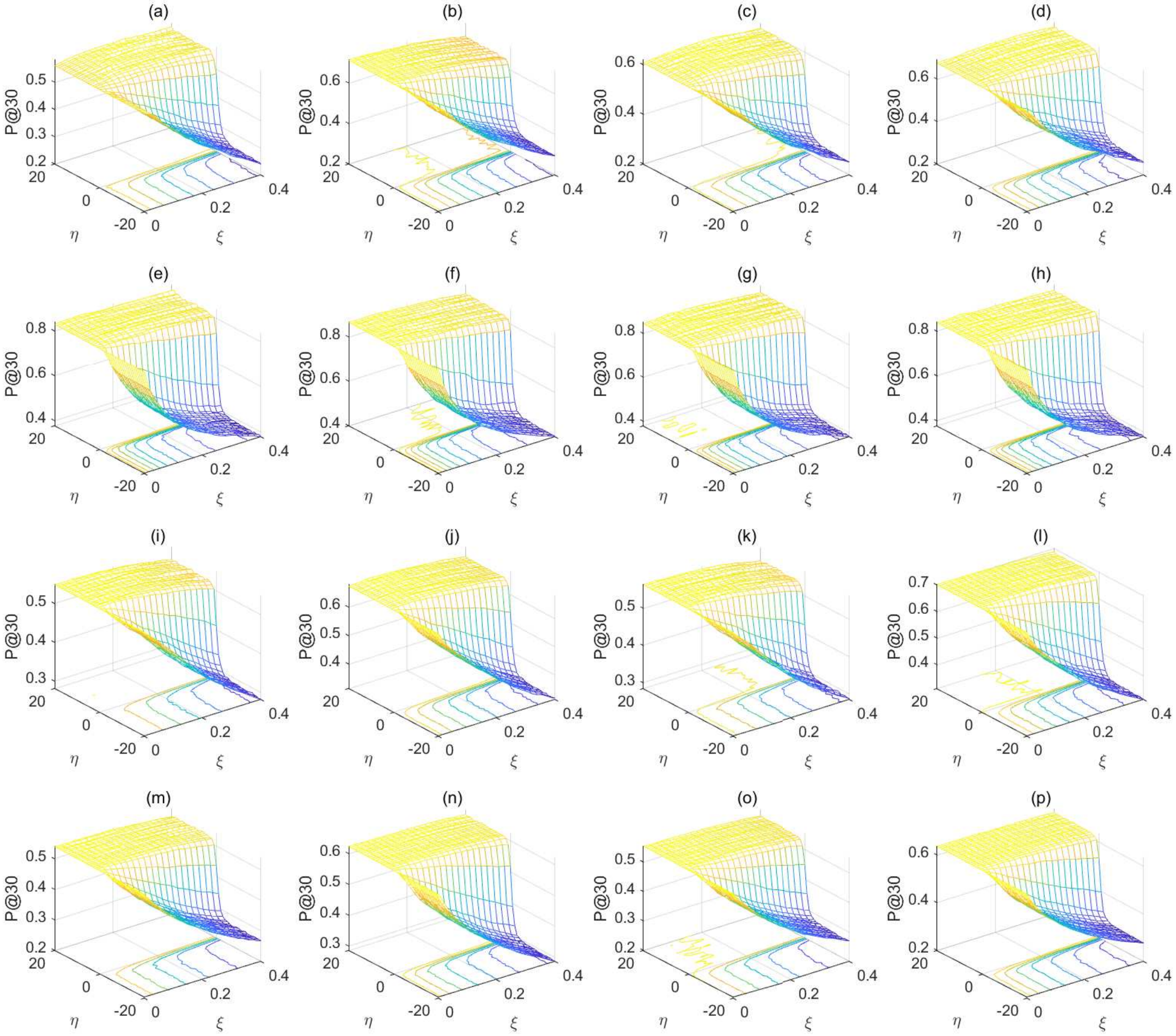}
    \caption{$P@30$ of different interlayer link prediction algorithms on the multiplex networks perturbed by the global perturbation strategy. The datasets and prediction algorithms in different subfigures are the same as Fig.~\ref{pic:result_real_diffeta_dmeth1}.}
    \label{pic:result_real_3d_dmeth1}
\end{figure*}
\begin{figure*}
    \centering
    \includegraphics[width=12cm,height=10.5cm]{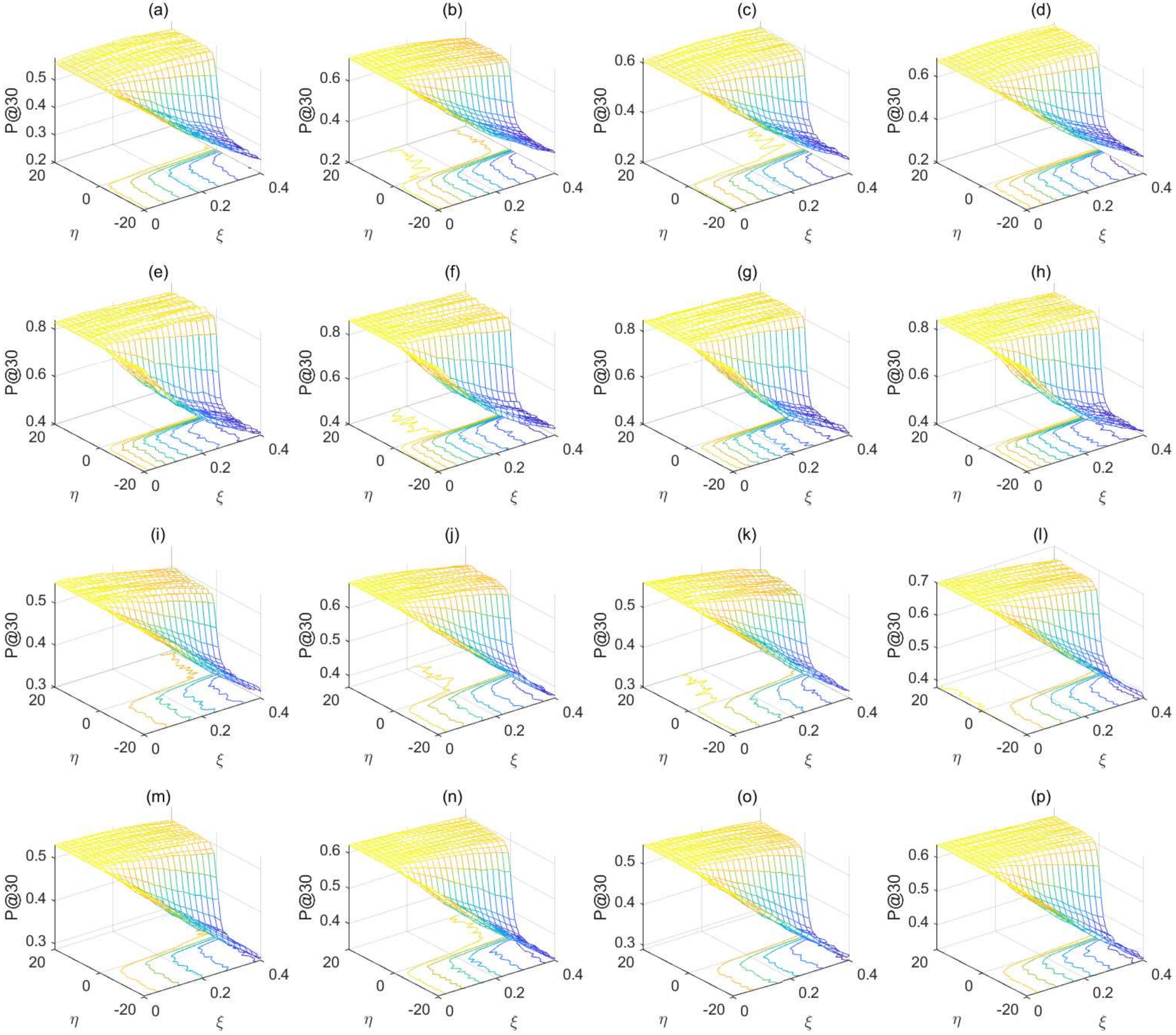}
    \caption{$P@30$ of different interlayer link prediction algorithms on the multiplex networks perturbed by the local perturbation strategy. The datasets and prediction algorithms in different subfigures are the same as Fig.~\ref{pic:result_real_diffeta_dmeth1}.}
    \label{pic:result_real_3d_dmeth2}
\end{figure*}

\begin{figure}
    \centering
    \includegraphics[width=0.8\textwidth]{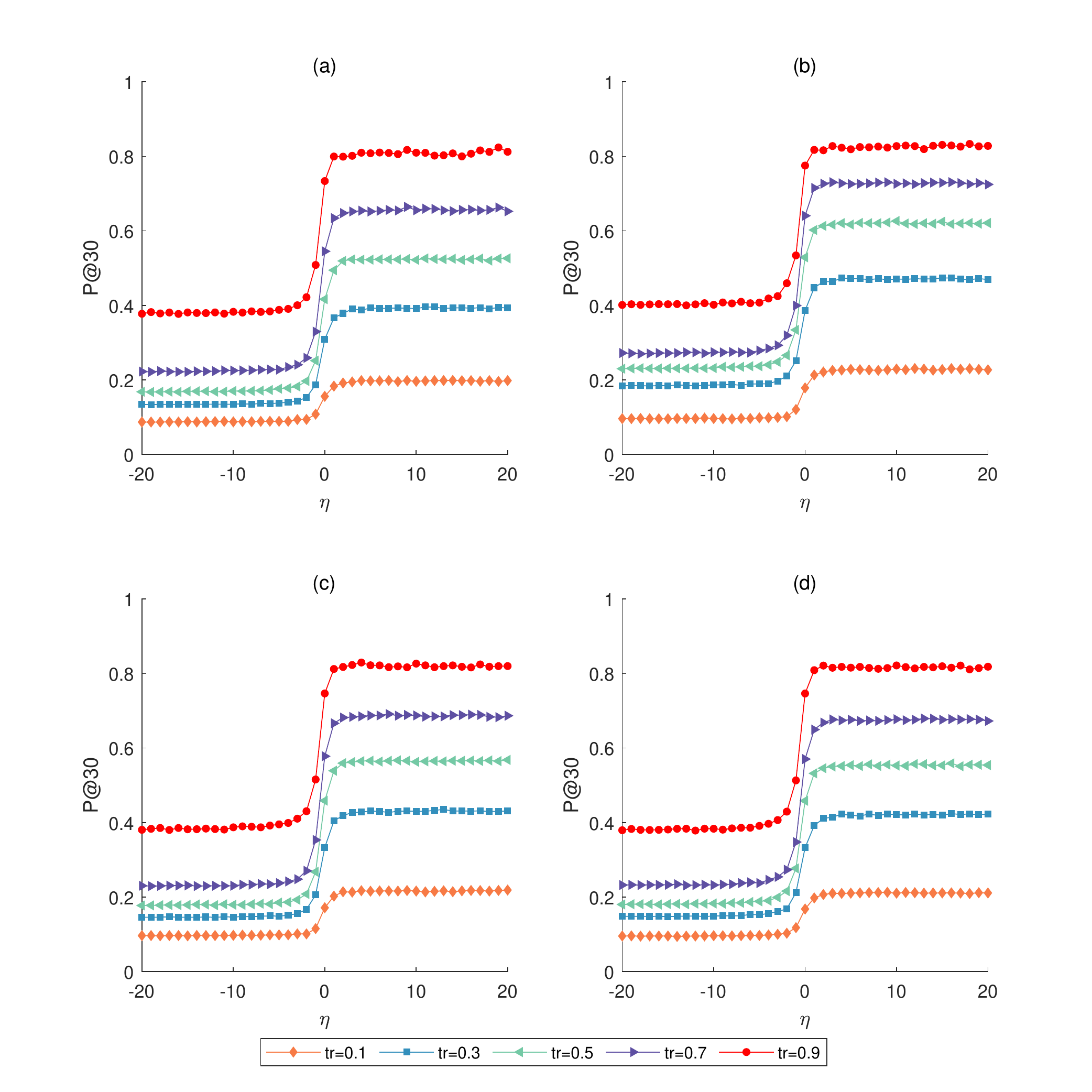}
    \caption{$P@30$ of different interlayer link prediction algorithms on the FT1 multiplex networks perturbed by the global perturbation strategy with different training ratios. (a) $P@30$ of algorithm CN. (b) $P@30$ of algorithm NS. (c) $P@30$ of algorithm FRUI. (d) $P@30$ of algorithm IDP.}
    \label{pic:result_real_difftr_dmeth1}
\end{figure}

Figure ~\ref{pic:result_real_diffxi_dmeth1} is the $P@30$ of different prediction algorithms on the four kinds of real-world multiplex networks perturbed by the global perturbation strategy under the above settings. From the figure, we can see that for a given $\eta$, $P@30$ of different prediction algorithms exhibit a similar trend with an increase in $\xi$. This phenomenon indicates that the effect of the global perturbation strategy on interlayer link prediction has a certain regularity. The curves of $P@30$ when $\eta=-20$, $\eta=-10$, and $\eta=0$ in all subfigures exhibit a trend of decreasing with an increase in $\xi$ because a larger $\xi$ means that more intralayer links have been removed. Less information on the multiplex networks leads to worse interlayer link prediction performance. This is consistent with our intuition. However, the curves of $P@30$ when $\eta=10$ and $\eta=20$ in all subfigures have not decreased with an increase in $\xi$. This is counterintuitive, as removing more intralayer links does not decrease interlayer link prediction performance. On datasets FT1, FT2, and Higgs-SCMT, the curves of $P@30$ when $\eta=10$ and $\eta=20$ of some prediction algorithms exhibit a trend of first increasing and then decreasing with an increase in $\xi$. Particularly, on dataset FT1, $P@30$ of CN is 0.5789 when $\xi=0.14,\eta=20$. It is nearly two percent higher than $P@30$ of CN when $\xi=0.14,\eta=0$, which is 0.5610. We already know from previous analysis that when $\eta=20$ and $\eta=10$, the global perturbation strategy tends to remove the intralayer link connected to the nodes with a high degree. These two curves first increase with the increase in $\xi$ because the intralayer link connected to the nodes with a high degree makes some side effects to the interlayer link prediction. These intralayer links are noise information for interlayer link prediction algorithms. After removing the noise information, the interlayer link prediction algorithms can achieve better performance. When the percentage of these removed interlayer links reaches about $14\%$, i.e., $\xi=0.14$, almost all the noise information has been removed. In this situation, different prediction algorithms reach the maximum value. When $\xi>0.14$, the non-noisy information begins to be removed so that the $P@30$ decreases with the increase in $\xi$.

Figure~\ref{pic:result_real_diffxi_dmeth2} is the $P@30$ of different prediction algorithms on the four kinds of real-world multiplex networks perturbed by the local perturbation strategy under the above settings. From the figure, we can see that for a given $\eta$, $P@30$ of different prediction algorithms exhibit a similar trend with an increase in $\xi$. This phenomenon indicates that the effect of the local perturbation strategy on interlayer link prediction has a certain regularity. The curves of $P@30$ when $\eta=-20$, $\eta=-10$, and $\eta=0$ in all subfigures exhibit a trend of decreasing with an increase in $\xi$. On datasets, FT1 and FT2, the curves of $P@30$ when $\eta=10$ and $\eta=20$ of some prediction algorithms exhibit a trend of first increasing and decreasing with an increase in $\xi$. Particularly, on dataset FT1, $P@30$ of CN is 0.5714 when $\xi=0.14,\eta=20$. It is nearly one percent higher than $P@30$ of CN when $\xi=0.14,\eta=0$, which is 0.5610. The reasons are the same as those in Fig.~\ref{pic:result_real_diffxi_dmeth1}.
\begin{figure}
    \centering
    \includegraphics[width=0.8\textwidth]{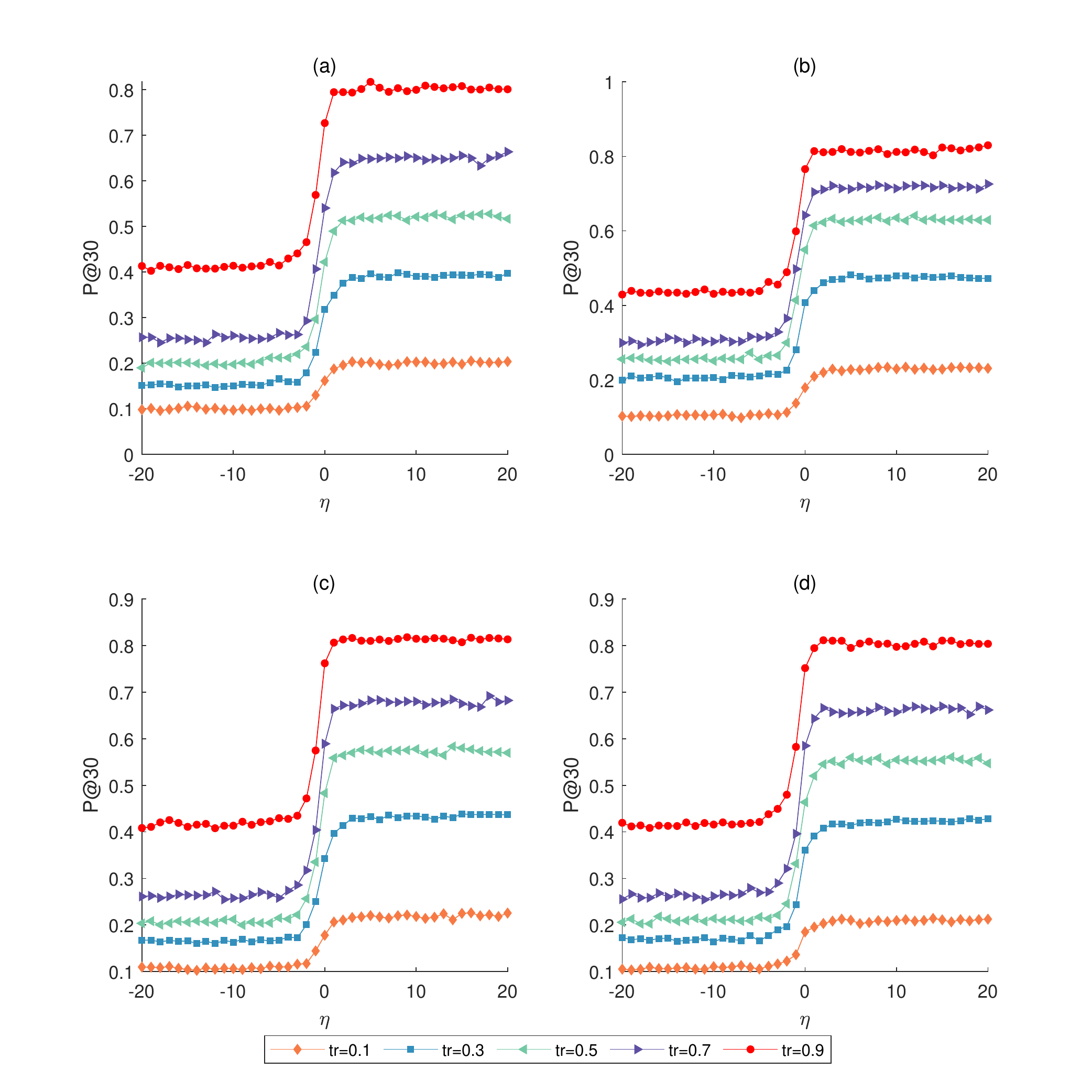}
    \caption{$P@30$ of different interlayer link prediction algorithms on the FT1 multiplex networks perturbed by the local perturbation strategy with different training ratios. (a) $P@30$ of algorithm CN. (b) $P@30$ of algorithm NS. (c) $P@30$ of algorithm FRUI. (d) $P@30$ of algorithm IDP.}
    \label{pic:result_real_difftr_dmeth2}
\end{figure}

Moreover, we execute experiments on more $\eta$ and $\xi$ settings. Figs.~\ref{pic:result_real_3d_dmeth1} and~\ref{pic:result_real_3d_dmeth2} show the results of global and local perturbation strategies, respectively. We can see that the trends observed on Figs.~\ref{pic:result_real_diffeta_dmeth1}, \ref{pic:result_real_diffeta_dmeth2},~\ref{pic:result_real_diffxi_dmeth1}, and~\ref{pic:result_real_diffxi_dmeth2} are similar with the section of Figs.~\ref{pic:result_real_3d_dmeth1} and~\ref{pic:result_real_3d_dmeth2}.

\subsubsection{Effects of training ratio}
Training ratio is an important factor for the performance of interlayer link prediction, which is the proportion of observed interlayer links in all of them. A larger training ratio means that more prior information can be used to make a prediction. We set $\xi=0.4$ and use global and local strategies with different $\eta$ to perturb FT1 datasets. Then, we set a training ratio equaling 0.1, 0.3, 0.5, 0.7, and 0.9 to execute the four interlayer link prediction algorithms on these perturbed multiplex networks. Figs.~\ref{pic:result_real_difftr_dmeth1} and ~\ref{pic:result_real_difftr_dmeth2} are the experimental results of global and local strategies on above settings, respectively.
\begin{figure*}
\centering
\subfigure[TF1]{
\begin{minipage}[t]{0.25\linewidth}
\centering
\includegraphics[width=3.5cm]{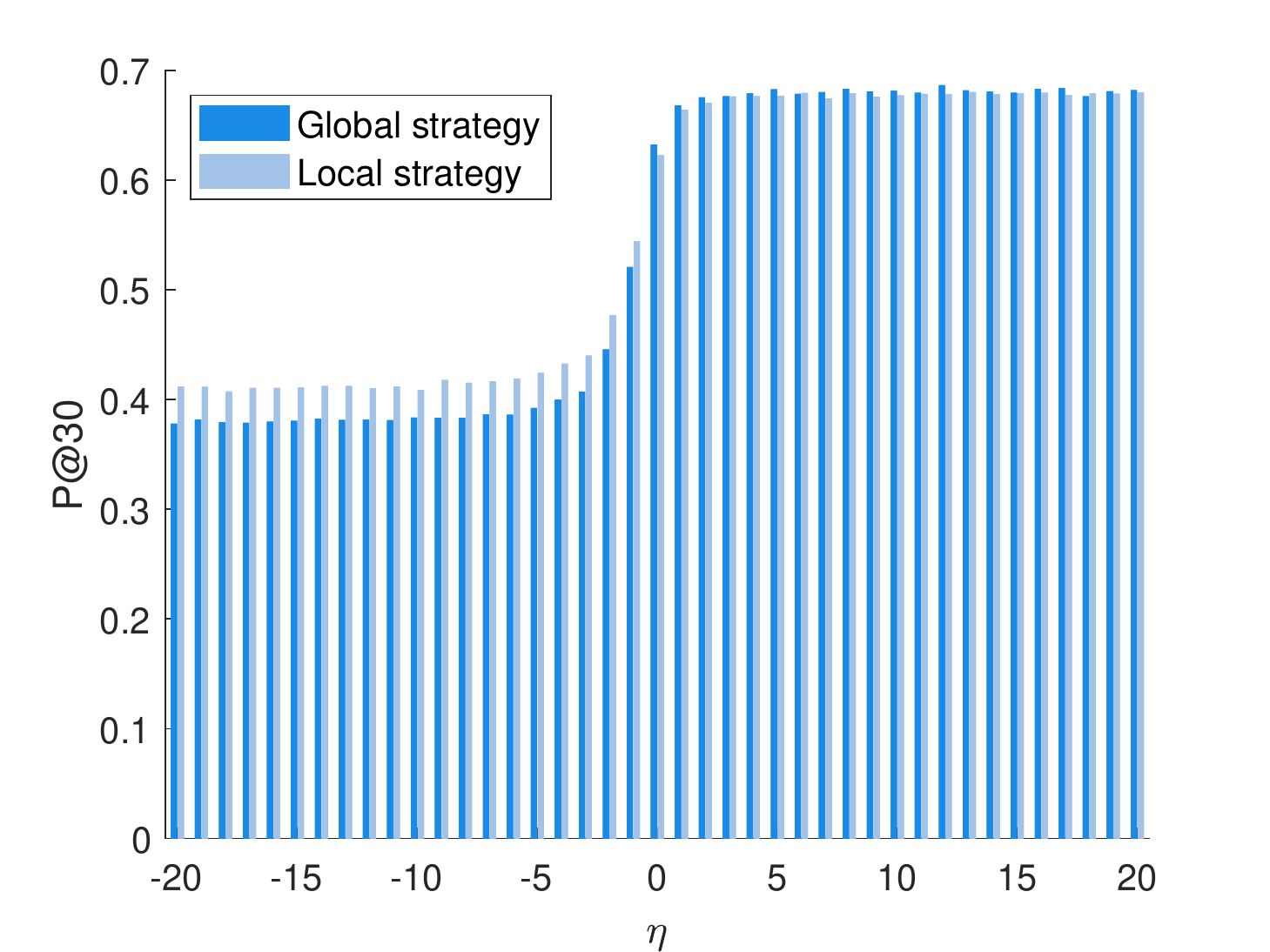}
\end{minipage}%
}%
\subfigure[TF2]{
\begin{minipage}[t]{0.25\linewidth}
\centering
\includegraphics[width=3.5cm]{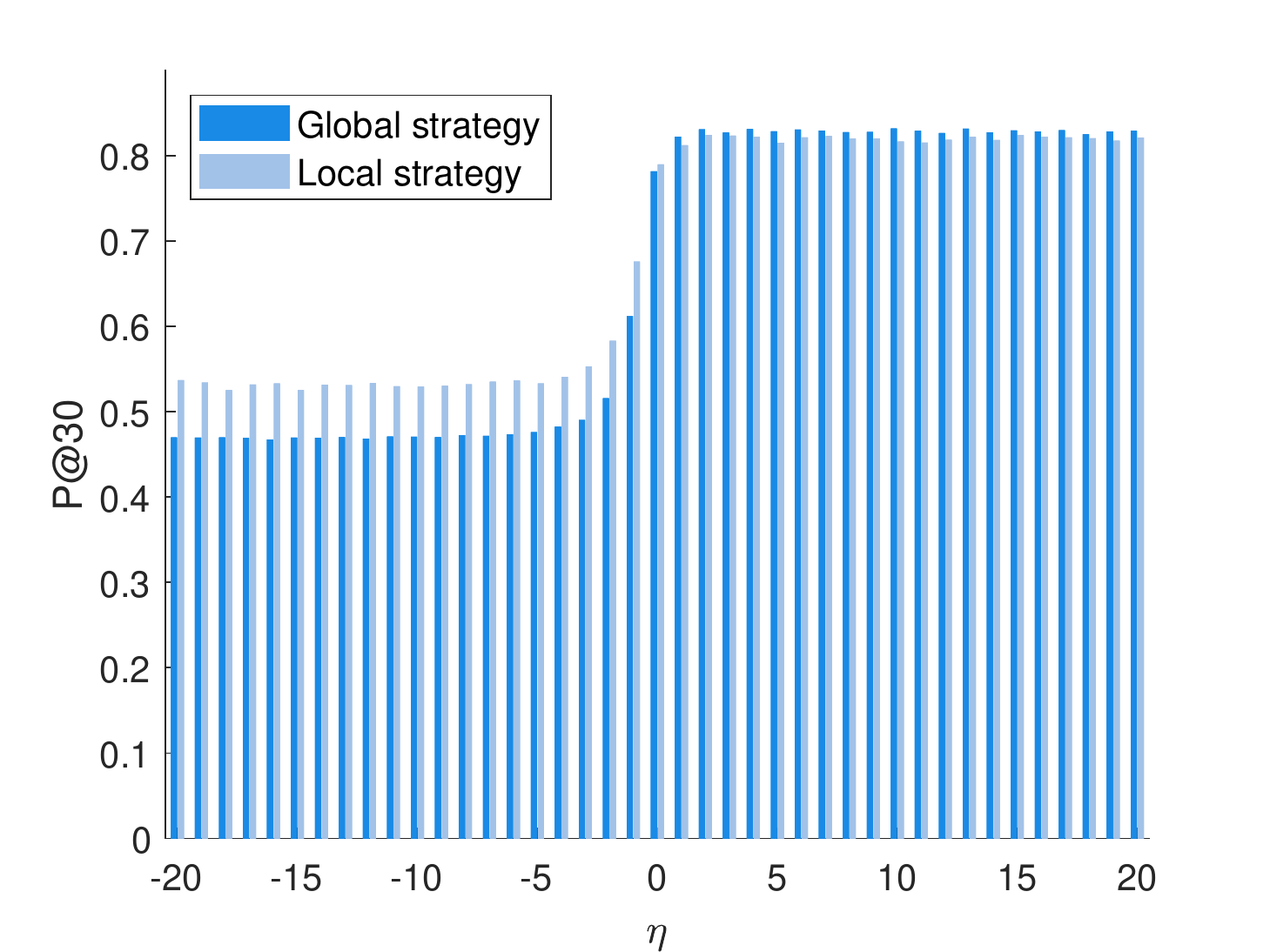}
\end{minipage}%
}%
\subfigure[Higgs-FSMT]{
\begin{minipage}[t]{0.25\linewidth}
\centering
\includegraphics[width=3.5cm]{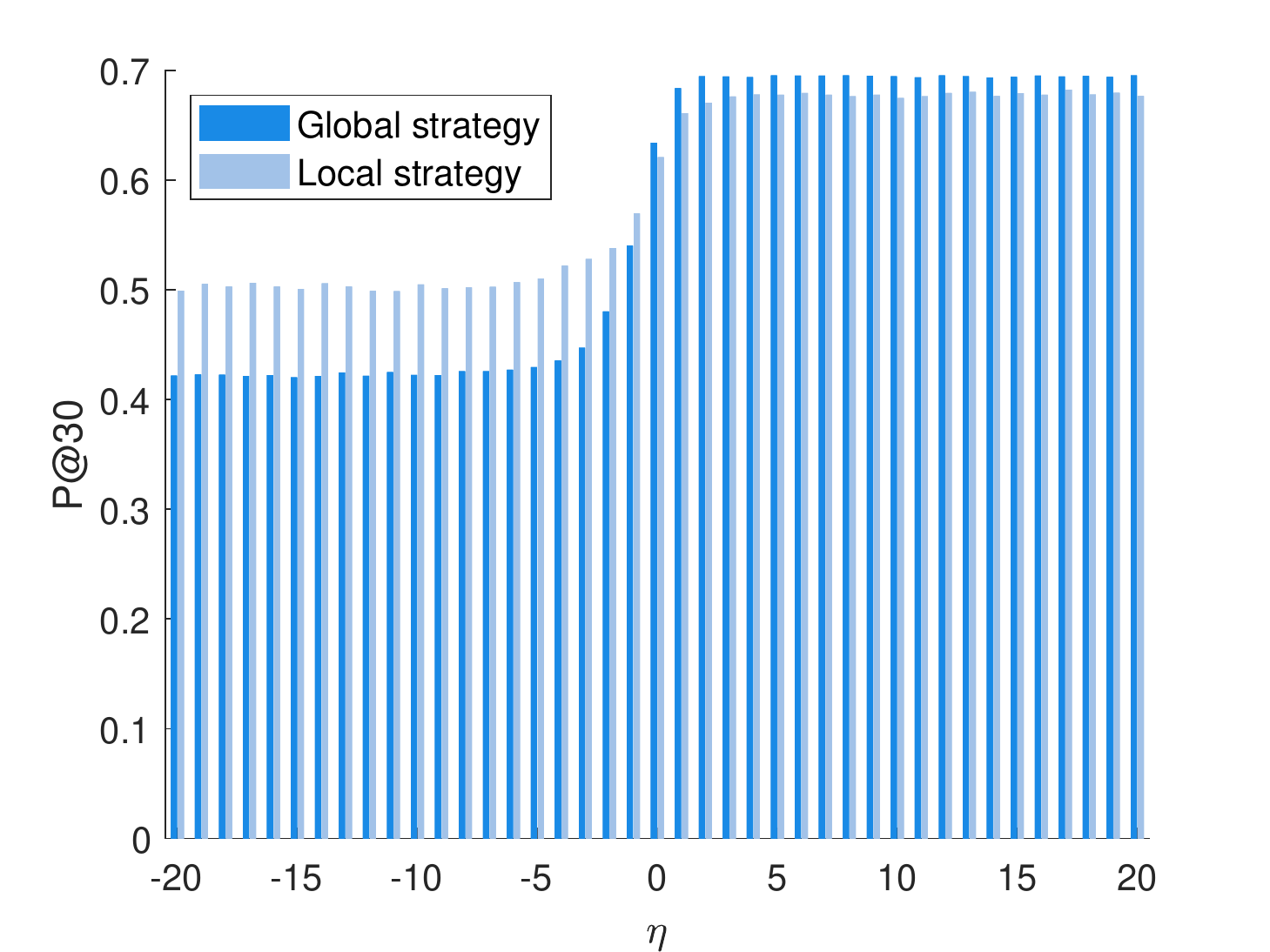}
\end{minipage}%
}%
\subfigure[Higgs-FSRT]{
\begin{minipage}[t]{0.25\linewidth}
\centering
\includegraphics[width=3.5cm]{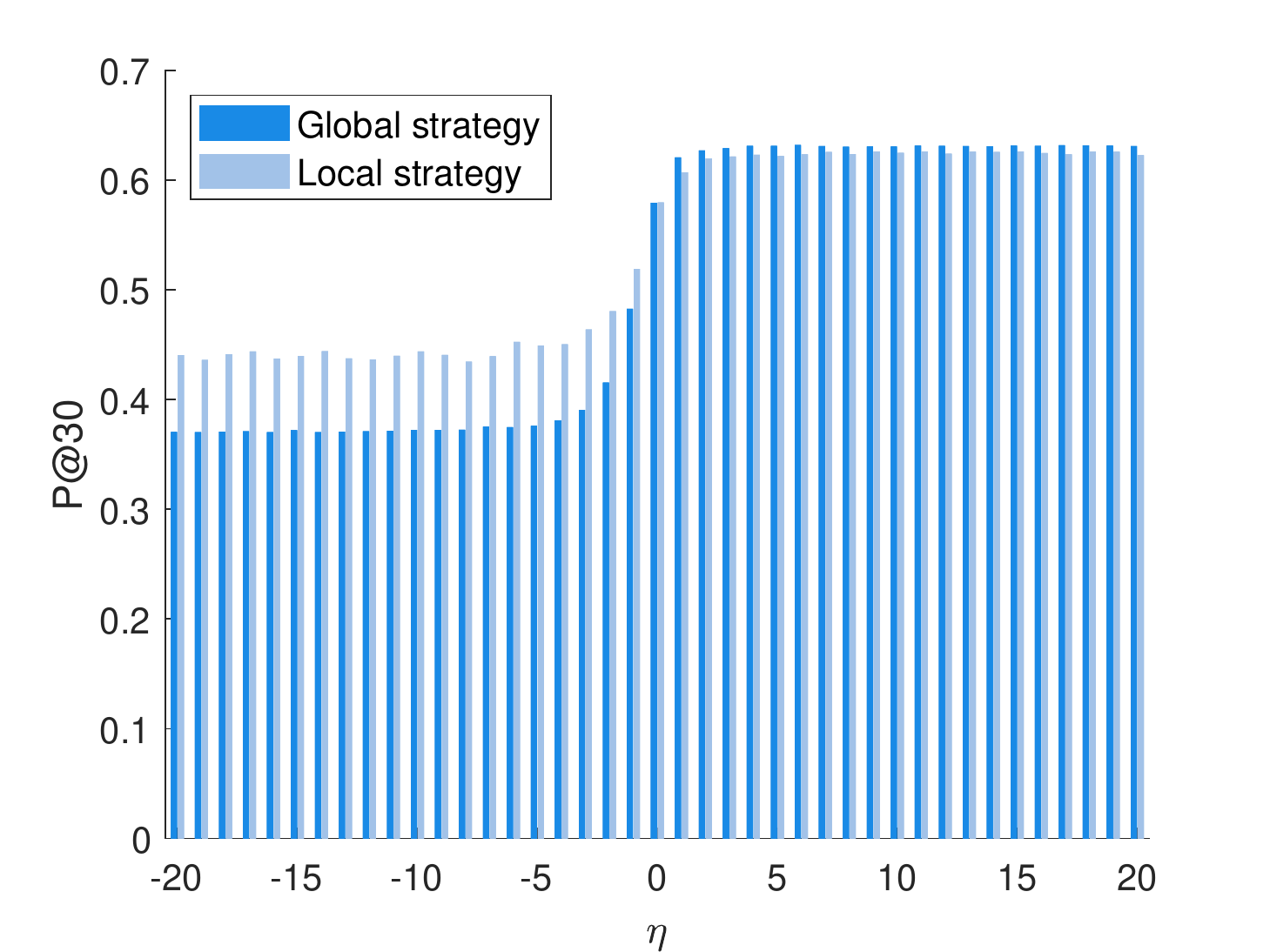}
\end{minipage}%
}%
\centering
\caption{Comparison between global perturbation strategy and local perturbation strategy on different $\eta$}.
\label{pics_glo_comp_local_diffeta}

\centering
\subfigure[Results in different average degree]{
\begin{minipage}[t]{0.33\linewidth}
\centering
\includegraphics[width=4.5cm]{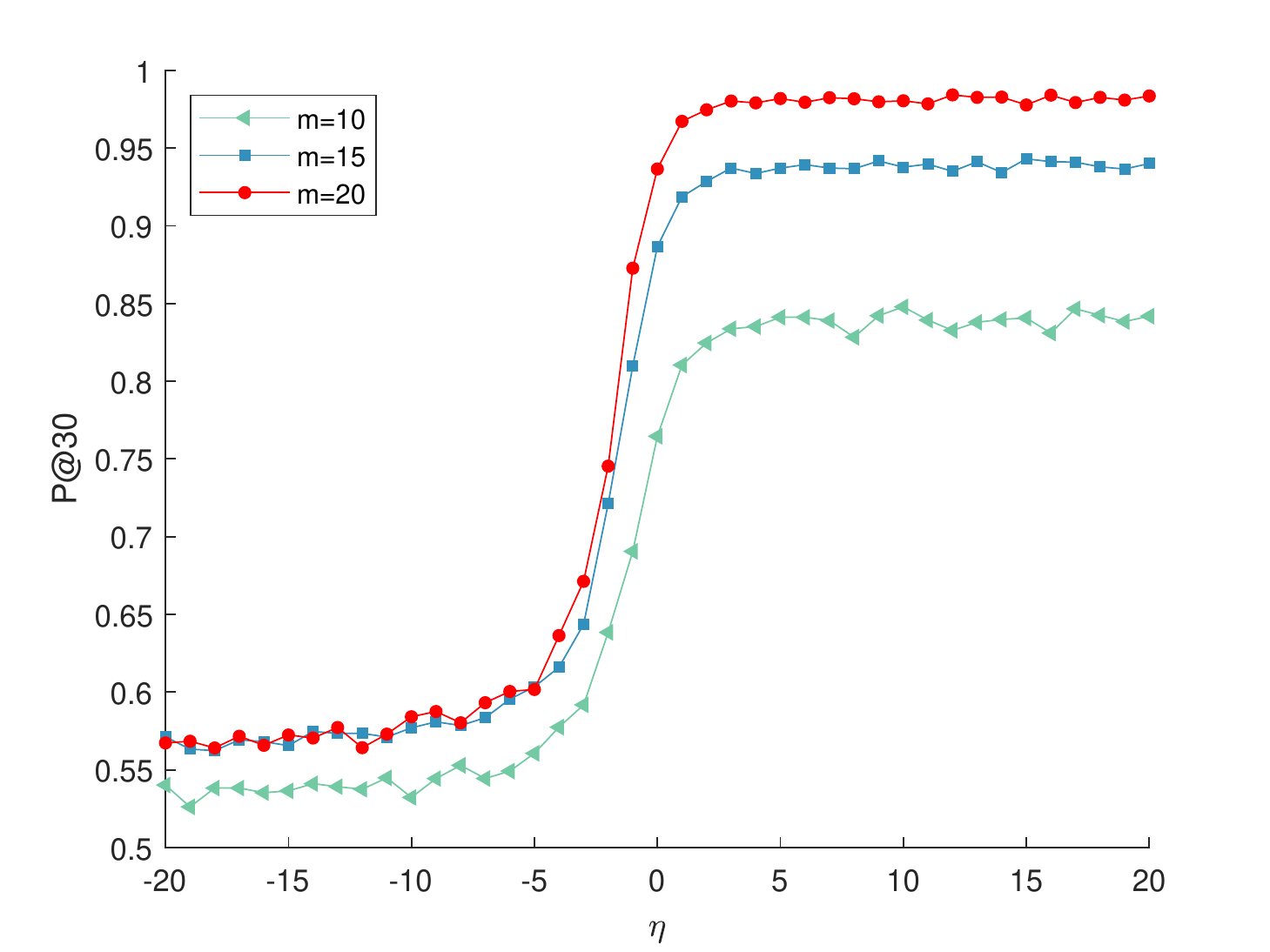}
\end{minipage}%
}%
\subfigure[Results in different node overlap]{
\begin{minipage}[t]{0.33\linewidth}
\centering
\includegraphics[width=4.5cm]{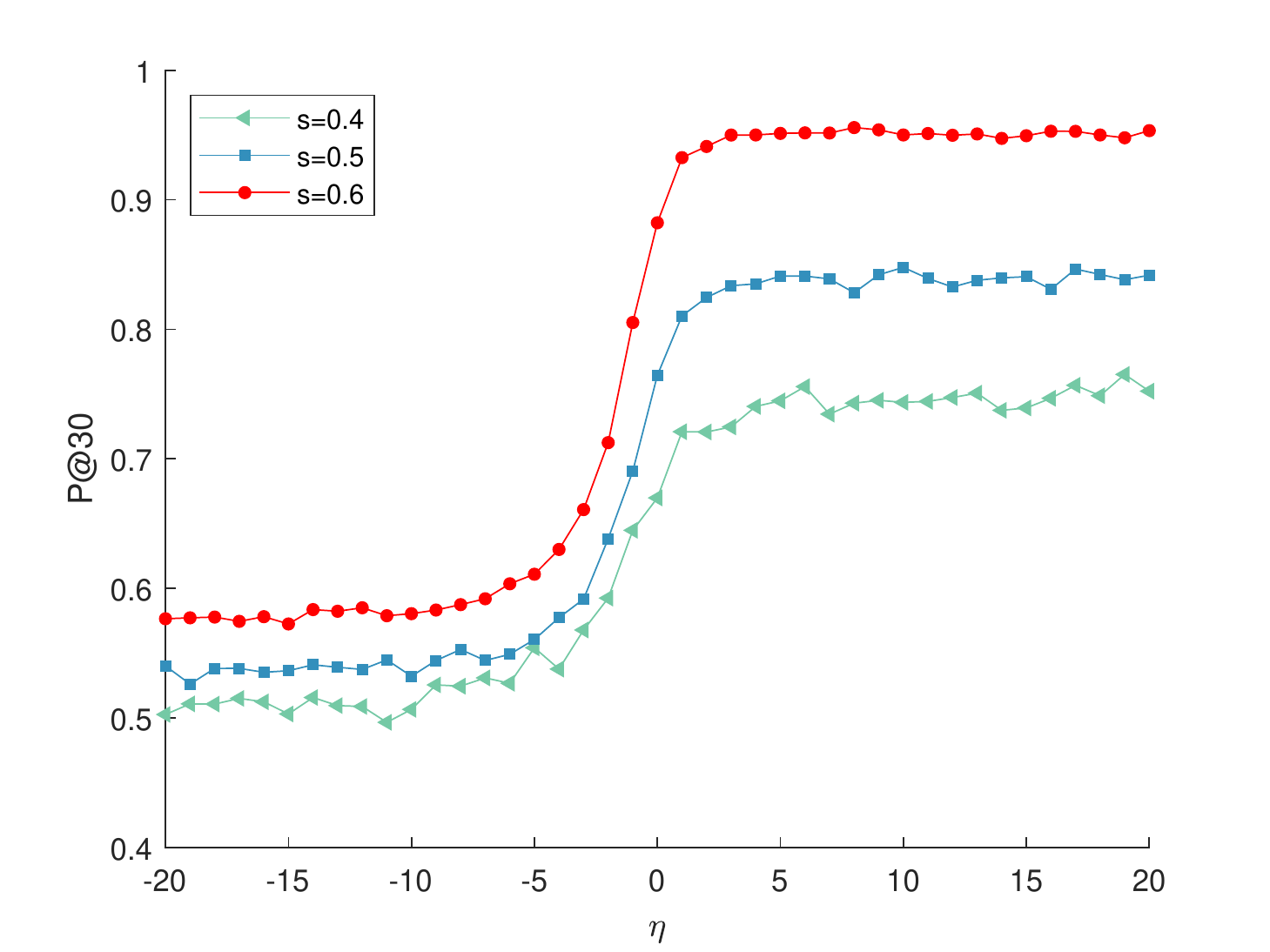}
\end{minipage}%
}%
\subfigure[Results in different network size]{
\begin{minipage}[t]{0.33\linewidth}
\centering
\includegraphics[width=4.5cm]{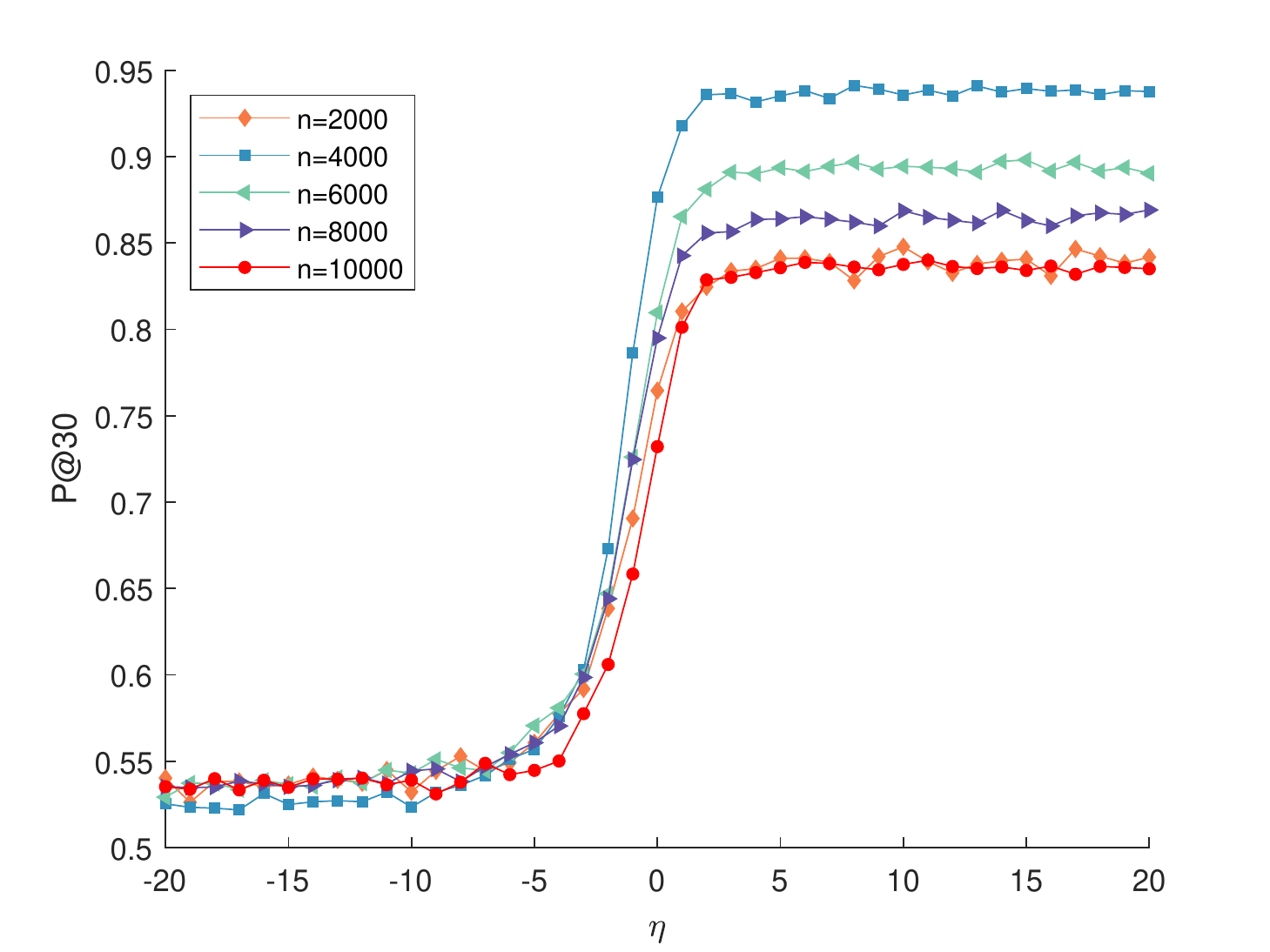}
\end{minipage}%
}%
\centering
\caption{The average values of $P@30$ of different interlayer link prediction algorithms on the BA artificial multiplex networks perturbed by the global perturbation strategy with different $\eta$.}
\label{pics_BA_d1}
\centering
\subfigure[Results in different average degree]{
\begin{minipage}[t]{0.33\linewidth}
\centering
\includegraphics[width=4.5cm]{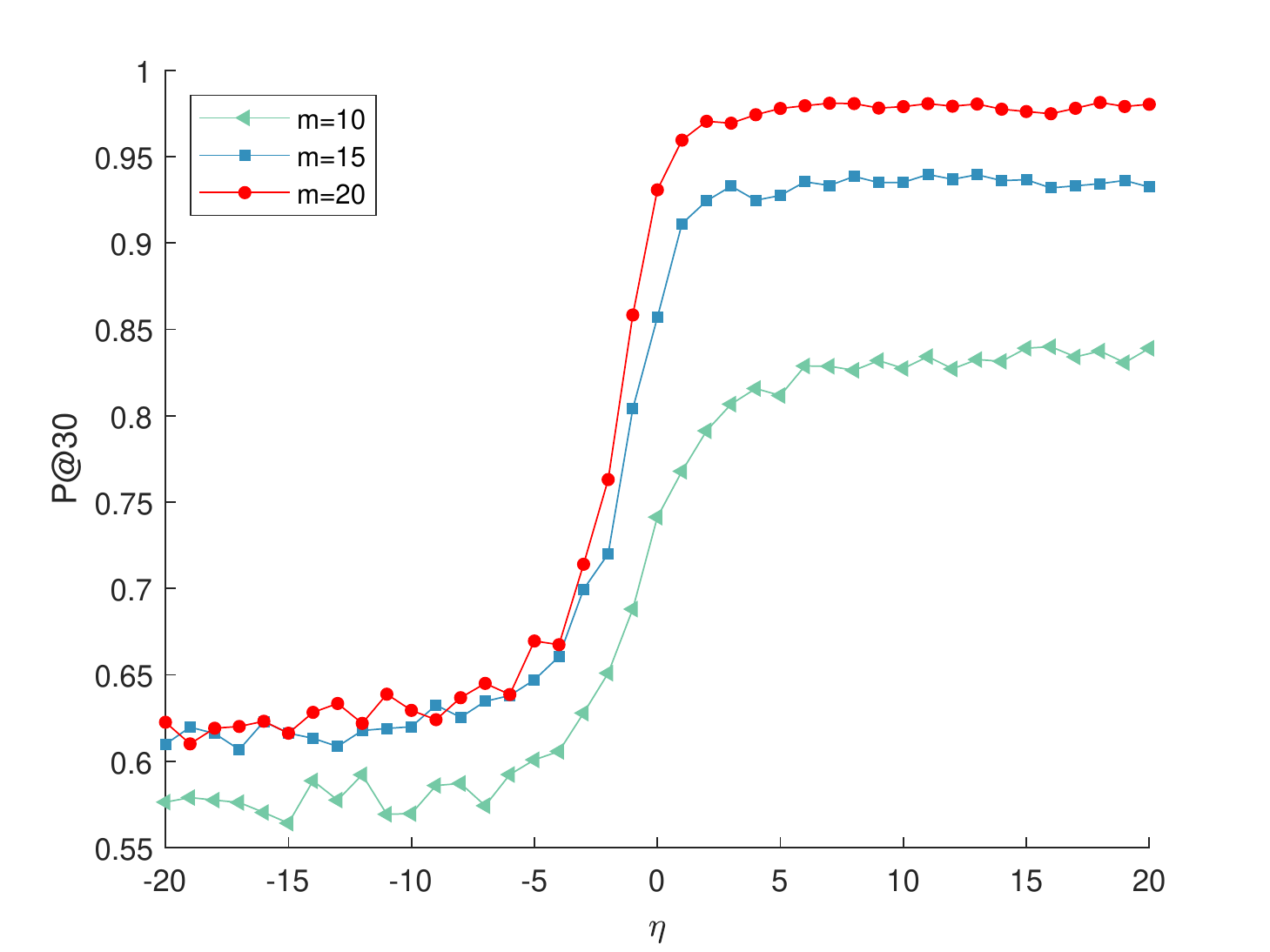}
\end{minipage}%
}%
\subfigure[Results in different node overlap]{
\begin{minipage}[t]{0.33\linewidth}
\centering
\includegraphics[width=4.5cm]{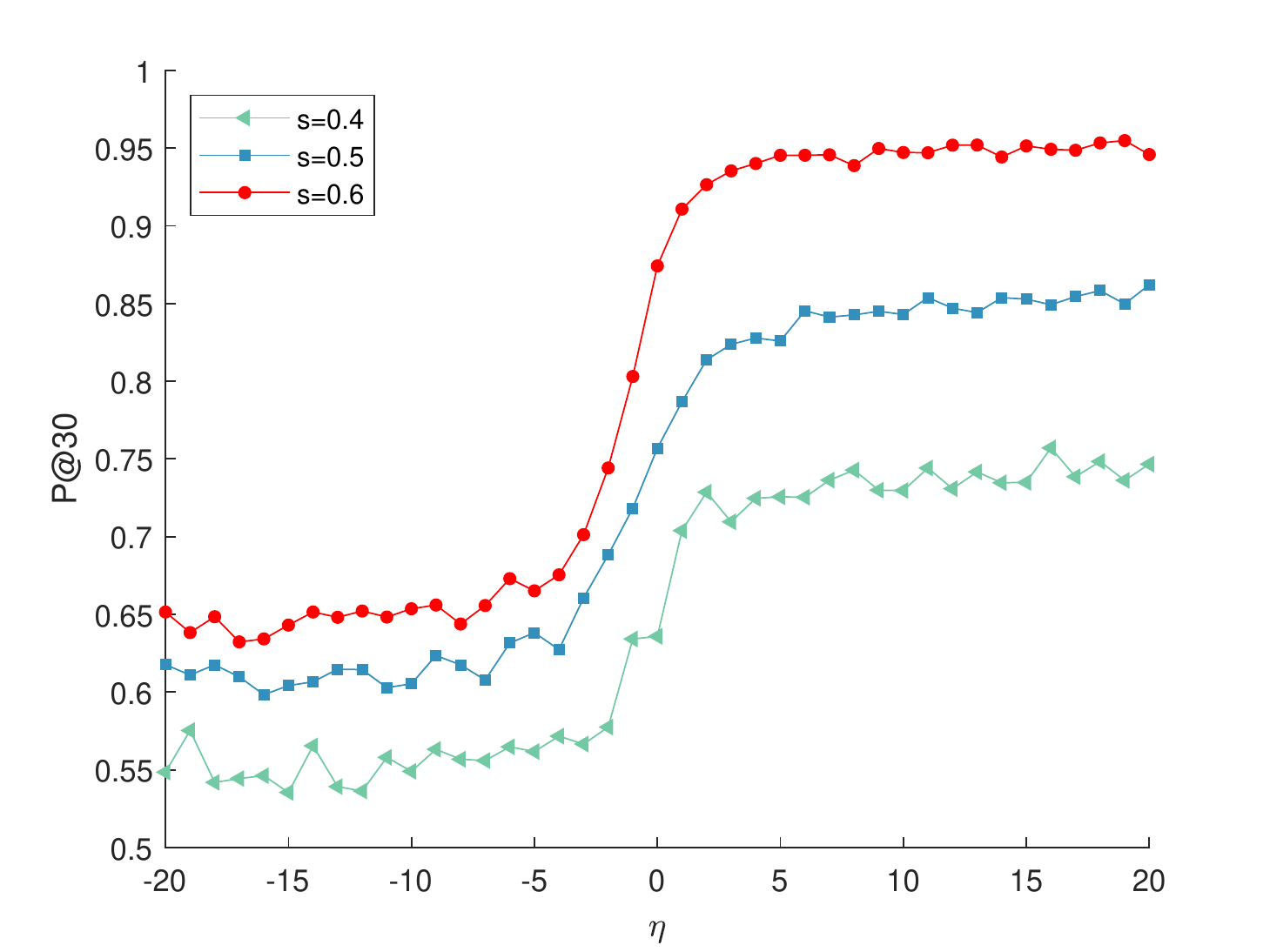}
\end{minipage}%
}%
\subfigure[Results in different network size]{
\begin{minipage}[t]{0.33\linewidth}
\centering
\includegraphics[width=4.5cm]{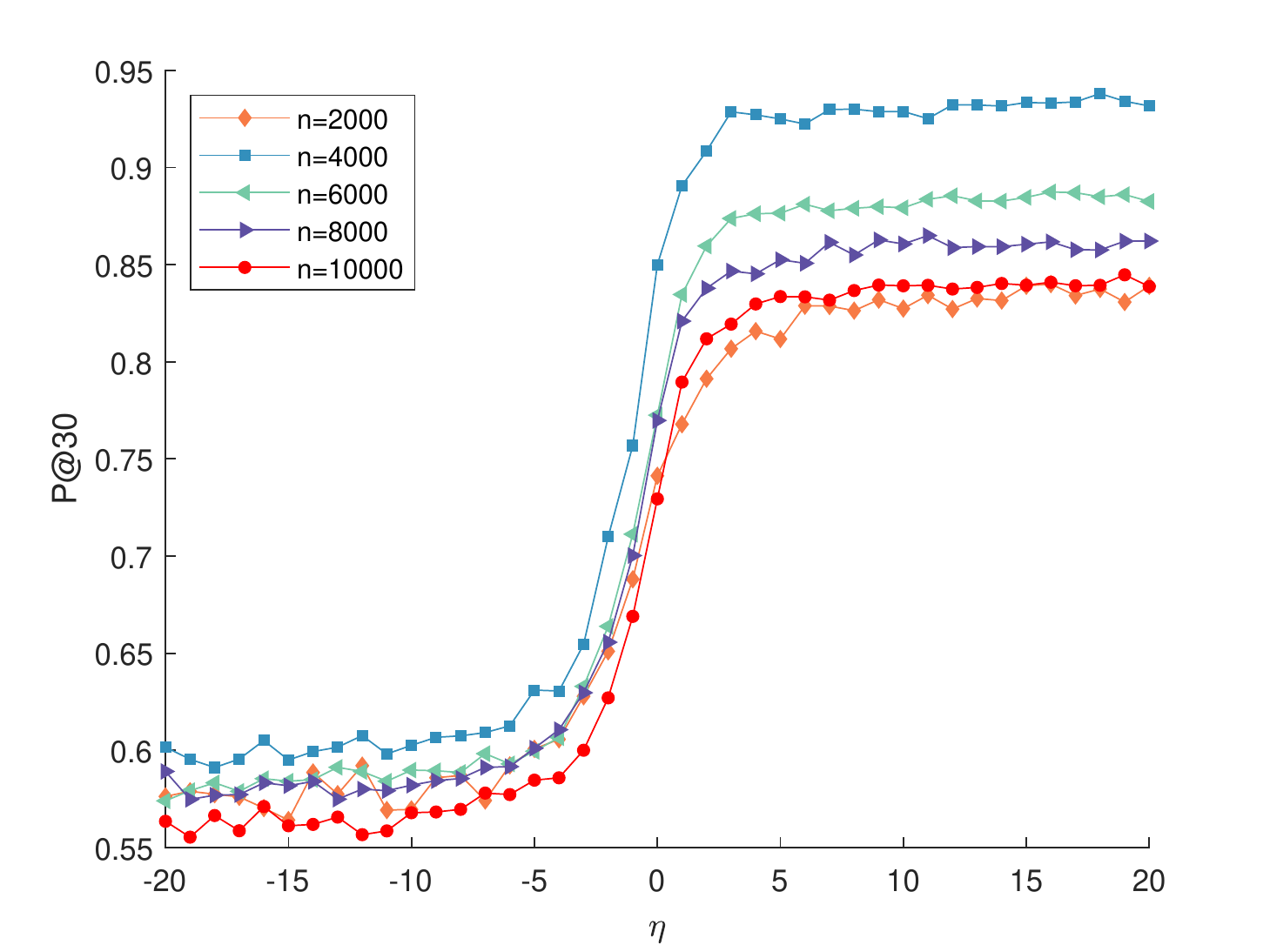}
\end{minipage}%
}%
\centering
\caption{The average values of $P@30$ of different interlayer link prediction algorithms on the BA artificial multiplex networks perturbed by the local perturbation strategy with different $\eta$.}
\label{pics_BA_d2}
\end{figure*}

From the two figures, we can see that the curves of $P@30$ with different training ratios exhibit the same trend. The curves have almost no change firstly, then increase rapidly, and finally becomes stable with an increase in $\eta$. This trend is the same as that in Figs.~\ref{pic:result_real_diffeta_dmeth1} and ~\ref{pic:result_real_diffeta_dmeth2}. These phenomena support the conclusion we have drawn in the previous section. Meanwhile, we can see that under the same $\eta$, the greater the training ratio, the greater the $P@30$. This is because the larger the training ratio, the more observed interlayer links provide more predictive clues for prediction algorithms.
\subsubsection{Comparison of global and local perturbation strategy}
We also compare the perturbation capability of the global and local strategies. Fig.~\ref{pics_glo_comp_local_diffeta} shows the comparison results. The dark blue bars are the average values of $P@30$ at $\xi=0.1, 0.2, 0.3,$ and 0.4 by the four prediction algorithms on the four multiplex networks perturbed by the global strategy, while the light blue bares are those by the local strategy. As can be seen from the figure, when $\eta$<0, the average $P@30$ of the local perturbation strategy is greater than that of the global perturbation strategy. It means that the global perturbation strategy has a greater effect on interlayer link prediction in this interval. The global perturbation strategy directly calculates the perturbation weights for all intralayer links according to the value of $\eta$. When $\eta<0$, the weights of intralayer links connected to nodes with low degree nodes are larger than other intralayer links. The other type of intralayer links, such as the intralayer link between a node with a high degree and a high degree, will seldom be removed.
In contrast, the local perturbation strategy takes a biased random walk procedure. The walker will randomly select any node as starting node many times. If the starting node is a node with a high degree and all neighbors of the starting node are nodes with a high degree, the intralayer link between a node with a high degree and a high degree will be removed in this situation. Therefore, when $\eta<0$, the average $P@30$ of the local perturbation strategy is greater than that of the global perturbation strategy.

When $\eta>0$, the average $P@30$ of the local perturbation strategy is less than that of the global perturbation strategy, the regional perturbation strategy has a more significant effect on the prediction of interlayer links in this interval.
Similarly, because the global perturbation strategy directly calculates the perturbation weights for all intralayer links according to the value of $\eta$, the weights of intralayer links connected to nodes with high degree nodes are larger than other intralayer links when $\eta>0$. The other type of intralayer links, such as intralayer links between a node with a low degree and a low degree, will seldom be removed. In contrast, the local perturbation strategy takes a biased random walk procedure. The walker will randomly select any node as starting node many times. If the starting node is a node with a low degree and all neighbors of the starting node are nodes with a low degree, the intralayer link between a node with a low degree and a node with a low degree will be removed in this situation. Therefore, when $\eta>0$, the average $P@30$ of the global perturbation strategy is greater than the local perturbation strategy. The above phenomena further indicate that the intralayer links connected to nodes with low degree nodes significantly affect the interlayer link prediction.

\subsection{Results on artificial datasets}
To further verify the accuracy of the conclusions drawn from the four real network data sets, we carry out experiments with nine kinds of BA~\cite{Barabasi1999-BA} artificial multiplex networks in different average degrees, node overlap, and network size.

For experiments with a different average degree, we set $m=10, 15,$ and $20$, network size to 2000, and node overlap to 0.5 to generate the BA artificial multiplex networks. For experiments with different node overlaps, we set node overlap to $0.4, 0.5,$ and $0.6$, network size to 2000, $m$ to 10 to generate the multiplex networks. For experiments with different network sizes, we set network size to 2000, 4000, 6000, 8000, and 10000, node overlap to $0.5$, and $m$ to 10 to generate the multiplex networks. The training ratio is 0.9 and $\xi=0.4$ for these experiments.

Figure~\ref{pics_BA_d1} and~\ref{pics_BA_d2} are the experimental results of global and local strategies on the above settings, respectively. From the two figures, we can see that the curves of $P@30$ with different settings exhibit the same trend. The curves have almost no change firstly, then increase rapidly, and finally becomes stable with an increase in $\eta$. This trend is the same as that in Figs.~\ref{pic:result_real_diffeta_dmeth1} and~\ref{pic:result_real_diffeta_dmeth2}. These phenomena support the conclusion we have drawn in the previous section again.

\subsection{Compared with other perturbation methods}
To evaluate the effectiveness of the proposed two strategies, we compare them with four perturbation methods focusing on single network analysis tasks since there are still few researches on the network structural perturbation for interlayer link prediction tasks. These four baselines are as follows.
\begin{figure} [!t]
    \centering
    \includegraphics[width=0.8\textwidth]{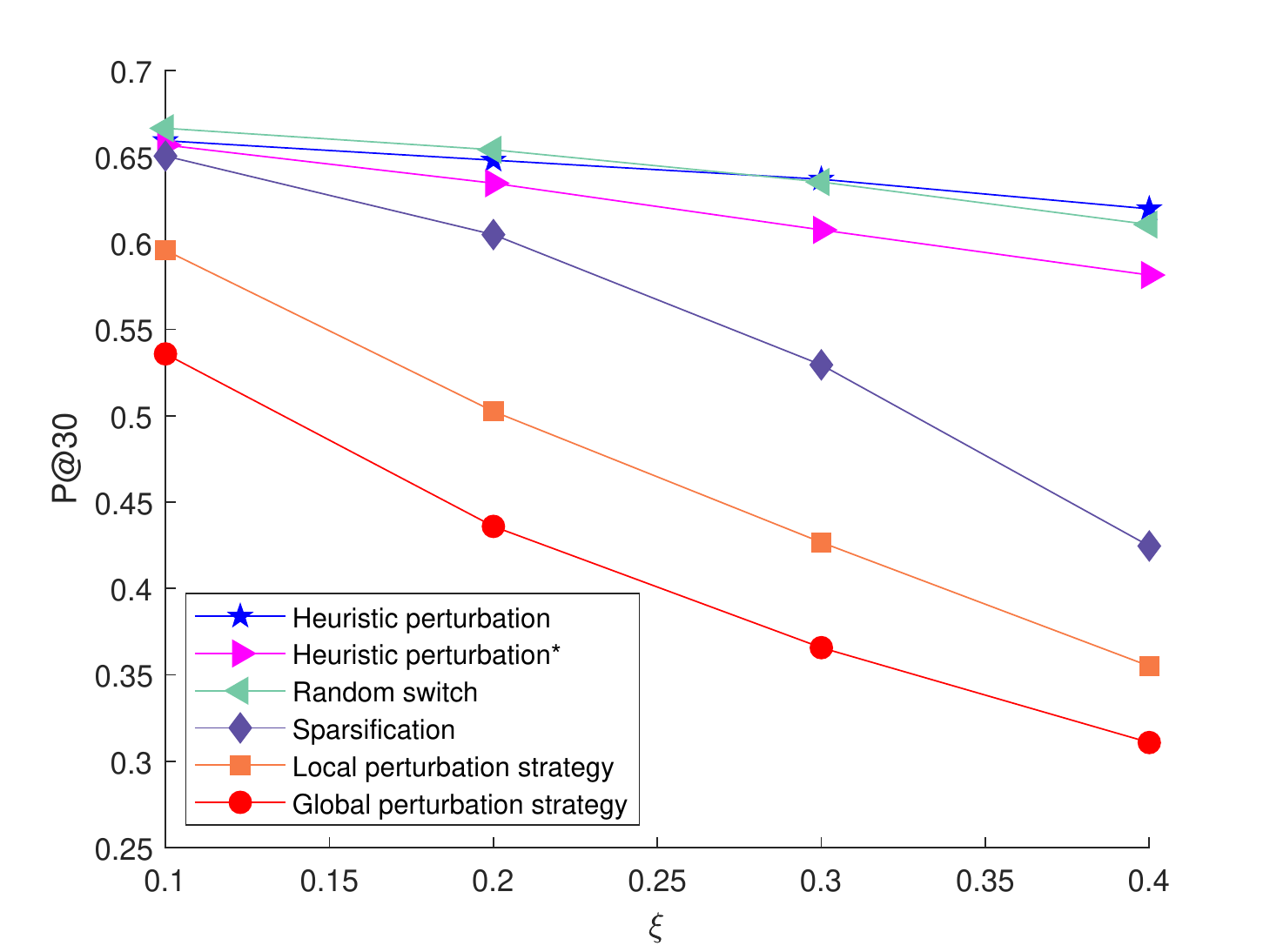}
    \caption{Comparing with the other perturbation methods on different $\xi$. A smaller P@30 indicates a better performance of structural perturbation.}
    \label{pic:comparewithbaseline}
\end{figure}

\begin{figure*}
\centering
\subfigure[Different operations for global perturbation strategy]{
\begin{minipage}[t]{0.45\linewidth}
\centering
\includegraphics[width=6cm]{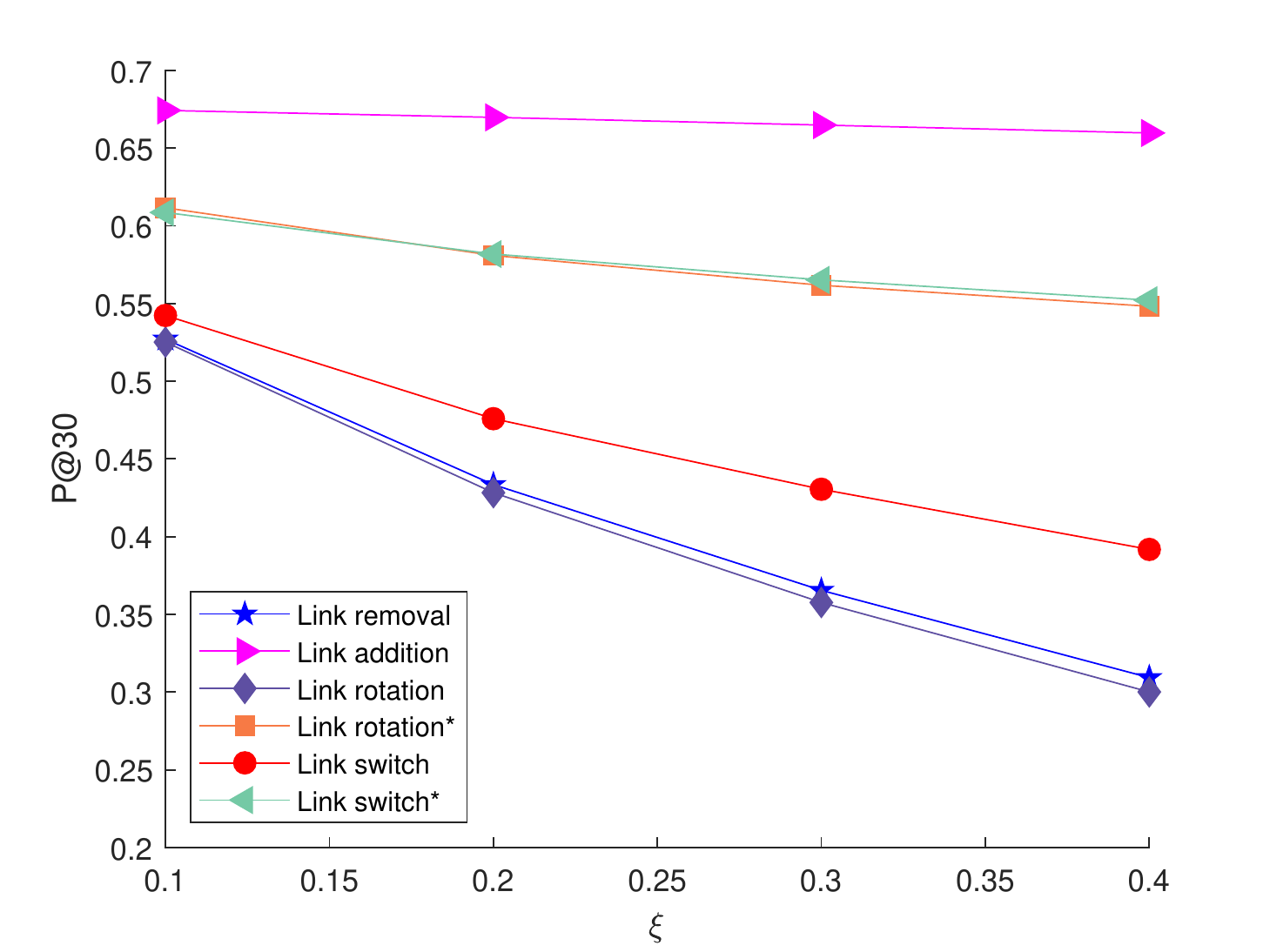}
\end{minipage}%
}%
\subfigure[Different operations for local perturbation strategy]{
\begin{minipage}[t]{0.45\linewidth}
\centering
\includegraphics[width=6cm]{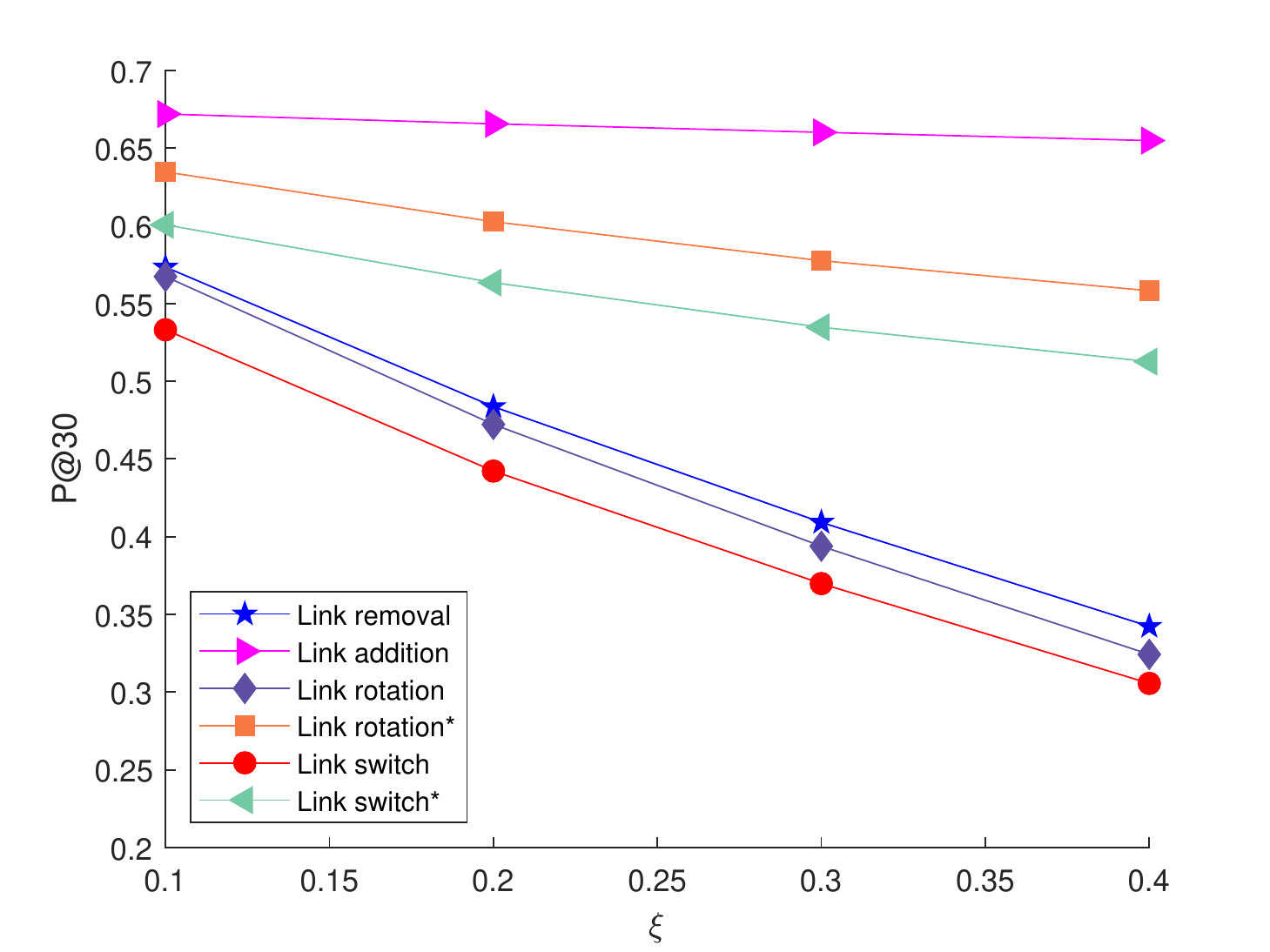}
\end{minipage}%
}%
\centering
\caption{Average $P@30$ of different basic operations.}
\label{pics_morebasicopr}
\end{figure*}

\begin{itemize}
\item \textbf{Heuristic perturbation~\cite{yu2019target}}: This method is used to perturb link prediction within a single network. It divide node pairs into three cases, i.e., node pairs in the training set, node pairs in the validation set, and node pairs in the non-existent node pair set. To hide the links in the training set, this method degrade the performance of link prediction through decreasing the RA~\cite{zhou2009predicting} scores of node pairs in the training set and increasing the scores of non-existent node pairs. 
\item \textbf{Heuristic perturbation*}: This method is a simplified version of the Heuristic perturbation which only remove the links in the training set.
\item \textbf{Sparsification method~\cite{bonchi2014identity}}: This method aims to protect the privacy of the network by selecting an anonymization level $p \in [0,1]$ to remove links. For each link in a network, it calculates an independent Bernoulli trial with probability $p$ to determine whether or not to remove this link.
\item \textbf{Random switch algorithm~\cite{hay2007anonymizing}}: This method randomly selects two exist links $(u_i,u_j)$ and $(v_a,v_b)$ from a network. If $(u_i,v_b)$ and $(v_a,u_j)$ are not exist, it switch the links by removing $(u_i,u_j)$ and $(v_a,v_b)$, and adding $(u_i,v_b)$ and $(v_a,u_j)$.
\end{itemize}

\begin{figure*}
\centering
\subfigure[$P@30$ on different $\xi$]{
\begin{minipage}[t]{0.33\linewidth}
\centering
\includegraphics[width=4.5cm]{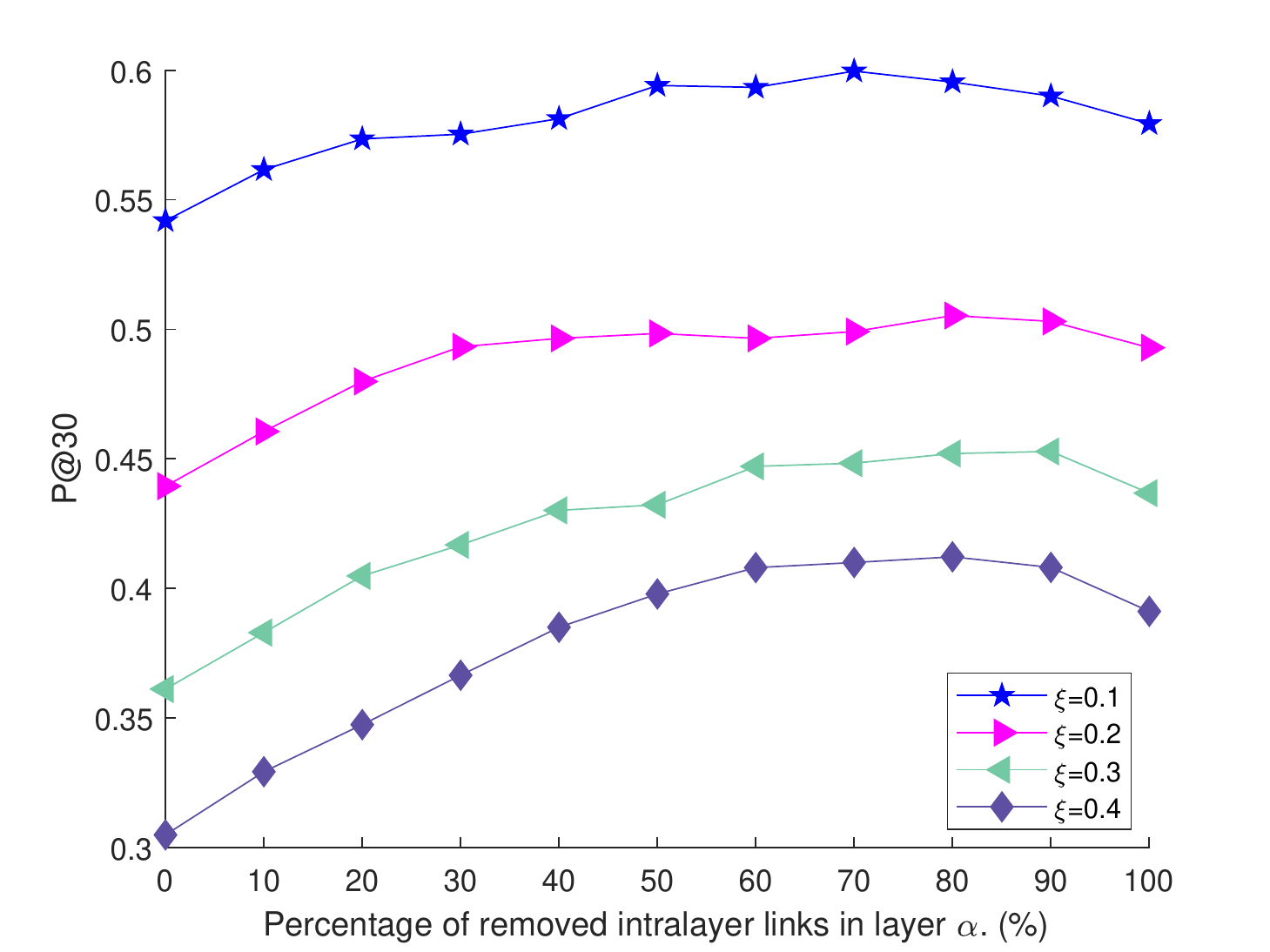}
\end{minipage}%
}%
\subfigure[$P@30$ on different interlayer link prediction algorithms]{
\begin{minipage}[t]{0.33\linewidth}
\centering
\includegraphics[width=4.5cm]{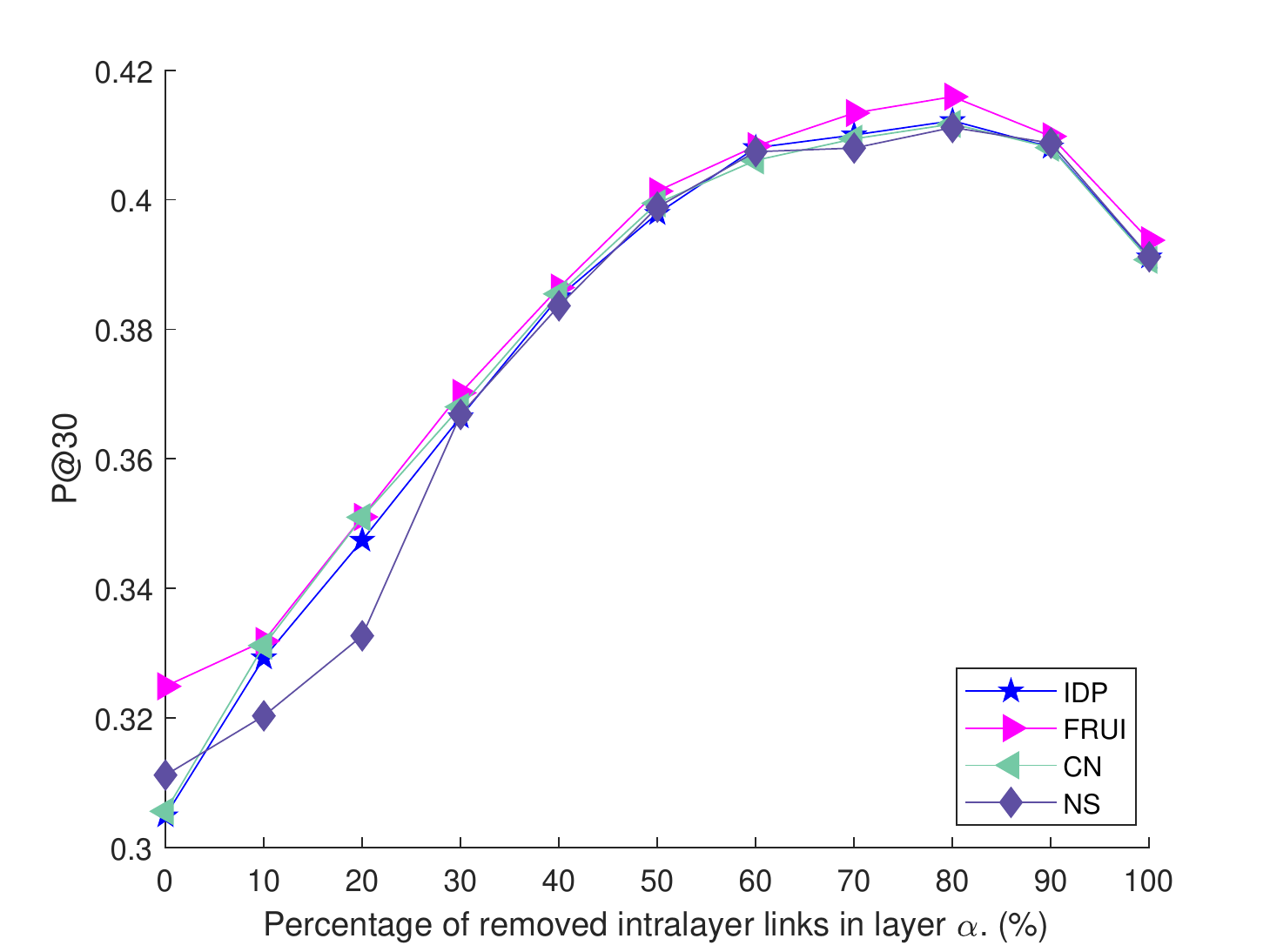}
\end{minipage}%
}%
\subfigure[$P@30$ on different types of perturbation strategies]{
\begin{minipage}[t]{0.33\linewidth}
\centering
\includegraphics[width=4.5cm]{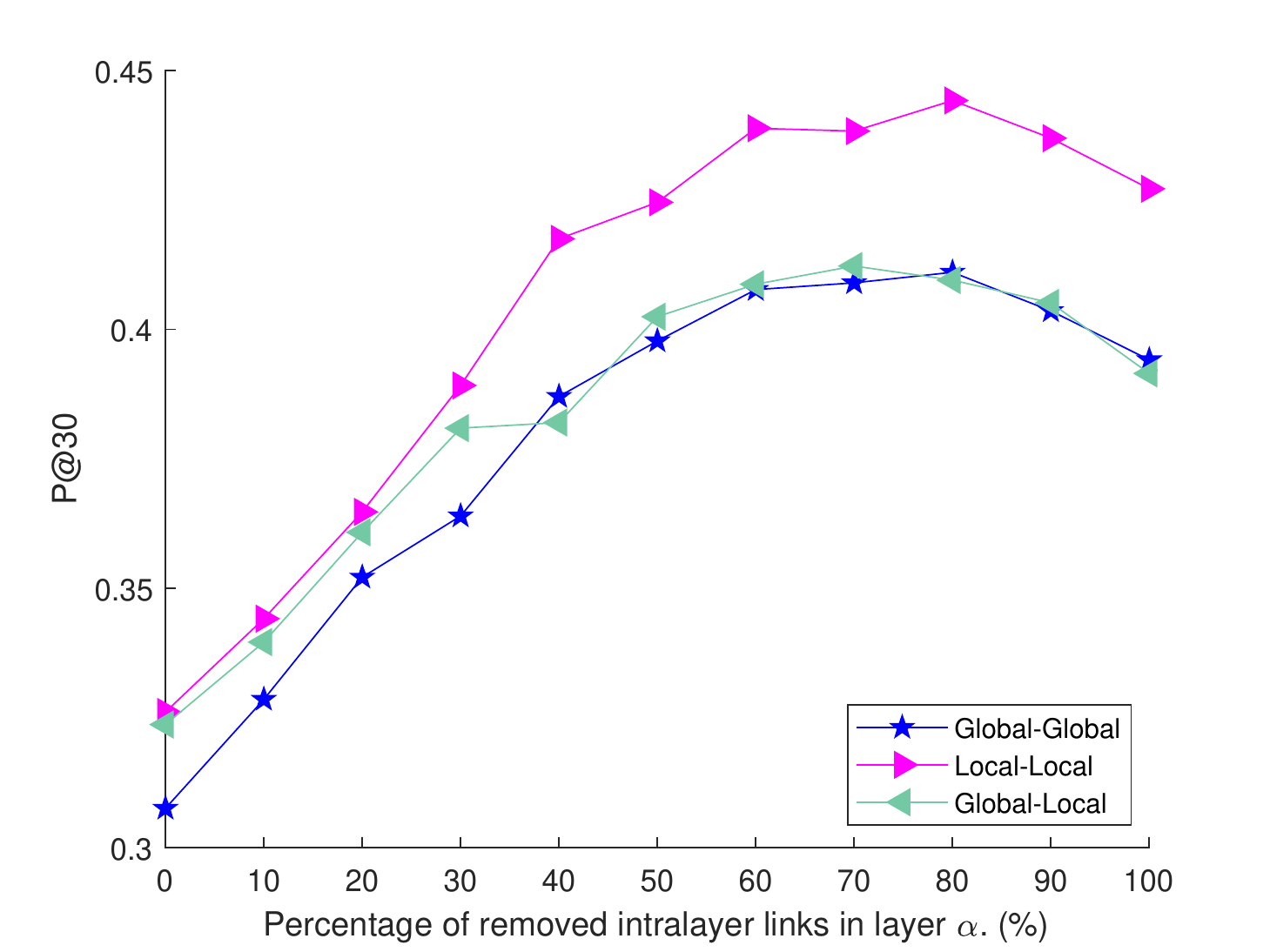}
\end{minipage}%
}%
\centering
\caption{$P@30$ of perturbing multiple layers simultaneously.}
\label{pics_interlayerlink}
\end{figure*}

Figure~\ref{pic:comparewithbaseline} shows the comparison results. The lines are the average values of $P@30$ at $\xi=0.1, 0.2, 0.3,$ and 0.4 by the four prediction algorithms on the four multiplex networks perturbed by different perturbation methods. A smaller P@30 in the figure indicates a better performance of structural perturbation. The value of $\eta$ for global and local perturbation strategy is $-10$ based on the above experimental results. As can be seen from the figure, the global perturbation strategy achieves the lowest $P@30$ for all $\xi$ settings. This phenomenon demonstrates that the global perturbation strategy has the greatest perturbation effect for interlayer link prediction. The local perturbation strategy achieves the second-lowest $P@30$ for all $\xi$ settings. It performs worse than the global perturbation strategy. The rankings of the perturbation performance of these two strategies and the reasons are the same as those illustrated in Fig.~\ref{pics_glo_comp_local_diffeta}. The Sparsification method shows a worse perturbation performance than the local perturbation strategy. It is a random perturbation method. All intralayer links will be removed with the same probability since the Sparsification method selects intralayer links through Bernoulli trail. It means that the probability of removing the intralayer links connected to low degree nodes by Sparsification method is less than that of the local perturbation strategy. Therefore, the Sparsification method shows a worse perturbation performance than the local perturbation strategy. The Heuristic perturbation* method sorts node pairs in descending order according to their RA scores and then remove the intralayer links between nodes with high RA scores. It performs worse than the random perturbation method. This may indicate that the perturbation method for a single network analysis task may not be suitable for the interlayer link prediction task in multiplex networks. Heuristic perturbation and Random switch are the methods based on replacing the intralayer links. They achieve the highest $P@30$ so that they are the worst two perturbation methods for interlayer link prediction. This indicates that the replacement strategies are more difficult to achieve good perturbation performance.

\subsection{Other experiments}
In the subsection, We conduct experiments to compare the different basic operations of perturbation, evaluate the performance of perturbing multiple networks simultaneously, analyze the influence of interlayer links, and compare the quasi-local perturbation strategy with global and local perturbation strategies discussed in subsection 4.3.

\subsubsection{Comparison of different basic operations for perturbation}
Figure~\ref{pics_morebasicopr} is the average values of $P@30$ of different basic operations for perturbation at $\xi=0.1, 0.2, 0.3,$and 0.4 by the four prediction algorithms on the four multiplex networks. Since the link rotation and link switch involve the two operations of removing selected links and adding new links, they both have higher cost than directly deleting intralayer links. The two curves of link rotation and link switch in Figure~\ref{pics_morebasicopr} represent the experimental results when the number of operated intralayer links is the same as the number of operated intralayer links of link removal. And the curves of link rotation* and link switch* are experimental results when the cost of them are the same as that of link removal, assuming that both removing and adding  intralayer links have the same cost. As can be seen from the figure, as for the global perturbation strategy, the link rotation operation exhibits better perturbation performance than merely removing intralayer links when operating the same number of intralayer links. And for the local perturbation strategy, both link rotation and link switch operations show better performance than link removal operations when operating the same number of intralayer links. The link rotation and link switch operations should theoretically have better perturbation performance than the link removal operation, because they both add perturbation information on the basis of removing the selected intralayer links. However, the link switch operation shows worse performance than link removal operation in the global perturbation strategy. This is may be because the global perturbation strategy selects intralayer links by probability and the perturbed intralayer links still have a high probability to be selected again.
\begin{figure*}
\centering
\subfigure[$P@30$ on different $\xi$]{
\begin{minipage}[t]{0.33\linewidth}
\centering
\includegraphics[width=4.5cm]{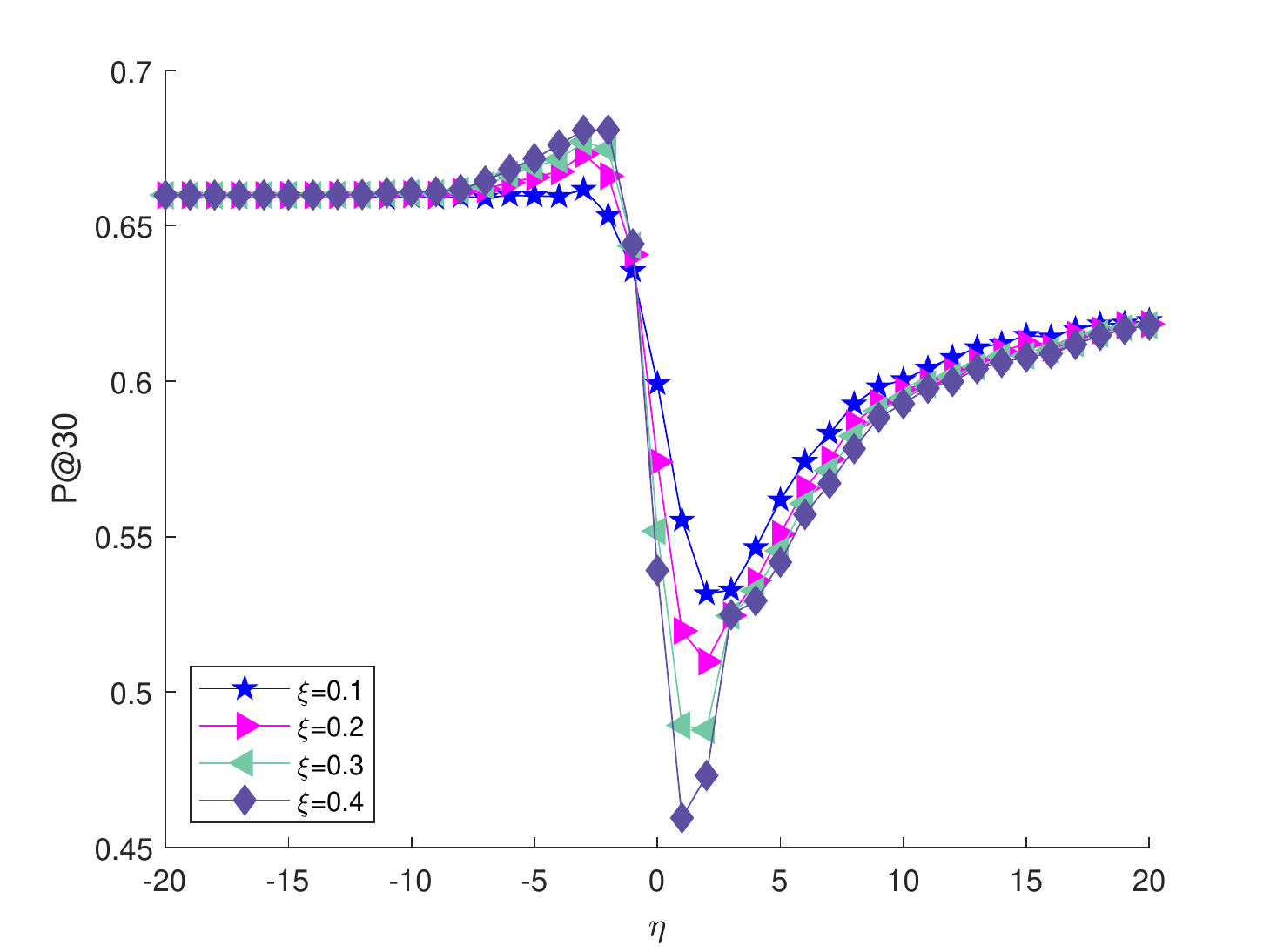}
\end{minipage}%
}%
\subfigure[$P@30$ on different interlayer link prediction algorithms]{
\begin{minipage}[t]{0.33\linewidth}
\centering
\includegraphics[width=4.5cm]{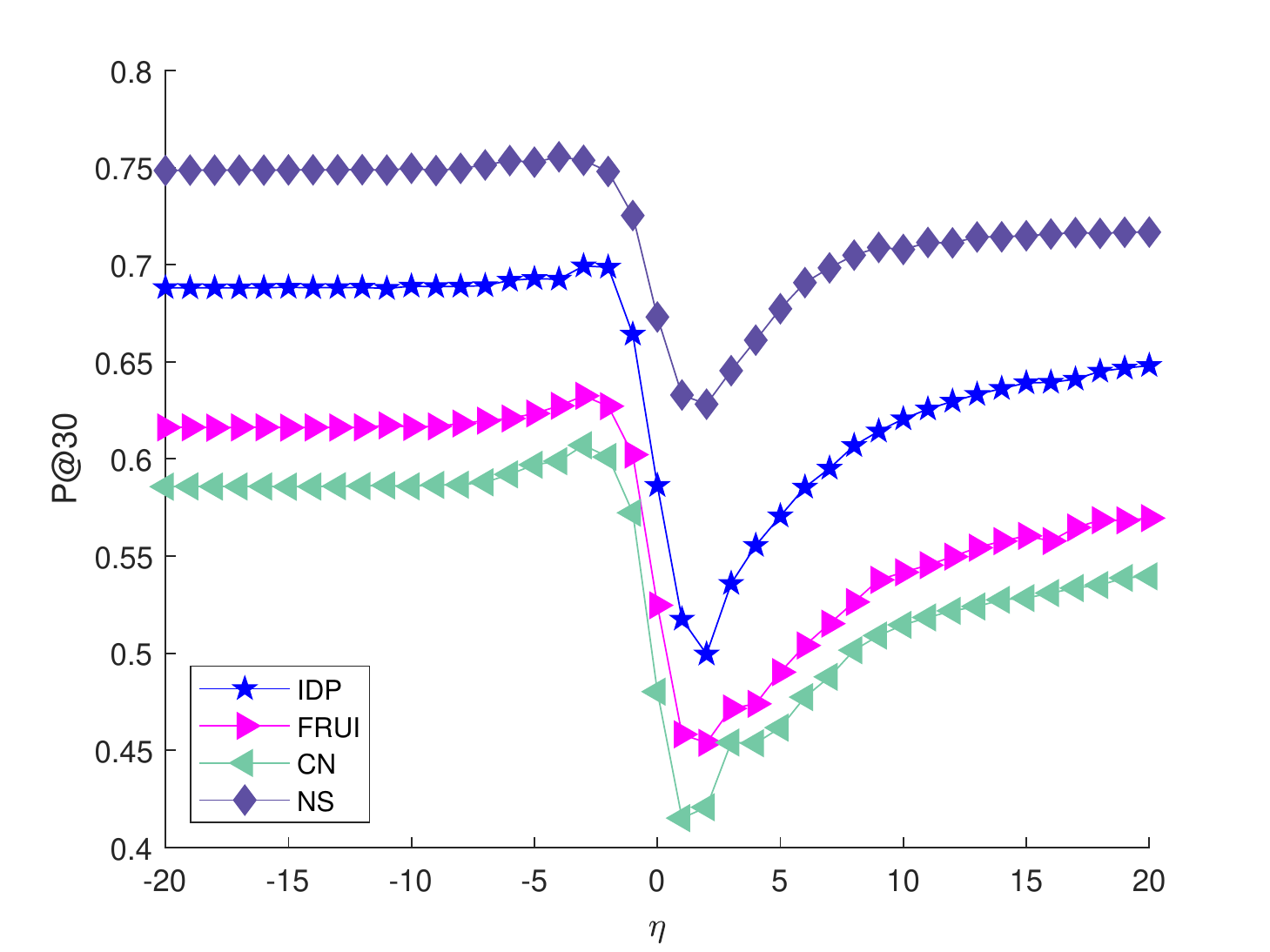}
\end{minipage}%
}%
\subfigure[$P@30$ on different datasets]{
\begin{minipage}[t]{0.33\linewidth}
\centering
\includegraphics[width=4.5cm]{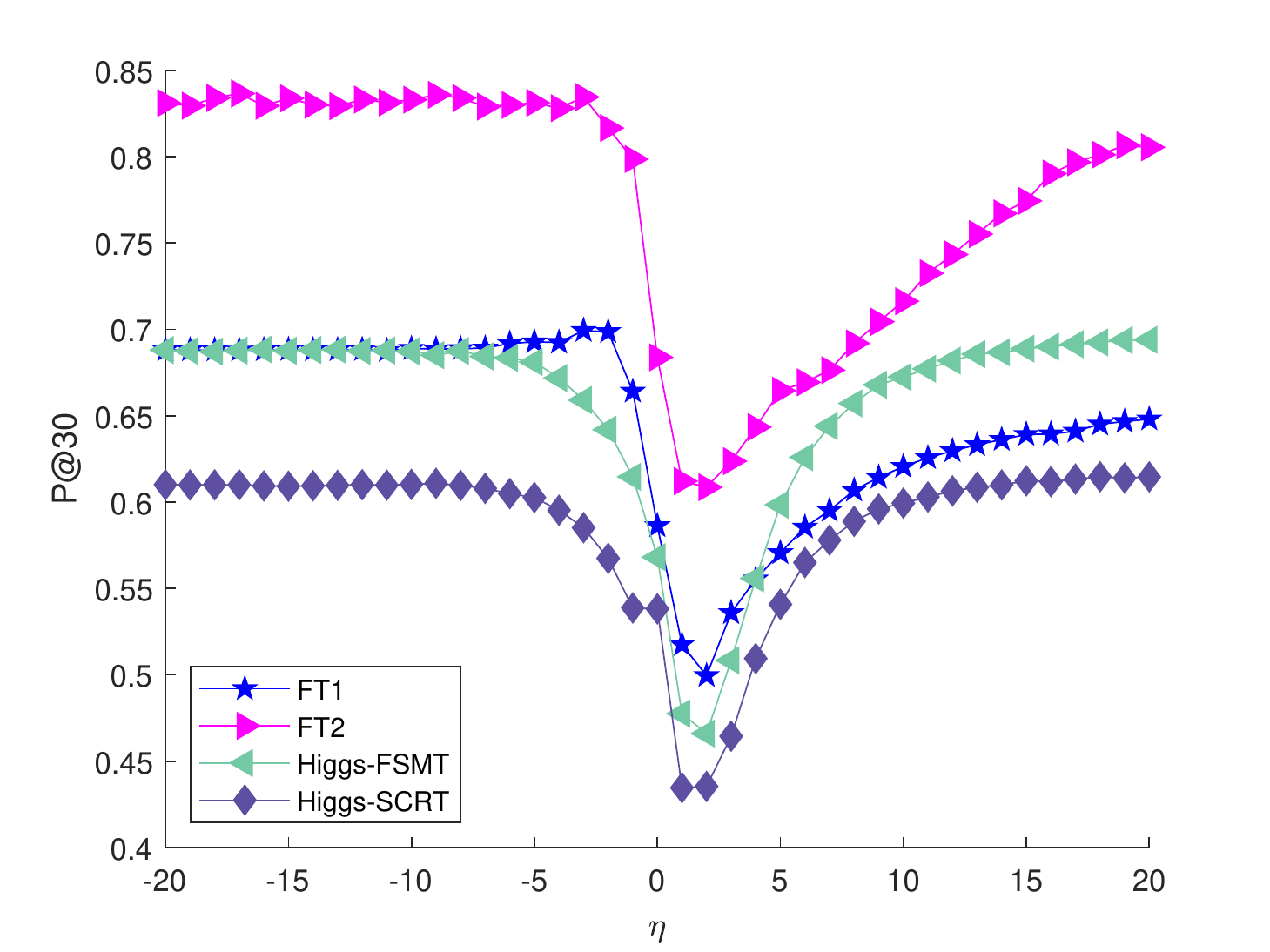}
\end{minipage}%
}%
\centering
\caption{$P@30$ of different interlayer link prediction algorithms on the multiplex networks by perturbing the observed interlayer links.}
\label{pics_pertmultilayers}
\end{figure*}

\begin{figure} [!t]
    \centering
    \includegraphics[width=0.8\textwidth]{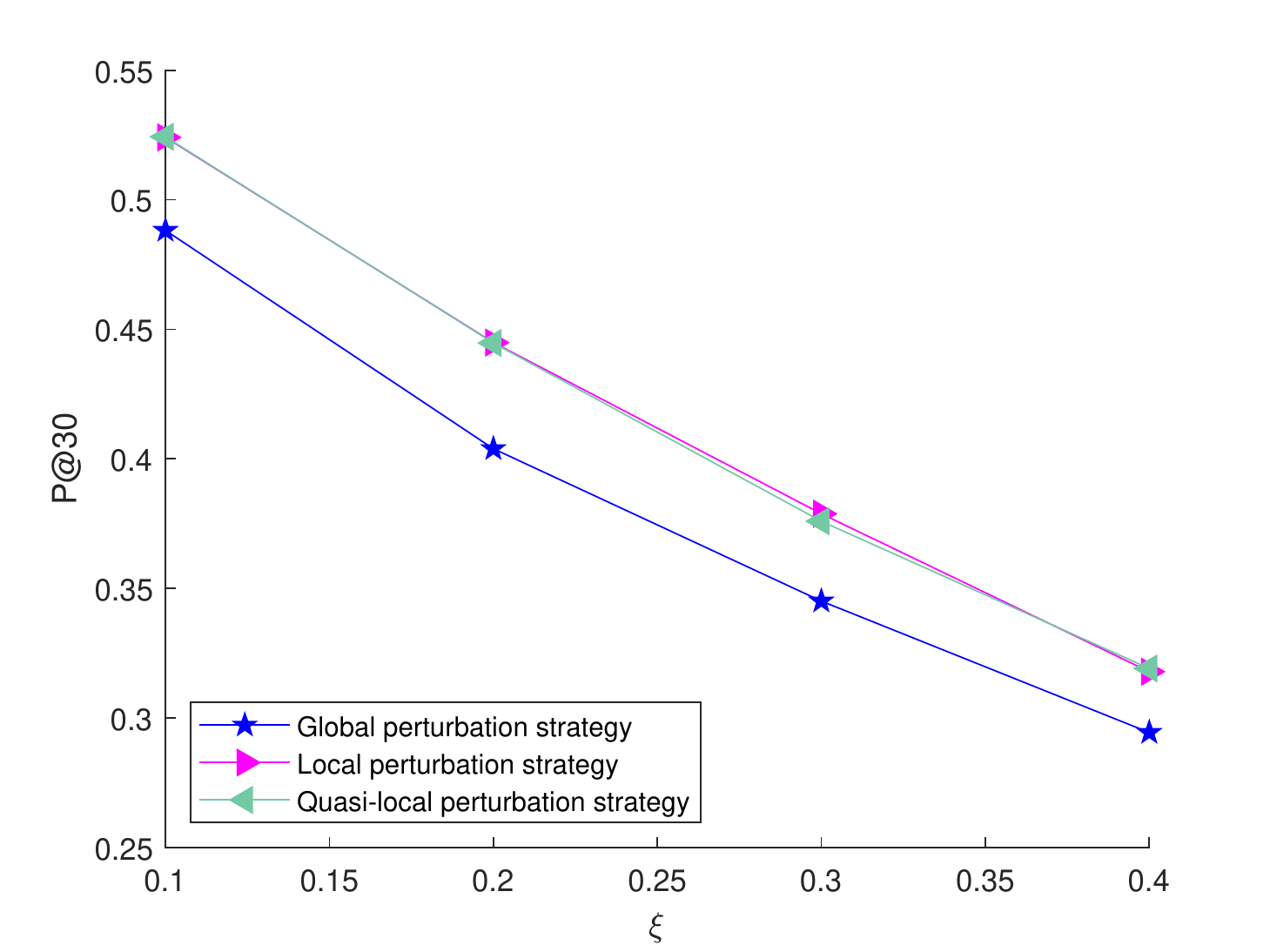}
    \caption{Comparison of the quasi-local perturbation strategy with the global and local perturbation strategies.}
    \label{pic_gloandlocw}
\end{figure}

\subsubsection{Results of perturbing multiple layers simultaneously}
Figure~\ref{pics_pertmultilayers} is the $P@30$ of perturbing multiple layers simultaneously and its abscissa represents the proportion of the perturbed intralayer links in layer $\alpha$ under a certain number of total perturbed links. All the curves in the figure show a trend of increasing first and then decreasing. This demonstrates that only perturbing one layer of the multiplex network will achieve better perturbation performance when the number of perturbed intralayer links is fixed. This is reasonable because removing all intralayer links in one layer clearly outperforms removing half of the intralayer links in each layer. It is worth noting that although simultaneously perturbing two layers of the network is not as effective as perturbing only one layer when the total numbers of perturbed intralayer links are the same. But this does not mean that there is no value in simultaneously perturbing multiple layers, because  perturbation also should take into account the availability of the social network itself. Experimental results of using different perturbation strategies and the same perturbation strategy in multiple layers are shown in Figure~\ref{pics_pertmultilayers} (c). The performance of perturbing both layers with a global perturbation strategy outperforms that of one layer with a global perturbation strategy and the other with a local perturbation strategy. Among the different types of perturbation strategies in two layers, perturbing both layers with a local perturbation strategy performs the worst. The reason for this phenomenon is the same as that of Fig.~\ref{pics_glo_comp_local_diffeta}.

\subsubsection{Results of perturbing the observed interlayer links}
Figure~\ref{pics_interlayerlink} is the $P@30$ of different interlayer link prediction algorithms on the multiplex networks by perturbing the observed interlayer links. Different form perturbing the intralayer links, $P@30$ of perturbing the observed interlayer links exhibit a trend that it has almost no change firstly, then decreases rapidly and finally increases with an increase in $\eta$. The differences in datasets, prediction algorithms, and $\xi$ did not change this trend. The values of $\eta$ for the minimum $P@30$ in the subfigures are almost the same, always 2 or 3. These phenomena illustrate that the interlayer links connected to nodes with middle degree have a greater impact than the other types of interlayer links. According to the experimental conclusions in this subsection and subsection 5.3, we can combine intralayer links and interlayer links to execute the perturbation by taking $\eta$ to -10 to select intralayer links and taking $\eta$ to 3 to select interlayer links.

\subsubsection{Results of quasi-local perturbation strategy}
Figure~\ref{pic_gloandlocw} is the average $P@30$ of interlayer link prediction algorithms on the four datasets under quasi-local, global, and local perturbation strategies when $\eta=-10$. As can be seen from the figure, the performance of quasi-local perturbation strategy is similar with the local perturbation strategy while worse than the global perturbation strategy. The ranking of the performance of global and local strategies and the reasons are the same as those illustrated in Fig.~\ref{pics_glo_comp_local_diffeta}. Quasi-local perturbation strategy is similar with the local perturbation strategy. The difference of them is that the quasi-local strategy calculates the perturbation weights for the biased random walk based on one- and two-hop neighbors. The essence of these two strategies is similar. Therefore, their performances are similar. It is worthwhile noting that this paper focuses on study the influence of structural information on the interlayer link prediction task by global and local strategies. Other possible quasi-local strategies can be further studied in the future.

\section{Conclusion}
In this study, we proposed two kinds of structural perturbation strategies to study the effects of intralayer links on interlayer link prediction according to whether the structural information of the whole network is completely known or not by the perturbation executor. By changing the parameter $\eta$, the backbone structure for the interlayer link prediction can be detected. Experiments demonstrated that the intralayer links connected with small degree nodes have the most significant impact on the prediction accuracy of interlayer links. The intralayer links connected with large degree nodes may have side effects on the interlayer link prediction. Our study can investigate what types of intralayer links are most important for a correct prediction, are there any intralayer links whose presence leads to worse predictive performance than their absence, and how to attack the prediction algorithms at the minimum cost, etc. In the future, we plan to explore more ways to implement perturbations, such as add noise intralayer links or replace significant intralayer links to detect backbone structures more accurately.

\section{Acknowledgments}
This work was supported by the National Natural Science Foundation of China (No. U19A2081.), Science and Engineering Connotation Development Project of Sichuan University (No. 2020SCUNG129), and Joint Research Fund of China Ministry of Education and China Mobile Company (No. CM20200409).

%

\end{document}